\documentstyle[12pt]{article}
\makeatletter
\@addtoreset{equation}{section}
\renewcommand{\theequation}{\thesection.\arabic{equation}}
\makeatother
\makeatletter
\@addtoreset{equation}{subsection}

\def\R{\displaystyle \mathop{\bf R}}
\def\RR{\displaystyle \mathop{R}}

\def\V{\displaystyle \mathop{V}}
\def\G1{\displaystyle \mathop{G}}
\def\SG1{\displaystyle \mathop{SG}}
\def\hSG1{\displaystyle \mathop{\hat{SG}}}
\def\e1{\displaystyle \mathop{e}}
\def\U1{\displaystyle \mathop{U}}
\def\P{\displaystyle \mathop{P}}
\def\n1{\displaystyle \mathop{\nu}}
\def\ps1{\displaystyle \mathop{\psi}}
\def\ph1{\displaystyle \mathop{\phi}}
\def\BB{\displaystyle \mathop{\Phi}}
\def\p1{\displaystyle \mathop{p}}

\def\O1{\displaystyle \mathop{O}}

\def\hG{\displaystyle \mathop{\hat{G}}}
\def\gam{\displaystyle \mathop{\gamma}}

\def\pr{\displaystyle \mathop{\partial}}
\def\lpr{\displaystyle \mathop{\overleftarrow{\partial}}}
\def\lD1{\displaystyle \mathop{\overleftarrow{D}}}

\def\hgam{\displaystyle \mathop{\hat{\gamma}}}
\def\hpr{\displaystyle \mathop{\hat{\partial}}}
\def\Dlt{\displaystyle \mathop{\Delta}}
\def\hN{\displaystyle \mathop{\hat{N}}}
\def\he{\displaystyle \mathop{\hat{e}}}
\def\hp1{\displaystyle \mathop{\hat{p}}}
\def\hps{\displaystyle \mathop{\hat{\Psi}}}

\def\bp{\displaystyle \mathop{\bar{\Psi}}}

\def\S{\displaystyle \sum}
\def\IIn{\displaystyle \int}
\def\FFr{\displaystyle \frac}
\def\Lm{\displaystyle \lim}

\def\F{\displaystyle \mathop{F}}
\def\sig1{\displaystyle \mathop{\sigma}}

\def\h1{\displaystyle \mathop{h}}
\def\H1{\displaystyle \mathop{H}}
\def\A1{\displaystyle \mathop{A}}
\def\Asp{\displaystyle \mathop{\hat{A}}}

\def\D1{\displaystyle \mathop{D}}
\def\B1{\displaystyle \mathop{B}}
\def\L1{\displaystyle \mathop{L}}
\def\J1{\displaystyle \mathop{J}}
\def\A1{\displaystyle \mathop{A}}
\def\F1{\displaystyle \mathop{F}}
\def\M1{\displaystyle \mathop{M}}
\def\g1{\displaystyle \mathop{g}}
\def\q1{\displaystyle \mathop{q}}
\def\x1{\displaystyle \mathop{x}}
\def\s1{\displaystyle \mathop{s}}
\def\1Q{\displaystyle \mathop{Q}}
\def\E1{\displaystyle \mathop{E}}
\def\Q{\displaystyle \mathop{Q_{0}}}
\def\ch1{\displaystyle \mathop{\chi}}
\def\PH1{\displaystyle \mathop{\Phi}}
\def\X1{\displaystyle \mathop{X}}
\def\m1{\displaystyle \mathop{m}}
\def\Y1{\displaystyle \mathop{Y}}
\def\Z1{\displaystyle \mathop{Z}}
\def\W1{\displaystyle \mathop{W}}
\def\Bx{\displaystyle \mathop{\Box}}

\begin{document}

\begin{flushright} 
BAO/2000-37
\end{flushright}
\vskip 0.5truecm
\begin{center}
{\Large {\bf Supersymmetric Microscopic Theory}} 
\vskip 0.2truecm
{\Large {\bf of the Standard Model}}
\vskip 1truecm
G.T.Ter-Kazarian
\\
{\em Byurakan Astrophysical Observatory, Byurakan, 378433 Armenia \\
E-mail:gago@bao.sci.am}
\end{center}
\begin{abstract}
We promote the microscopic theory of standard model 
(MSM) [1,2] into supersymmetric framework in order to solve its technical 
aspects of vacuum zero point energy and hierarchy problems, and 
attempt, further, to develop its realistic viable minimal SUSY 
extension. Among other things that - the MSM provides a natural unification 
of geometry and the field theory,  has clarified the physical conditions in 
which the geometry and particles come into being, in microscopic sense 
enables an insight to key problems of particle phenomenology and answers 
to some of its nagging questions -  a present approach also leads to quite a 
new realization of the SUSY yielding a physically realistic particle 
spectrum. It stems from the special subquark algebra, from which the 
nilpotent supercharge operators are derived. 
The resulting theory makes plausible following testable 
implications for the current experiments at LEP2, at the Tevatron and at LHC
drastically different from those of the conventional MSSM models: 

$\bullet$ {\em All the sparticles and the Higgs bosons never could emerge in 
spacetime continuum, thus, they cannot be discovered in any experiment nor at 
any energy range}.

$\bullet$ {\em For each of the three SM families of quarks and 
leptons there are corresponding heavy family partners with the same 
quantum numbers 
\footnote{This prediction 
was directly  ensued from the MSM [1,2], as well the similar one was 
made in phenomenological consideration by S.L.Adler [3]).}
and common mass-shift coefficients $(1+k)$
given for the low-energy poles  $k_{1}>\sqrt{2}, \quad k_{2}=\sqrt{{8/3}}$ 
and $k_{3}= 2$, lying far above the electroweak scale, respectively, at the 
energy threshold values:} 
$
E_{1}>(419.6 \pm 12.0)GeV,\quad E_{2}= (457.6 \pm 13.2)GeV$ and 
$
E_{3}=(521.4 \pm 15.0)GeV.
$
\end{abstract} 

 \section {Introduction}
\label {int}
A phenomenological standard model (SM) of high energy physics [4-22] with 
enormous success settles order in entangled experimental data. 
Although it has proven to be in spectacular agreement with experimental 
measurements and highly successful in a description and predicting a wide 
range of phenomena, however, it suffers from some vexing problems and many 
key questions of both the phenomenological and SUSY aspects have yet to be 
answered.

$\bullet$ In phenomenological aspect the mechanism of the electroweak 
symmetry breaking is a complete mystery. 
The most problematic ingredient of such a 
breaking is the Higgs boson (in simplest version), which has not yet been 
discovered experimentally. If a weakly interacting Higgs boson exists, 
it will then appear below the TeV scale. Many possible extension of the 
Higgs sector have been considered (see e.g. [20,21,23,24] and references 
therein). Along with the verifying experimentally of the Higgs sector, 
the most important open questions of the SM are as follows: 
We have no understanding why the SM is as it is? 
Why is the gauge symmetry? Why is this the particle spectrum? 
Why the electroweak symmetry breaking sector consists of just one 
$SU(2)_{L}$ doublet of Higgs 
bosons as it is in SM? The untested aspects of SM are the mass spectrum of 
the particles, the mixing patterns and the CP violation. 
The latter is introduced through complex Yukawa couplings 
of fermions to Higgs bosons, resulting in complex parameters in the CKM 
matrix. The SM contains a large number of arbitrary parameters, while a 
consistent complete theory would not have so many free parameters.

$\bullet$ There is another line of reasoning which supports the side of 
supersymmetrization of the SM, i.e., there are two well known principal 
issues which remain open in the SM. The first is the vacuum zero point 
problem standing before any quantum field theory. Second is often referred to 
as the problem of quadratic divergences 
or the hierarchy and naturalness problem (the dimensional analysis problem) 
arisen as the quadratic growth of the Higgs boson mass beyond tree level in 
perturbation theory, namely, the extreme difference in energy scales in the 
theory is inconsistent in the fundamental scalar sector. This is strong 
indication for the physics beyond SM. These last two problems can be 
solved by extending the symmetry of the theory to supersymmetry [25-49], 
which is believed in conventional physics to be manifest at energies in 
the TeV range. Given the SUSY requiring doubling the number of all the 
particles by their SUSY partners (sparticles), the quantum radiative 
corrections may cancel because some loop diagrams vanish due to cancellation 
between bosons and fermions since they have opposite signs. Then, if the 
SUSY is present in the TeV range, the masses of the Higgs bosons are no more 
unstable than fermion masses, whose smallness is natural and hold due to the 
approximate chiral symmetries. In this manner, its simplest form, SUSY 
solves the technical aspects of the hierarchy problem as well as the zero 
point energy problem, when due to power of the boson-fermion 
cancellation the zero point energy of the fermions exactly cancels that of 
the bosons and the degeneracy is not arisen. Therefore, in usual, the SM 
should be regarded as an effective low energy field theory valid up to 
the energy range smaller than a few hundred GeVs. 

$\bullet$ However, the SUSY in turn introduces its own set of difficulties. 
Despite the beautiful mathematical features of SUSY theories and that 
the SUSY has been theoretically invented almost three decades ago, but 
a physically realistic realization of SUSY had not been achieved yet and 
this principal problem was ever since much the same as now. In all 
suggested SUSY theories the supercharges have been inserted in 
{\em ad hoc} manner directly into the four-dimensional spacetime 
continuum adding a new structure, i.e., a new four odd fermionic dimensions. 
In fact, a physical essence of the basic concept of supercharge remains 
unknown and, therefore, the physical theory is 
beset by various difficulties. Perhaps the most discouraging and disturbing 
feature of the general class of proposed SUSY theories is the absence at the 
moment of a solid experimental motivation of supersymmetry, i.e., there is 
not a direct experimental evidence for the existence of any of the numerous 
new sparticles predicted by such theories. It is clear, then, that SUSY 
cannot be an exact symmetry in nature but has to be realized at least in 
broken phase. The last one is the least understood aspect of such 
theories. The spontaneously broken SUSY should be ruled out at once since 
it runs into phenomenological difficulties [31-49]. One of the viable way out 
from this situation is an explicit breaking of the global SUSY. A generic 
parametrization of this phenomenon introduces the much larger free 
parameter space ($\simeq$ 124) in the models of minimal supersymmetric 
extension of the SM (MSSM-124 [32]). Thus, it 
is important to develop the other schemes that attempt to reduce the number 
of free parameters. The conventional SUSY theories predict that the 
sparticles must reside in the TeV range. All such arguments that 
nature is supersymmetric, and that SUSY is broken at scales not too 
different than the weak scale, are theoretical. The next generation of 
experiments at Fermilab and CERN [35,39,45,49-60] will explore this energy 
range, where at least some of sparticles are expected to be found.

$\bullet$ All this variety prompts us, further, to adopt the idea that perhaps a 
more deeper level of organization of physical world may be existed. 
In the light of current status of particle physics, any new more elaborated 
outlook seems worthy of investigation. To solve in microscopic sense some of 
the above mentioned  problems of phenomenological aspect
the MSM is built up in [1,2]. 
The operator manifold (OM) formalism (part I) [1] enables to develop an 
approach to the unification of the geometry and the field theory, as well
the quantization of geometry different from all existing schemes. 
Here we explore the query how did the geometry and fields, as they are, come
into being? In the first a major purpose is to prove our physical outlook 
embodied in the idea that the geometry and fields, with the internal 
symmetries and all interactions, as well the four major principles of 
relativity (special and general), quantum, gauge and colour confinement are 
derivative, and they come into being simultaneously (sec.2). The substance 
out of which the geometry and fields are made is the ``primordial structures'' 
involved into reciprocal ``linkage'' establishing processes (subsec.2.3, 2.4)
We generalize this formalism via the  concept of operator multimanifold 
(OMM) (sec.3), which yields the MW geometry (subsec.3.2)
involving the spacetime continuum and internal worlds of given number.
All this is not merely an exercise in abstract reasoning but presumably 
bears directly on the geometry of the universe in which we live.
In an enlarged framework of the OMM we define and 
clarify the conceptual basis of subquarks and their
characteristics stemming from the various symmetries of the internal worlds
(subsec.3.2).
The OMM formalism has the following features: 

$\bullet$ {\em It provides a natural unification of the geometry-
yielding the special and general relativity principles (subsec.2.2), and 
the fermion fields serving as the basis for the constituent subquarks 
(subsec.3.2). 

$\bullet$ It has cleared up the physical conditions in which the geometry 
and particles come into being (subsec.2.2, 3.1).

$\bullet$ The subquarks
emerge in the geometry only in certain permissible combinations
utilizing the idea of the subcolour (subquark) confinement 
principle (subsec.3.2), and have undergone the transformations yielding the 
internal symmetries and gauge principle (subsec.2.6)}.
\\
We developed the MSM (part II) [2] based on the OMM formalism , which 
attempts to answer to some of the above mentioned questions of particle 
phenomenology. 
All the fields along with the spacetime component have nontrivial 
composite internal MW structure (subsec.12.1) such that
the possible elementary particles are thought to be 
composite dynamical systems in analogy to quantum mechanical stationary 
states of compound atom, but, now a dynamical treatment built up on the 
MW geometry is quite different and more amenable to qualitative 
understanding. The microscopic structure of leptons, quarks and other 
particles will be governed by the only possible conjunctions of constituent
subquarks implying concrete symmetries.
Although within considered schemes the subquarks are defined on the internal 
worlds, however the resulting spacetime components of particles, which we 
are going to deal with to describe the leptons and quarks defined on 
the spacetime continuum, are affected by them (sec.12) in such a way 
that they carry exactly all the quantum numbers of the various constituent 
subquarks of the given composition.
The hypothesis of existence of the MW structures manifests 
its virtue by solving some key problems of particle phenomenology, when we
attempt to suggest a microscopic approach to the properties of particles and 
interactions. First of all the theoretical significance of the MSM resides in 
the microscopic interpretation of all physical parameters.
\\
Continuing this program towards the supersymmetrization, in this 
article we shall attempt, further,  to promote the MSM into the SUSY 
framework by elaborating its realistic manifestly minimal SUSY extension 
(M\rlap/SMSM) (sec.15).
The major difference of outlined here supersymmetric approach (MW-SUSY) from 
those of conventional SUSY theories is as follows: 

{\em
$\bullet$ The MW-SUSY has an algebraic origin in the sense that it stems 
from a special subquark algebra defined on the internal worlds,
while the  nilpotent supercharge operators  are derived (sec.4).
Therefore, the MW-SUSY has realized only on the internal worlds, but not on 
the spacetime continuum, which are all the ingredients of the broken 
super-multimanifold (\rlap/SMM).

$\bullet$ Defined on the \rlap/SMM (sec.8) it implies the super-algebra 
different from the conventional SUSY algebra.

$\bullet$ Here we are led to the principal point of drastic 
change of the standard SUSY scheme to specialize the superpotential to be in 
such a form of eq.(11.28)-eq.(11.34), which allows within this framework, 
further, to build up the MSM (sec.12).} 

We develop the microscopic approach to the isospinor Higgs boson with 
self-interaction and Yukawa couplings (subsec.12.9-sec.13), wherein the two 
complex self-interacting isospinor-scalar Higgs doublets 
$\left(H_{u},\,H_{d}\right)$ as well as their spin-$\FFr{1}{2}$ SUSY 
partners $\left(\widetilde{H}_{u},\,\widetilde{H}_{d}\right)$ Higgsinos  
have arisen on the $W$-world as the Bose-condensate.
In contrast to SM, the MSM predicts the electroweak symmetry breaking in 
the $W$-world by the vacuum expectation value (VEV) of spin zero Higgs bosons 
and its transmission from the $W-$world 
to the spacetime continuum (subsec.12.14). This is the most remarkable 
feature of suggested approach, especially, in the view of existing great 
belief of the conventional theories for a discovery of the Higgs boson 
with other new particles at next round of experiments at LEP2, at the 
Tevatron, at LHC and other colliders, which will explore the TeV energy 
range (e.g. see [50]). 
The LEP2 data (is currently running at 189 GeV) provide a lower 
limit $m_{H}>89.3 GeV$ on its mass in simplest version. 
Furthermore, there is a tight upper limit $(m_{h^{0}}< 150 GeV)$ on the mass 
of the lightest Higgs boson $h^{0}$ among the 5 physical Higgs bosons 
predicted by the MSSM models. The current direct search limits 
from LEP2 give $m_{h^{0}}> 75 GeV.$ Therefore, the future searches for this 
boson (if the mass is below 150 GeV or so) would be a crucial point in 
testing the efforts made in the conventional models building as well in the 
present MSM based on a quite different approaches. 
Actually, reflecting upon the results far obtained in 
the sec.12, in strong contrast to conventional theories, the MSM rejects 
drastically any expectation of discovery of any Higgs boson, but in the same 
time it expects to include a rich spectrum of new particles at higher 
energies. Namely, if the MSM proves viable it becomes an crucial issue to hold 
in experiments the following two solid tests: 

{\em
$\bullet$   
The Higgs bosons never could emerge in spacetime continuum since they have 
arisen only on the internal $W$-world, i.e., thus, the unobserved effects 
produced by such bosons cannot be discovered in experiments nor at any 
energy range. 

$\bullet$  For each of the three SM families of quarks and 
leptons, there are corresponding heavy family partners with the same quantum 
numbers lying far above the electroweak scale}.

Regarding to the last phenomenological implication of the MSM, it is 
remarkable that the similar in many respects prediction is made in somewhat 
different context by S.L.Adler [3] within a phenomenological scheme of a 
compositeness of the quarks and leptons. It based on the generic group 
theoretical framework of rishon type models exploring the preon constituents. 
But, therein a present, a bit premature, state of the theory does not 
allow the exact estimate of this scale. 
Although one admits that such a scale could be much higher than 
electroweak scale, however, it is necessary special argumentation   
in support of validity of this prediction in a case if this scale has 
turned out to be low enough, namely, if these heavy partners lie not too far 
above the electroweak scale. Even thus, as it is notified in [3], one must not 
worry for the existence of 6 heavy flavors, which is then marginally 
compatible with the current LEP data [18]. A complete analysis of this 
question, naturally, is now possible in suggested microscopic approach. 
The MSM enables oneself to study in detail the phenomenology 
associated with such extra heavy families and to estimate the value of 
energy threshold of their creation (sec.12.14). While, the low energy scale 
could not be realized since it lies far below the energy threshold of the 
next pole for appearing of the heavy partners. The 
estimate gives the common mass-shift coefficients $(1+k)$, where $k$ reads 
for the next few low energy poles with respect to the lowest one: 
$k_{0}=0,\quad k_{1}>\sqrt{2}, \quad k_{2}=\sqrt{{8/3}}$ and $k_{3}= 2$.
The first one obviously does not produce the extra families, but the energy 
thresholds corresponded to the next non-trivial poles can be respectively 
written:
$
E_{1}> (419.6 \pm 12.0)GeV,\quad E_{2}= (457.6 \pm 13.2)GeV$ and 
$
E_{3}=(521.4 \pm 15.0)GeV.
$
\\
These predictions above together with a new one given in the sec.17 that

$\bullet$ {\em the sparticles could never emerge in spacetime continuum 
since they have 
arisen only on the internal $W$-world, thus, they cannot be discovered in 
experiments nor at any energy range,} 
\\
are the three solid implications of the resulting 
M\rlap/SMSM for the experiments at LEP2, at the Tevatron and at LHC,
which are drastically different from those of MSSM models. 
Which of these schemes, if any, is realized either exactly or 
at least approximately in nature remains to be seen in the years 
to come.       
 
\section {Preliminaries}
\label{Prin}
To facilitate the physical picture and provide sufficient background, this 
section contains some of the necessary preliminaries on generic of the OMM 
formalism, which one to know in order to understand a structure of suggested 
SUSY approach. 
Since it is too technical for present article, we outline only relevant 
steps in concise schematic form hoping to 
supplement this shortage of insufficient rigorous treatment by referring to 
[1,2] for more detailed justification of some of the procedure and complete 
exposition.
The present article is a direct continuation of [1,2], so we adopt its all 
ideas and conventions.
In the next section and further we shall deal with the MW geometry, 
except for the change of the concept of quark inserted schematically here
to subquark defined on the given internal world. 
To be brief we often suppress the indices without notice.
\\
We start by tracing at elementary level the relevant steps of motivation of 
the OM formalism:

$\bullet$ First step is an extension of the Minkowski space 
${\M1_{x}}{}_{4}\rightarrow M_{8}={\M1_{x}}{}_{4}\oplus {\M1_{u}}{}_{4}$ in 
order to introduce the particle mass operator defined on the internal wold 
${\M1_{u}}{}_{4}$ of the inner degrees of freedom. 
For example, in a case of Dirac's particle one proceeds at once:
$$(\underbrace{\gamma p}_{x}-m)\ps1_{x}=0
\quad\rightarrow \quad
\gamma p\psi=0,$$ 
provided by
$\psi=\ps1_{x}\ps1_{u}, \quad
\gamma p=\underbrace{\gamma p}_{x} - \underbrace{\gamma p}_{u},
\quad
m\ps1_{u}\equiv\underbrace{\gamma p}_{u} \ps1_{u}
$
and
$$d\,x^{2}=inv \quad\rightarrow \quad
d\,x^{2}_{8}=d\,x^{2}-d\,u^{2}=0, \quad x_{8}\in M_{8}. 
$$
The same holds for the other fields of arbitrary spin.

$\bullet$ Next, a two-steps passage 
$M_{4}\rightarrow M_{6}  \stackrel{45^{0}}{\rightarrow} 
G_{6}$ will be performed for each sample of the $M_{4}$.
\\
a) A passage $M_{4}\rightarrow M_{6}$ restores the complete equivalence 
between the three spacial and three time components:
$$
\begin{array}{l}
e_{4}=(\vec{e},e_{0})\quad\rightarrow\quad 
\vec{e}_{6}=(\vec{e},\vec{e}_{0})\in M_{6},\quad
x_{4}=(\vec{x},x_{0})\quad\rightarrow \quad
x_{6}=(\vec{x},\vec{x}_{0})\in M_{6}.
\end{array}
$$
b) A rotation $M_{6} \stackrel{45^{0}}{\rightarrow} G_{6}$ of the basis 
vectors  on the angle $45^{0}$ provides an adequate algebra for 
quantization of the geometry (subsec.2.1):
$$
\begin{array}{l}
\vec{e}_{6} \stackrel{45^{0}}{\rightarrow} e_{(\lambda\alpha)},
\quad \lambda = \pm, \quad \alpha=1,2,3, \\ 
e_{\pm\alpha}=\FFr{1}{\sqrt{2}}(e_{0\alpha}\pm e_{\alpha})=
O_{\pm}\otimes\sigma_{\alpha}, \quad 
<O_{\lambda},O_{\tau}>=1-\delta_{\lambda\tau}, 
\quad <\sigma_{\alpha},\sigma_{\beta}>=\delta_{\alpha\beta}.
\end{array}
$$ 
Accordingly one gets $M_{8}\rightarrow G_{12}$.
Thus, within a simplified scheme (one $u$-channel) of the following it is 
convenient to deal in terms of smooth differentiable manifold 
$$G=\G1_{\eta}\oplus \G1_{u},$$  
$ Dim \,G=12,\quad Dim\,\G1_{i}=6 \quad (i=\eta, u)$. Note that presumably 
we are allowed to perceive directly only the $\G1_{\eta}$ which will be 
related to the spacetime continuum, but not 
the $\G1_{u}$ which will be displayed as a space 
of inner degrees of freedom (see below).

$\bullet$ Finally, in suggested approach we will be dealing in terms of 
first degree of the line element, which entails an additional phase 
multiplier $\Phi(\zeta$) for the vector defined on $G$:\\
$$d\,\zeta^{2}\quad\rightarrow \quad d\,\vec{\zeta}\,\,e^{iS}, 
\quad \vec{\zeta}\quad\rightarrow \quad\vec{\Phi}(\zeta)=
\vec{\zeta}\,\Phi(\zeta), \quad
\Phi(\zeta)\equiv e^{iS},$$ 
where 
$\vec{\zeta}=\vec{e}\, \,\zeta, \quad \vec{e}=
(\vec{\e1_{\eta}}, \,\vec{\e1_{u}}),\quad 
S(\zeta)$ is the invariant action defined on $G$.\\
The 
$\{e_{(\lambda,\mu,\alpha)}=O_{\lambda,\mu}\otimes
\sigma_{\alpha}\} \subset G$
$(\lambda,\mu=1,2; \quad \alpha=1,2,3)$
are linear independent $12$ unit vectors at the point $p$ of the 
manifold $G$, provided by the linear unit bipseudovectors $O_{\lambda,\mu}$
implying
$$<O_{\lambda,\mu},O_{\tau,\nu}>={}^{*}\delta_{\lambda,\tau}
{}^{*}\delta_{\mu,\nu}\quad
<\sigma_{\alpha}, \sigma_{\beta}>= \delta_{\alpha\beta},\quad
{}^{*}\delta=1-\delta,
$$
where $\delta$ is Kronecker symbol, the $\{ O_{\lambda,\mu}= 
O_{\lambda}\otimes  O_{\mu} \}$ is
the basis for tangent vectors of $2 \times 2$ dimensional linear 
pseudospace ${}^{*}{\bf R}^{4}={}^{*}{\bf R}^{2}\otimes
{}^{*}{\bf R}^{2}$, the $ \sigma_{\alpha}$ refers to three dimensional 
ordinary space ${\bf R}^{3}$. 
Henceforth we always let the first two subscripts in the parentheses
to denote the pseudovector components, while the third refers to the
ordinary vector components.
The metric on $G$ is 
$\hat{\bf g}:{\bf T}_{p}\otimes {\bf T}_{p}\rightarrow C^{\infty}(G)$ 
a section of conjugate vector bundle $S^{2}{\bf T}$.
Any vector ${\bf A}_{p}\in{\bf T}_{p}$ reads ${\bf A}=
e A$, provided with
components $A$ in the basis
$\{e\}$.
In holonomic coordinate basis $\left(\partial/\partial\,
\zeta\right)_{p}$ one gets 
$A=\left.\FFr{d\,
\zeta}{d\,t}\right|_{p}$ and
$\hat{g}=g d\zeta\otimes d\zeta $.
The manifold $G$ decomposes as follows:
$$G={}^{*}{\bf R}^{2}
\otimes {}^{*}{\bf R}^{2} \otimes {\bf R}^{3}=\G1_{\eta}\oplus
\G1_{u}=\displaystyle\sum^{2}_{\lambda,\mu=1} 
\oplus {\bf R}^{3}_{\lambda \mu}=
{\R_{x}}^{3}\oplus 
{\R_{x_{0}}}^{3}\oplus
{\R_{u}}^{3}\oplus 
{\R_{u_{0}}}^{3}$$ 
employing corresponding basis vectors  
${\e1_{i}}{}_{(\lambda\alpha)}={\O1_{i}}{}_{\lambda}\otimes 
\sigma_{\alpha}
\subset \G1_{i}$ $(\lambda =\pm,\quad 
i=\eta, u)$ of tangent sections, where 
$${\O1_{i}}{}_{+}=
\displaystyle \frac{1}{\sqrt{2}}(O_{1,1} +\varepsilon_{i} O_{2,1}),\quad
{\O1_{i}}{}_{-}=
\displaystyle \frac{1}{\sqrt{2}}(O_{1,2} +\varepsilon_{i} O_{2,2}),\quad
\varepsilon_{\eta}=1,\quad\varepsilon_{u}=-1,$$ 
and
$<{\O1_{i}}{}_{\lambda},{\O1_{j}}{}_{\tau}>=
\varepsilon_{i}\delta_{ij}{}^{*}\delta_{\lambda \tau}$.
The $\G1_{\eta}$ decomposes into three dimensional
ordinary and time  flat 
spaces 
$$\G1_{\eta}=
{\R_{x}}^{3}\oplus {\R_{x_{0}}}^{3}$$ 
with signatures 
$sgn({\R_{x}}^{3})=(+++)$ 
and $sgn({\R_{x_{0}}}^{3})=(---)$. The same holds for $\G1_{u}$ with 
the opposite signatures $sgn({\R_{u}}^{3})=(---)$ 
and $sgn({\R_{u_{0}}}^{3})=(+++)$. \\
The passage to Minkowski space is a further step as follows:
Since all directions in ${\R_{x_{0}}}^{3}$ are
equivalent, then by notion {\em time} one implies the projection of
time-coordinate on fixed arbitrary universal direction in ${\R_{x_{0}}}^{3}$.
This clearly respects the physical ground.
By the reduction ${\R_{x_{0}}}^{3}\rightarrow {\R_{x_{0}}}^{1}$ the passage
$$\G1_{\eta}\rightarrow M_{4}={\R_{x}}^{3}\oplus {\R_{x_{0}}}^{1}$$ 
may be performed whenever it will be necessary.\\
In the case of gravity, the passage from six dimensional curved manifold 
$\widetilde{G}$ to four dimensional Riemannian geometry $R_{4}$ is 
straightforward by making use of reduction of three time components 
$e_{0\alpha}=
\displaystyle \frac{1}{\sqrt{2}}(e_{(+\alpha)}+e_{(-\alpha)})$ of basis 
sixvector $e_{(\lambda \alpha)}$ to the single one $e_{0}$ in the given
universal direction, which merely fixed a time 
coordinate. Actually, since Lagrangian of the fields defined on
$\widetilde{G}$ is a function of scalars such as
$A_{(\lambda \alpha)}B^{(\lambda \alpha)}=
A_{0 \alpha}B^{0 \alpha}+A_{\alpha}B^{\alpha}$, thus taking into account that
$A_{0 \alpha}B^{0 \alpha}=A_{0 \alpha}<e^{0\alpha},e^{0\beta}>B_{0 \beta}
=A_{0}<e^{0},e^{0}>B_{0}=A_{0}B^{0}$, one readily may perform the
required passage.
In this case one has
$$
d\,\zeta^{2}=d\,\eta^{2}-d\,u^{2}=0,
\quad \left.d\,\eta^{2} \right|_{6\rightarrow 4} \equiv
d\,s^{2}=g_{\mu\nu}d\,x^{\mu}d\,x^{\nu}=d\,u^{2}=inv.
$$
For more discussion see [1].

\subsection{Quantization of geometry}
We proceed at once with the secondary quantization of geometry by 
substituting the pseudo vectors $O_{\lambda}$ for the  
operators supplied by additional index ($r$) referring to the quantum 
numbers of corresponding state
$$
\hat{O}^{r}_{1}=O^{r}_{1}\alpha_{1},\quad
\hat{O}^{r}_{2}=O^{r}_{2}{\alpha}_{2},\quad
\{ \hat{O}^{r}_{\lambda},\hat{O}^{r'}_{\tau} \}=
\delta_{rr'}{}^{*}\delta_{\lambda\tau}I_{2}, 
$$
where
$\{ {\alpha}_{\lambda},{\alpha}_{\tau} \}={}^{*}\delta_
{\lambda\tau}I_{2},
\quad
{\alpha}^{\lambda}={}^{*}\delta^{\lambda\mu}
{\alpha}_{\mu}={({\alpha}_{\lambda})}^{+},
$
For example
${\alpha}_{1}=\left( \matrix{
0 &1 \cr
0 &0 \cr
}\right), \quad
{\alpha}_{2}=\left( \matrix{
0 &0 \cr
1 &0 \cr
}\right).$
Then
$$
\hat{O}^{r}_{1}\mid 0>={O}^{r}_{1}\mid 1>.
\quad
\hat{O}^{r}_{2}\mid 1>={O}^{r}_{2}\mid 0>.
$$
Hence
$\hat{O}^{r}_{1}\mid 1>=0, \quad
\hat{O}^{r}_{2}\mid 0>=$0.
A matrix realization of the states
is
$$\mid 0>\equiv\chi_{1}=\left( \matrix{
0 \cr
1\cr
}\right), \\
\mid 1>\equiv\chi_{2}=\left( \matrix{
1 \cr
0 \cr
}\right).$$
Also, instead of ordinary basis vectors we introduce
the operators
$\hat{\sigma}^{r}_{\alpha}\equiv\delta_{\alpha\beta\gamma}
\sigma^{r}_{\beta}\widetilde{\sigma}_{\gamma}$,
where $\widetilde{\sigma}_{\gamma}$ are Pauli's matrices, and
$$
<\sigma_{\alpha}^{r},\sigma_{\beta}^{r'}>=\delta_{rr'}\delta_{\alpha\beta},
\quad
\hat{\sigma}^{\alpha}_{r}=\delta^{\alpha\beta}\hat{\sigma}^{r}_{\beta}=
{(\hat{\sigma}_{\alpha}^{r})}^{+}=\hat{\sigma}_{\alpha}^{r},
\quad
\{\hat{\sigma}_{\alpha}^{r},\hat{\sigma}_{\beta}^{r'}\}=2
\delta_{rr'}\delta_{\alpha\beta}I_{2}.
$$
Than, the states $\mid 0>\equiv{\varphi}_{1(\alpha)}$
and $\mid 1_{(\alpha)}>
\equiv{\varphi}_{2(\alpha)}$ are as follows:
$${\varphi}_{1(\alpha)}\equiv\chi_{1}, \quad
{\varphi}_{2(1)}=\left( \matrix{
1 \cr
0 \cr
}
\right), \quad
{\varphi}_{2(2)}=\left( \matrix{
-i \cr
0\cr
}
\right), \quad
{\varphi}_{2(3)}=\left( \matrix{
0 \cr
-1\cr
}
\right),$$
and
$$
{\hat{\sigma}}_{\alpha}^{r}\varphi_{1(\alpha)}=\sigma_
{\alpha}^{r}\varphi_{2(\alpha)}=(\sigma_{\alpha}^{r}\widetilde{\sigma}_
{\alpha})\varphi_{1(\alpha)},
\quad
{\hat{\sigma}}_{\alpha}^{r}\varphi_{2(\alpha)}=\sigma_
{\alpha}^{r}\varphi_{1(\alpha)}=(\sigma_{\alpha}^{r}\widetilde{\sigma}_
{\alpha})\varphi_{2(\alpha)}.
$$
Hence, the single eigenvalue
$(\sigma_{\alpha}^{r}\widetilde{\sigma}_{\alpha})$
associates with different $\varphi_{\lambda(\alpha)}$,
namely it is degenerated with 
degeneracy degree equal 2. Thus, among quantum numbers $r$
there is also the quantum number of the half integer spin 
$\vec{\sigma}$ 
$(\sigma_{3}=\FFr{1}{2}s,\quad s=\pm1)$.
This consequently gives rise to the spins of particles.
Next we introduce the operator
$$
{\hat{\gamma}}^{r}_{(\lambda,\mu,\alpha)}\equiv{\hat{O}}^{r_{1}}_{\lambda}
\otimes{\hat{O}}^{r_{2}}_{\mu}\otimes{\hat{\sigma}}^{r_{3}}_{\alpha}$$
and the state vector
$$
\chi_{\lambda,\mu,\tau(\alpha)}\equiv\mid\lambda,\mu,\tau(\alpha)>=
\chi_{\lambda}\otimes\chi_{\mu}\otimes\varphi_{\tau(\alpha)}, 
$$
where $\lambda,\mu,\tau,\nu=
1,2;\quad \alpha,\beta=1,2,3$ and $r\equiv (r_{1},r_{2},r_{3})$.
Omitting two valuedness of state vector we apply
$\mid\lambda,\tau,\delta(\beta)>\equiv\mid\lambda,\tau>$,
and remember that always the summation 
must be extended over the double degeneracy of the spin states $(s=\pm 1)$.
The matrix elements read
$$
<\lambda,\mu\mid{\hat{\gamma}}^{r}_{(\tau,\nu,\alpha)}\mid \tau,\nu>=
{}^{*}\delta_{\lambda\tau}{}^{*}\delta_{\mu\nu}
e^{r}_{(\tau,\nu,\alpha)},\quad
<\tau,\nu\mid{\hat{\gamma}}_{r}^{(\tau,\nu,\alpha)}\mid\lambda,\mu >=
{}^{*}\delta_{\lambda\tau}{}^{*}\delta_{\mu\nu}
e_{r}^{(\tau,\nu,\alpha)}.
$$
for given $\lambda,\mu.$
The operators of occupation numbers are
$$
{\hN_{1}}{}^{rr'}_{\alpha\beta}=
{\hat{\gamma}}^{r}_{(1,1,\alpha)}{\hat{\gamma}}{}^{r'}_{(2,2,\beta)},
\quad
{\hN_{2}}{}^{rr'}_{\alpha\beta}=
{\hat{\gamma}}^{r}_{(2,1,\alpha)}{\hat{\gamma}}{}^{r'}_{(1,2,\beta)},
$$
with the expectation values implying Pauli's exclusion principle.
The set of operators $\{{\hat{\gamma}}^{r}\}$
is the basis for tangent operator vectors 
$\hat{\Phi}(\zeta)={\hat{\gamma}}^{r}
\Phi_{r}(\zeta)$ of the 12 dimensional flat operator manifold $\hat{G}$,
where we introduce the vector function
of ordinary class of functions of $C^{\infty}$ smoothness 
defined on the manifold $G$:
$\quad \Phi_{r}^{(\lambda,\mu,\alpha)}(\zeta)=
\zeta^{(\lambda,\mu,\alpha)} \Phi_{r}^{\lambda,\mu}(\zeta),\quad
\zeta \in G$.\\

The basis $\{{\hat{\gamma}}^{r}\}$
decomposes 
$(\lambda=\pm;\quad\alpha=1,2,3;\quad i=\eta,u)$:
$$
{\hgam_{i}}{}^{r}_{(+\alpha)}=\FFr{1}{\sqrt{2}}
(\gamma^{r}_{(1,1\alpha)}+\varepsilon_{i}
\gamma^{r}_{(2,1\alpha)}),
\quad {\hgam_{i}}{}^{r}_{(-\alpha)}=\FFr{1}{\sqrt{2}}
(\gamma^{r}_{(1,2\alpha)}+\varepsilon_{i}
\gamma^{r}_{(2,2\alpha)})
.$$ 
The expansions of operator vectors $\hps_{i}\in\hG_{i}$ and 
operator covectors  $\bar{\hps_{i}}$ are written 
$\hps_{i}={\hgam_{i}}{}^{r}{\ps1_{i}}{}_{r},
\quad\bar{\hps_{i}}=
{\hgam_{i}}{}_{r}{\ps1_{i}}{}^{r}.$
As to the vector functions, they  are 
defined on the manifold $\G1_{i}$: 
$$
{\ps1_{\eta} }{}_{r}^{(\pm\alpha)}(\eta,p_{\eta})=\eta^{(\pm\alpha)}
{\ps1_{\eta} }{}_{r}^{\pm}(\eta,p_{\eta}),\quad 
{\ps1_{u}}{}_{r}^{(\pm\alpha)}(u,p_{u})=u^{(\pm\alpha)}
{\ps1_{u}}{}_{r}^{\pm}(u,p_{u}).
$$
Due to the spin states, the ${\ps1_{i}}{}_{r}^{\pm}$ is the Fermi field 
of the positive and negative frequencies 
${\ps1_{i}}{}^{\pm}_{r}={\ps1_{i}}{}^{r}_{\pm p}$.

\subsection{Realization of the flat manifold $G$} 
The bispinor $\Psi(\zeta)$ defined on manifold $G=\G1_{\eta}
\oplus\G1_{u}$ is written  
$$\Psi(\zeta)=\ps1_{\eta}(\eta)\ps1_{u}(u),$$
where the $\ps1_{i}$ is a bispinor defined on 
the manifold $\G1_{i}.$
The free state of $i$-type fermion with definite values of momentum
$p_{i}$ and spin projection $s$ is described by means of plane waves.
We make use of localized wave packets of operator vector
fields ${\hps_{i}}$ and $\hat{\Phi}(\zeta)$.
In this manner we get the important relation
$$
\begin{array}{l}
\S_{\lambda=\pm}<\chi_{\lambda}\mid\hat{\Phi}(\zeta)
\bar{\hat{\Phi}}(\zeta)\mid
\chi_{\lambda}>= 
\S_{\lambda=\pm}<\chi_{\lambda}\mid
\bar{\hat{\Phi}}(\zeta)\hat{\Phi}(\zeta)\mid\chi_{\lambda}>= \\ 
=-i\,\zeta^{2}\,\G1_{\zeta}(0)=-i\,\left(\eta^{2}\,\G1_{\eta}(0)-
u^{2}\,\G1_{u}(0)\right),
\end{array}
$$
where
$\G1_{i}(0)\equiv \Lm_{i\rightarrow i'}\G1_{i}(i-i'), \quad 
(i=\zeta,\eta, u,)$, etc., the Green's function 
$
\G1_{i}(i-i')=-(i\hpr_{i}+m)\Dlt_{i}(i-i')
$
is provided by the usual invariant singular functions 
$\Dlt_{i}(i-i')$ $(i=\eta,u)$, the state vectors $\chi_{\lambda}$ are given 
in App. A.
Realization of the manifold $G$ ensued from the  
constraint imposed upon the matrix element of bilinear form,
which is, as a geometric object, required to be finite
$$
\S_{\lambda=\pm}<\chi_{\lambda}\mid\hat{\Phi}(\zeta)
\bar{\hat{\Phi}}(\zeta)\mid
\chi_{\lambda}> < \infty ,
$$
which gives rise to
$$
\zeta^{2}{\G1_{\zeta}}{}_{F}(0) < \infty ,
$$
and
$$
\begin{array}{l}
{\G1_{\zeta}}{}_{F}(0)={\G1_{\eta}}{}_{F}(0)={\G1_{u}}{}_{F}(0)=\\
= \Lm_{u\rightarrow u'}\left[ -i\S_{{\vec{p}}_{u}}{\ps1_{u}}{}_{{p}_{u}}(u)
{\bp_{u}}{}_{{p}_{u}}(u')\theta (u_{0}-u'_{0})+
i\S_{{\vec{p}}_{u}}{\bp_{u}}{}_{{-p}_{u}}(u'){\ps1_{u}}{}_{{-p}_{u}}(u)
\theta (u'_{0}-u_{0}) \right],
\end{array}
$$
where ${\G1_{\zeta}}{}_{F},{\G1_{\eta}}{}_{F}$ and ${\G1_{u}}{}_{F}$ are
the causal Green's functions 
characterized by the boundary condition that only positive frequency
occur for $\eta_{0}>0\quad(u_{0}>0)$, only negative for
$\eta_{0}<0\quad(u_{0}<0)$. Here $\eta_{0}=\mid \vec{\eta}_{0}\mid $, 
$\eta_{0\alpha}=\FFr{1}{\sqrt{2}}
(\eta_{(+\alpha)}+\eta_{(-\alpha)})$ and the same holds for $u_{0}$.
Then, satisfying the condition eq.(2.2.3) a length of each vector
${\bf \zeta}=e\zeta\in G$
compulsory should be equaled zero
$$
\zeta^{2}=\eta^{2}-u^{2}=0.
$$
Relativity principle holds
$$
\left. d\,\eta^{2}\right|_{6\rightarrow 4}\equiv d\,s^{2}=
d\,u^{2}=inv.
$$

\subsection{Primordial structures and link establishing processes}
In [1] we have 
chosen a simple setting and considered the primordial 
structures designed to possess certain physical properties satisfying the 
stated general rules. These structures are thought to be the substance 
out of which the geometry and particles are made.
We distinguish $\eta$- and $u$-types
primordial structures involved in the linkage establishing processes 
occurring between the structures of different types.\\
The $\eta$-type structure may accept the linkage
only from $u$-type structure, which is described by the link
function 
${\ps1_{\eta}(\eta) }$
belonging to the ordinary class of functions of $C^{\infty}$ smoothness, 
where $\eta={\e1_{\eta}}{}_{(\lambda\alpha)}
\eta^{(\lambda\alpha)},\quad (\lambda = \pm; \alpha = 1,2,3,$ see subsec.2.1),
$\eta$ is the link coordinate.
Respectively the $u$-type structure may accept the linkage only
from $\eta$-type structure described by the link function
$
{\ps1_{u}}(u)$ (u-channel,
$u=\e1_{u} u) $,
where 
$$
{\ps1_{\eta} }{}^{(\pm\alpha)}(\eta,p_{\eta})=
\eta^{(\pm\alpha)}
{\ps1_{\eta} }{}^{\pm}(\eta,p_{\eta}),\quad 
{\ps1_{u} }{}^{(\pm\alpha)}(u,p_{u})=
u^{(\pm\alpha)}
{\ps1_{u} }{}^{\pm}(u,p_{u}),
$$
a bispinor ${\ps1_{i} }{}^{\pm}$ is the invariant state wave function of 
positive or negative frequencies, $p_{i}$ is the corresponding link momentum.
Thus, a primordial structure can be considered as
a fermion. 
A simplest system made of two structures of different types becomes
stable only due to the stable linkage
$$
\left|\p1_{\eta}\right|={({\p1_{\eta}}{}^{(\lambda\alpha)},
{\p1_{\eta}}{}_{(\lambda\alpha)})}^{1/2}=
\left|\p1_{u}\right|={({\p1_{u}}{}^{(\lambda\alpha)},
{\p1_{u}}{}_{(\lambda\alpha)})}^{1/2}.
$$
Otherwise they are unstable.
There is not any restriction on the number of primordial structures
of both types involved in the link establishing processes simultaneously.
Only, in the stable system the link stability condition must be
held for each linkage separately.
Suppose that persistent processes of creation and annihilation of the 
primordial 
structures proceed in different states $s, s',s'',...$ The "creation"
of structure in the given state $(s)$ is due to its transition to 
this state from other states $(s',s'',...)$, while the "annihilation"
means a vice versa.
Satisfying the stability condition the primordial structures from 
arbitrary states can establish a stable linkage.
Among the states $(s,s',s'',...)$ there is a lowest one ($s_{0}$),
in which all structures are regular, i.e., they are in free (pure) state
and described by the plane wave functions
${\ps1_{\eta} }{}^{\pm }(\eta_{f},p_{\eta})$ or 
${\ps1_{u} }{}^{\pm }(u_{f},p_{u})$
defined respectively on flat manifolds
$\G1_{\eta}$ and $\G1_{u}$. The index (f) specifies the points of 
corresponding flat manifolds $\eta_{f}\in\G1_{\eta}$, $u_{f}\in
\G1_{u}$. Note that the processes of creation and annihilation of regular 
structures in lowest state are described by the OM formalism given above.

\subsection{Distorted primordial structures}
In all higher states the  primordial structures are distorted (interaction 
states) and described by distorted link functions defined on 
distorted manifolds $\widetilde{\G1_{\eta}}$ and $\widetilde{\G1_{u}}$.
The distortion 
$G\rightarrow \widetilde{G}$ 
with hidden Abelian local group
$G=U^{loc}(1)=SO^{loc}(2)$ and one dimensional trivial algebra
$\hat{g}=R^{1}$ is considered in [63].
Within that scheme the basis $e^{f}$ undergoes distortion transformation 
$
e(\theta)=D(\theta)\,e^{f}
$.  
The matrix $D(\theta)$ is in the form $D(\theta)=C\otimes R(\theta)$, 
where 
$O_{(\lambda\alpha)}=C^{\tau}_{(\lambda\alpha)}
O_{\tau}$ and
$\sigma_{(\lambda\alpha)}(\theta)=R^{\beta}_{(\lambda\alpha)}
(\theta)\sigma_{\beta}$. Here 
$R(\theta)$ is the matrix of the group $SO(3)$ of ordinary rotations of 
the planes involving two arbitrary basis vectors 
of the spaces $R^{3}_{\pm}$ around  the orthogonal third axes $(\pm k)$  
through the angle $(\theta_{\pm k})$. 
The relation between the  
wave functions of distorted and regular structures reads 
$$
{\ps1_{u}}^{\lambda}(\theta_{+k})=
f_{(+)}(\theta_{+k}){\ps1_{u}}{}^{\lambda}, 
\quad
{\ps1_{u}}{}_{\lambda}(\theta_{-k})=
{\ps1_{u}}{}_{\lambda}f_{(-)}(\theta_{-k}),
$$
where ${\ps1_{u}}{}^{\lambda}\,\,({\ps1_{u}}{}_{\lambda})$ 
is the plane wave function 
of regular ordinary structure (antistructure).
Next, we supplement the previous assumptions given in subsec.2.3 by the new 
one that now the $\eta$-type (fundamental) regular structure can not directly 
form a stable system with the regular $u$-type (ordinary) structures.
Instead of it the $\eta$-type regular structure 
forms a stable system with the infinite number of distorted ordinary 
structures, where the link stability 
condition held for each linkage separately. Such structures take part 
in realization of flat manifold $G$. 
We employ the wave packets constructed by superposition of these functions 
furnished by generalized operators of creation and annihilation as the 
expansion coefficients. Geometry realization condition now should be 
satisfied for each ordinary structure in terms of
$$
{\G1_{u}}{}^{\theta}_{F}(0)=\Lm_{\theta_{+} \rightarrow \theta_{-}}
{\G1_{u}}{}^{\theta}_{F}(\theta_{+} - \theta_{-})
={\G1_{\eta}}{}_{F}(0)= \Lm_{\eta'_{f} \rightarrow \eta_{f}}
{\G1_{\eta}}{}_{F}(\eta'_{f}-\eta_{f}).
$$ 
Then
$$
\S_{k}{\ps1_{u}}(\theta_{+k})
{\bar{\ps1_{u}}}(\theta_{-k})
=\S_{k}{\ps1_{u}}'(\theta'_{+k})
{\bar{\ps1_{u}}}'(\theta'_{-k})=
\cdots = inv. 
$$
Namely, the distorted ordinary structures emerge in geometry only 
in permissible combinations forming a stable system.
Below, in simplified schematic way we exploit the background of the
known colour confinement and gauge principles.
Naive version of such construction still should be considered as a 
preliminary one, which will be further elaborated to make sense in the 
sec.3.

\subsection{Quarks and colour confinement}
At the very first to avoid irrelevant complications, here, for illustrative 
purposes, we will attempt to 
introduce temporarily skeletonized ``quark'' and 
``antiquark'' fields emerged in confined phase in the simplified  
geometry with the one-u channel given in the previous subsections. 
The complete picture of such a dynamics is beyond the scope of this 
subsection, but some relevant discussions on this subject will also be 
presented in the subsec.3.2.
We may think of the function
${\ps1_{u}}{}^{\lambda}(\theta_{+k})$ at fixed $(k)$ as being 
$u$-component of bispinor field of "quark" ${q}_{k}$, and of
${\bar{\ps1_{u}}}{}_{\lambda}(\theta_{-k})$ - an $u$-component of 
conjugated bispinor field of "antiquark" ${\bar{q}}_{k}$.
The index $(k)$ refers to colour degree of freedom in the 
case of rotations through the angles  $\theta_{+k}$ and anticolour 
degree of freedom in the case of $\theta_{-k}$.
The $\eta$-components of quark fields are plane waves.
There are exactly three colours.
The rotation through the angle $\theta_{+k}$ yields
a total quark field defined on the flat manifold $G=\G1_{\eta}\oplus\G1_{u}$
$$
{q}_{k}(\theta)=\Psi(\theta_{+k})={\ps1_{\eta}}{}^{0}
\ps1_{u}(\theta_{+k})
$$
where ${\ps1_{\eta}}^{0}$ is a plane wave defined on $\G1_{\eta}$.
This allows an other interpretation of quarks, which is absolutely 
equivalent to the former one and will be widely used throughout this article, 
i.e.,
$$
{q}_{k}(\theta)={\ps1_{\eta}}{}^{0}{\q1_{u}}{}_{k}(\theta)=
{\q1_{\eta}}{}_{k}(\theta){\ps1_{u}}{}^{0},
\quad {\q1_{\eta}}{}_{k}(\theta)\equiv 
f_{(+)}(\theta_{+k}){\ps1_{\eta}}{}^{0}, 
$$
where ${\ps1_{u}}{}^{0}$ is a plane wave, ${\q1_{u}}{}_{k}(\theta)$ and
${\q1_{\eta}}{}_{k}(\theta)$ may be considered as the quark fields with the 
same quantum numbers 
defined respectively on flat manifolds $\G1_{u}$ and  $\G1_{\eta}$. 
Making use of the rules stated 
one may readily return to Minkowski space
$\G1_{\eta}\rightarrow M_{4}$. In the sequel, a conventional quark 
fields defined on $M_{4}$ will be ensued
${\q1_{\eta}}{}_{k}(\theta) \rightarrow q_{k}(x)$, $x \in M_{4}$.
They imply
$$
\S_{k}{q}_{kp}{\bar{q}}_{kp} =
\S_{k}{q'}_{kp}{\bar{q'}}_{kp}=
\cdots = inv.
$$
It utilizes the idea of colour confinement principle: 
the quarks emerge in the geometry only in special combinations
of colour singlets.
Only two colour singlets are available (see below)
$$
(q\bar{q})=\FFr{1}{\sqrt{3}}\delta_{kk'}{\hat{q}}_{k}{\bar{\hat{q}}}_{k'}=
inv, 
\quad
(qqq)=\FFr{1}{\sqrt{6}}\varepsilon_{klm}{\hat{q}}_{k}{\hat{q}}_{l}
{\hat{q}}_{m}=inv.
$$

\renewcommand{\theequation}{\thesubsection.\arabic{equation}}
\subsection{Gauge principle-internal symmetries}
Each regular structure in the lowest state
can be regarded as a result of transition from an
arbitrary state, in which they assumed to be distorted. 
Hence, the following transformations may be implemented upon
distorted ordinary structures
\begin{equation}
\label {eq: R2.6.1}
{\ps1_{u}}'{}^{\lambda}(\theta'_{+l})=
f^{(+)}_{lk}{\ps1_{u}}{}^{\lambda}(\theta_{+k})=
f(\theta'_{+l},\theta_{-k}){\ps1_{u}}{}^{\lambda}(\theta_{+k}),\quad
f(\theta'_{+l},\theta_{-k})=
f_{(+)}(\theta'_{l})f_{(-)}(\theta_{k}).
\end{equation}
The transformation functions are the operators in the space 
of internal degrees of freedom labeled by $(\pm k)$ corresponding to 
distortion rotations around the axes $(\pm k)$ by the
angles  $\theta_{\pm k}$. 
We make proposition that the distortion rotations are incompatible, namely 
the transformation operators 
$f^{(\pm)}_{lk}$ obey the incompatibility relations
\begin{equation}
\label {eq: R2.6.2}
\begin{array}{l}
f^{(+)}_{lk}f^{(+)}_{cd}-f^{(+)}_{ld}f^{(+)}_{ck}=\|f^{(+)}\|
\varepsilon_{lcm}\varepsilon_{kdn}f^{(-)}_{nm},\\ 

f^{(-)}_{kl}f^{(-)}_{dc}-f^{(-)}_{dl}f^{(-)}_{kc}=\|f^{(-)}\|
\varepsilon_{lcm}\varepsilon_{kdn}f^{(+)}_{mn},
\end{array}
\end{equation}
where $l,k,c,d,m,n=1,2,3$.
This relations hold in general for
both local and global rotations.
Then one gets the transformations 
implemented upon the quark field, which in matrix notation take the form
$
q'(\zeta)=U(\theta(\zeta))q(\zeta), 
\quad
\bar{q'}(\zeta)=\bar{q}(\zeta)U^{+}(\theta(\zeta)),
$
where $q = \{ {q}_{k} \}, \quad U(\theta)=\{ f^{(+)}_{lk}  \}$. 
Due to the incompatibility commutation relations
the transformation matrices $\{ U \}$  generate
the unitary group of internal symmetries $U(1), SU(2),$ 
$SU(3)$, while an action of physical system must be invariant under
such transformations (Gauge principle).

\renewcommand{\theequation}{\thesection.\arabic{equation}}
\section {Operator multimanifold $\hat{G}_{N}$}
\label {Oper}
The OM formalism  $\hat{G}=
\hG_{\eta}\oplus\hG_{u}$ is built up by assuming 
an existence only of ordinary primordial structures of one sort 
(one u-channel).
To develop the microscopic approach to field theory
based on MW geometry, 
henceforth instead of one sort of ordinary structures we are 
going to deal with different species of ordinary structures. That is,
before we enlarge the previous model we must make an additional
assumption concerning 
an existence of infinite number of ${}^{i}u$-type ordinary
structures of different species $i=1,2,\ldots ,N $ (multi-u channel).
These structures will be specified by the superscript
to the left. 
This hypothesis,
as it will be seen in the following, leads to 
the progress of understanding of the properties of particles.
At the very outset we consider the processes
of creation and annihilation of regular structures of $\eta$- and 
${}^{i}u$-types in the lowest state ($s_{0}$).
The general rules stated in subsec 2.1 regarding to this change apply 
a substitution of operator basis pseudo vectors and covectors by a new
ones supplied with additional superscript $(i)$ to the left referred
to different species $(i=1,2,\ldots, N)$
\begin{equation}
\label {eq: R3.1}
{}^{i}\hat{O}^{r_{1}r_{2}}_{\lambda,\mu}=
{}^{i}\hat{O}^{r_{1}}_{\lambda}\otimes
{}^{i}\hat{O}^{r_{2}}_{\mu}\equiv
{}^{i}\hat{O}^{r}_{\lambda,\mu}={}^{i}O^{r}_{\lambda,\mu}
(\alpha_{\lambda}\otimes \alpha_{\mu}),
\end{equation}
provided $ r=(r_{1},r_{2})$ and
$$
\begin{array}{l}
{}^{i}O^{r}_{1,1}=\FFr{1}{\sqrt{2}}(\nu_{i}{\O1_{\eta}}_{+}^{r}+
{}^{i}{\O1_{u}}_{+}^{r}), \quad
{}^{i}O^{r}_{2,1}=\FFr{1}{\sqrt{2}}(\nu_{i}{\O1_{\eta}}_{+}^{r}-
{}^{i}{\O1_{u}}_{+}^{r}), \\ 
{}^{i}O^{r}_{1,2}=\FFr{1}{\sqrt{2}}(\nu_{i}{\O1_{\eta}}_{-}^{r}+
{}^{i}{\O1_{u}}_{-}^{r}), \quad
{}^{i}O^{r}_{2,2}=\FFr{1}{\sqrt{2}}(\nu_{i}{\O1_{\eta}}_{-}^{r}-
{}^{i}{\O1_{u}}_{-}^{r}),
\end{array}
$$
where 
\begin{equation}
\label {eq: R3.2}
<\nu_{i},\nu_{j}>=\delta_{ij},\quad
<{}^{i}{\O1_{u}}_{\lambda}^{r},{}^{i}{\O1_{u}}_{\tau}^{r'}>=
-\delta_{ij}\delta_{rr'}{}^{*}\delta_{\lambda\tau},\quad
<{\O1_{\eta}}_{\lambda}^{r},{}^{i}{\O1_{u}}_{\tau}^{r'}>=0.
\end{equation}
In analogy with subsec.2.1 we consider the operators
$
{}^{i}{\hat{\gamma}}^{r}_{(\lambda,\mu,\alpha)}=
{}^{i}{\hat{O}}^{r_{1}r_{2}}_{\lambda, \mu}
\otimes{\hat{\sigma}}^{r_{3}}_{\alpha}.$ 
and calculate nonzero matrix elements
\begin{equation}
\label{eq: R3.3}
<\lambda,\mu\mid {}^{i}{\hat{\gamma}}^{r}_{(\tau,\nu,\alpha)}\mid \tau,\nu>=
{}^{*}\delta_{\lambda\tau}{}^{*}\delta_{\mu,\nu}
{}^{i}e^{r}_{(\tau,\nu,\alpha)},
\end{equation}
where ${}^{i}e^{r}_{(\lambda,\mu,\alpha)}=
{}^{i}O^{r}_{\lambda,\mu} \otimes\sigma_{\alpha}$.
The set of operators $\left\{ 
{}^{i}{\hat{\gamma}}^{r} \right\}$
is the basis for all operator vectors 
$\hat{\Phi}(\zeta)={}^{i}{\hat{\gamma}}^{r}\,\,
{}^{i}\Phi_{r}(\zeta)$ of tangent section of 
principle bundle with the base of operator multimanifold
$\hat{G}_{N}=\left( \S_{i}^{N}\oplus {}^{*}\hat{\RR_{i}}^{4} \right)
\otimes\hat{R}^{3}$.
Here $^{*}\hat{\RR_{i}}^{4}$ is the $2\times2$ dimensional linear 
pseudo operator space, with the set of the linear unit operator 
pseudo vectors eq.(3.1) as the basis of tangent vector section,
and $\hat{R}^{3}$ is the three dimensional real linear operator space with
the basis consisted of the ordinary unit operator vectors  
$\{ {\hat{\sigma}}^{r}_{\alpha}\}$. 
The 
$\hat{G}_{N}$ decomposes as follows:
\begin{equation}
\label{eq: R3.4}
\hat{G}_{N}=\hG_{\eta}\oplus\hG_{u_{1}}
\oplus \cdots \oplus\hG_{u_{N}},
\end{equation}
where $\hG_{u_{i}}$ is the six dimensional operator
manifold of the species $(i)$ with the basis \\
$\left\{ {}^{i}{\hgam_{u}}{}^{r}_{(\lambda\alpha)} =
{}^{i}{\hat{\O1_{u}}}{}_{\lambda}^{r}\otimes {\hat{\sigma}}^{r}_{\alpha}
\right\}$.
The expansions of operator vectors and 
covectors  are written 
$\hps_{\eta}={\hgam_{\eta}}{}^{r}
{\ps1_{\eta}}{}_{r}, \quad
\hps_{u}= {}^{i}{\hgam_{u}}{}^{r}\,\,
{}^{i}{\ps1_{u}}{}_{r}, 
\quad
\bar{\hps_{\eta}}=
{\hgam_{\eta}}{}_{r}
{\ps1_{\eta}}{}^{r},\quad
\bar{\hps_{u}}=
{}^{i}{\hgam_{u}}{}_{r}\,\,
{}^{i}{\ps1_{u}}{}^{r},$
where the components ${\ps1_{\eta}}{}_{r}(\eta)$ and
${}^{i}{\ps1_{u}}{}_{r}(u)$ are respectively the 
link functions of $\eta$-type and ${}^{i}u$-type structures.
The quantum field and differential geometric aspects of  OMM $\hat{G}_{N}$
can be discussed on the analogy
of $\hat{G}_{N=1}$. 
Recall that we consider the special system of 
regular structures made of fundamental structure of $\eta$-type and 
infinite number of ${}^{i}u$-type ordinary structures of different species 
$(i=1,\ldots, N)$. 
The  primordial structures establish a stable linkage to form stable system:
\begin{equation}
\label{eq: R3.5}
p^{2}=p^{2}_{\eta}- \S^{N}_{i=1}p^{2}_{u_{i}}=0.
\end{equation}
The state of free ordinary structure of ${}^{i}u$-type with the given values 
of link momentum $\p1_{u_{i}}$ and spin projection $s_{i}$ is described 
by means of plane wave (see App. A).
We also involve the solution of negative
frequencies with the normalized bispinor amplitude.
\renewcommand{\theequation}{\thesubsection.\arabic{equation}} 
\subsection {Realization of multimanifold (MM) $G_{N}$}
\label {manif}
On analogy of subsec.2.2  
we make use of localized wave packets by
means of superposition of plane wave solutions furnished 
by creation and annihilation operators in agreement with Pauli's 
principle.
The new proposal includes and generalizes an early version of the relation 
\begin{equation}
\label{eq: R3.1.1}
\begin{array}{l}
\S_{\lambda=\pm}<\chi_{\lambda}\mid\hat{\Phi}(\zeta)
\bar{\hat{\Phi}}(\zeta)\mid
\chi_{\lambda}>= 
\S_{\lambda=\pm}<\chi_{\lambda}\mid
\bar{\hat{\Phi}}(\zeta)\hat{\Phi}(\zeta)\mid\chi_{\lambda}>= \\ 
-i\,\zeta^{2}{\G1_{\zeta}}(0) = -i\,\left(\eta^{2}{\G1_{\eta}}(0)-
\S^{N}_{i=1}{u_{i}}^{2}{\G1_{u_{i}}}(0)\right).
\end{array}
\end{equation}
provided by the Green's function 
$
\G1_{u_{i}}(u_{i}-u'_{i})=-(i\hpr_{u_{i}}+m)\Dlt_{u_{i}}(u_{i}-u'_{i}),
$
thus
\begin{equation}
\label{eq: R3.1.2}
\zeta^{2}{\G1_{\zeta}}{}_{F}(0) = \eta^{2}{\G1_{\eta}}{}_{F}(0)-
\S^{N}_{i=1}{u_{i}}^{2}{\G1_{u_{i}}}{}_{F}(0),
\end{equation}
where ${\G1_{\eta}}{}_{F},
{\G1_{u}}{}_{F}$ and ${\G1_{\zeta}}{}_{F}$ are causal Green's functions
of the $\eta-,u-$ and $\zeta$-type structures.
This result is of particular interest because along the same line with 
sec.2, the realization of the MM stems 
from the condition imposed  upon the matrix element eq.(3.1.1),
that as the bilinear form it is 
required to be finite
\begin{equation}
\label{eq: R3.1.3}
\S_{\lambda=\pm}<\chi_{\lambda}\mid\hat{\Phi}(\zeta)
\bar{\hat{\Phi}}(\zeta)\mid
\chi_{\lambda}> < \infty,
\end{equation}
or
$$
\zeta^{2}{\G1_{\zeta}}{}_{F}(0)< \infty.
$$
Denote
$
u^{2}\G1_{u}(0)\equiv
\Lm_{u_{i}\rightarrow u'_{i}}\S_{i=1}^{N}(u_{i}u'_{i})
\G1_{u_{i}}(u_{i}-u'_{i})
$
and consider a stable system eq.(3.5), i.e., as to the Green's functions they 
satisfy the condition
\begin{equation}
\label{eq: R3.1.4}
{\G1_{u}}{}_{F}(0)=
{\G1_{\eta}}{}_{F}(0)=
{\G1_{\zeta}}{}_{F}(0),
\end{equation}
provided
$m\equiv \left| p_{u}\right|=\left( \S_{i=1}^{N}
{p_{u_{i}}}^{2} \right)^{1/2}=\left| p_{\eta}\right|.$
According to eq.(3.1.3) and eq.(3.1.4),
the length of each vector
${\bf \zeta}={}^{i}e\,
{}^{i}\zeta\in G_{N}$
should be equaled zero
$\zeta^{2}=\eta^{2}-u^{2}=\eta^{2}-\S_{i=1}^{N}({u^{G}_{i}})^{2}=0,$
where use is made of
$$
u^{G}_{i}\equiv u_{i}
\left[ 
\Lm_{u_{i}\rightarrow u'_{i}}{\G1_{u_{i}}}{}_{F}(u_{i}-u'_{i})
\left. \right/
\Lm_{\eta\rightarrow\eta'}{\G1_{\eta}}{}_{F}(\eta-\eta')
\right]^{1/2},
$$
and 
$u^{G}_{i}={}^{i}{\he_{u}}{}_{(\lambda,\alpha)}
u_{i}^{G\,(\lambda,\alpha)}$.
Thus, the MM $G_{N}$ comes into being, which 
decomposes as follows:
\begin{equation}
\label{eq: R3.1.5}
G_{N}=\G1_{\eta}\oplus\G1_{u_{1}}\oplus\cdots
\oplus \G1_{u_{N}}.
\end{equation}
It brings us to the conclusion: the major requirement eq.(3.1.3)
provided by stability condition eq.(3.1) or eq.(3.1.4) yields 
the flat MM $G_{N}$.
Meanwhile the  Minkowski flat space $M_{4}$ stems from the
flat submanifold $\G1_{\eta}$ (subsec.2.1), in which the line 
element turned out to be invariant. That is,
the principle of Relativity comes into being with $M_{4}$ ensued from
the MW geometry $G_{N}$.
In the following we shall use a notion of $i$-th 
internal world for the submanifold $\G1_{u_{i}}$.

\subsection {Subquarks and subcolour confinement}
\label {sub}
Since our discussion within this section in many respects is similar to 
that of sec.2, here we will be brief.
We assume that the distortion rotations through the angles
${}^{i}\theta_{+k}$ and ${}^{i}\theta_{-k}\quad k=1,2,3$ occur 
separately in the three dimensional internal spaces 
${\RR_{u_{i}}}{}_{+}^{3}$ and ${\RR_{u_{i}}}{}_{-}^{3}$ 
composing six dimensional
distorted submanifold 
${\G1_{u_{i}}}\stackrel{\theta}{\rightarrow} \widetilde{\G1_{u_{i}}}=
{\RR_{u_{i}}}{}_{+}^{3}\oplus{\RR_{u_{i}}}{}_{-}^{3}$. 
As it is exemplified in subsec.2.4, the laws apply 
in use the wave packets constructed by superposition of the link functions
of distorted ordinary structures furnished by generalized operators of
creation and annihilation as the expansion coefficients
\begin{equation}
\label{eq: R3.2.3}
{\hps_{u}}(\theta_{+})=
\S_{\pm s}\IIn\frac{d^{3}p_{u_{i}}}{{(2\pi)}^{3/2}}
\left( {}^{i}{\hgam_{u}}{}_{(+\alpha)}^{k}
{}^{i}{\ps1_{u}}{}^{(+\alpha)}({}^{i}\theta_{+k})+
{}^{i}{\hgam_{u}}{}_{(-\alpha)}^{k}
{}^{i}{\ps1_{u}}{}^{(-\alpha)}({}^{i}\theta_{+k})\right), 
\end{equation}
etc. The fields ${}^{i}{\ps1_{u}}(\theta_{+k})$
and ${}^{i}{\ps1_{u}}(\theta_{-k})$
are defined on the distorted internal spaces
${\RR_{u_{i}}}{}_{+}^{3}$ and ${\RR_{u_{i}}}{}_{-}^{3}$.
The generalized expansion coefficients in eq.(3.2.1) imply
\begin{equation}   
\label{eq: R3.2.4}
<\chi_{-}\mid \{ {}^{i}{\hgam_{u}}{}^{(+\alpha)}_{k}(p_{u_{i}},s_{i}),\,\,
{}^{j}{\hgam_{u}}{}_{(+\beta)}^{k'}(p'_{u_{j}},s'_{j})\}\mid\chi_{-}>=
-\delta_{ij}\delta_{kk'}\delta_{ss'}\delta_{\alpha\beta}\delta^{3}
({\vec{p}}_{u_{i}}- {\vec{p'}}_{u_{i}}).
\end{equation}
The condition eq.(3.1.4) of MW geometry realization
reduces to
\begin{equation}
\label{eq: R3.2.5}
\S_{i=1}^{N}\omega_{i}
\left[ \Lm_{{}^{i}\theta_{+}\rightarrow {}^{i}\theta_{-}}
{\G1_{u_{i}}}{}^{\theta}_{F}({}^{i}\theta_{+}-{}^{i}\theta_{-})\right]=
\Lm_{\eta_{f}\rightarrow\eta'_{f}}
\G1_{\eta}(\eta_{f}-\eta'_{f}),
\end{equation}
provided
$\omega_{i}=\FFr{u_{i}^{2}}{u^{2}}.$
Taking into account the expression of causal Green's function
for given $(i)$, in the case if
\begin{equation}
\label{eq: R3.2.5}
\Lm_{
u_{i_{1}}\rightarrow u_{i_{2}}
}
{\G1_{u_{i_{1}}}}{}_{F}(u_{i_{1}}-u_{i_{2}})
= 
\Lm_{u'_{i_{1}}\rightarrow u'_{i_{2}}}
{\G1_{u_{i'_{1}}}}{}_{F}(u'_{i_{1}}-u'_{i_{2}})
=\cdots=inv,
\end{equation}
one gets
\begin{equation}
\label{eq: R5.16}
\S_{k}\,
{}^{i}{\ps1_{u}}({}^{i}\theta_{+k})\,\,
{}^{i}{\bar{\ps1_{u}}}({}^{i}\theta_{-k})=
\S_{k}\,
{}^{i}{\ps1_{u}}'({}^{i}\theta'_{+k})\,\,
{}^{i}{\bar{\ps1_{u}}}'({}^{i}\theta'_{-k})=\cdots = inv.
\end{equation}
Thus, in the context of MW geometry it is legitimate to substitute 
a formerly introduced term of "quark" $(q_{k})$ (sec.2.5) by
"subquark" $({}^{i}q_{k})$ defined on the given internal world.
Everything said will then remain valid, provided we make also a simple 
change of colours into subcolours.
Hence, we may think of the function 
${}^{i}{\ps1_{u}}({}^{i}\theta_{+k})$ as being $u$-component of
bispinor field of subquark $({}^{i}q_{k})$ of species $(i)$
with subcolour $k$, and respectively 
${}^{i}{\bar{\ps1_{u}}}({}^{i}\theta_{-k})$ -conjugated bispinor field 
of antisubcolour $(k)$.
The subquarks and antisubquarks may be local $({}^{i}q_{k})$ or
global $({}^{i}q_{k}^{c})$.
Whence, the subquark $({}^{i}q_{k})$ is the fermion with the 
half integer spin and subcolour degree of freedom, and,
according to eq.(3.2.7), could emerge on mass shell only in confined phase 
\begin{equation}
\label{eq: R3.2.8}
\S_{k}\,
{}^{i}{q}_{kp}\,{}^{i}{\bar{q}}_{kp} =
\S_{k}\,
{}^{i}{q'}_{kp}\,{}^{i}{\bar{q'}}_{kp}=
\cdots=inv.
\end{equation}
To trace a resemblance with sec.3 in [1], the internal 
symmetry group
${}^{i}G=U(1), SU(2),$\\
$ SU(3)$ enables to introduce the gauge theory
in internal world with the subcolour charges as exactly conserved 
quantities. Thereto the subcolour transformation are implemented on
subquark fields right through local and global rotation matrices of
group ${}^{i}G$ in fundamental representation. 
Due to the Noether procedure the 
conservation of global charges ensued from the global 
gauge invariance of physical system, while reinforced requirement
of local gauge invariance may be satisfied as well by introducing the 
gauge fields with the values in Lie algebra ${}^{i}\hat{g}$ of group
${}^{i}G$.

\renewcommand{\theequation}{\thesection.\arabic{equation}}
\section {The subquark algebra and  supercharges}
\label {Supcharg}
According to eq.(2.5.1) the following transformations are implemented upon the 
subquarks (antisubquarks) on the given (i) internal world:
\begin{equation}
\label {eq: R4.1}
q'_{l}=f^{(+)}_{lk} \,\, q_{k}, \quad
\bar{q}'_{l}=\bar{q}_{k}\,\,f^{(-)}_{kl},  
\end{equation}
where as well for the next section we left implicit the MW-superscript $(i)$ 
to the left. Then, the following composition rules hold 
for the transformation functions
\begin{equation}
\label {eq: R4.2}
f^{(+)}_{lk} = f_{l} \circ f^{-1}_{k}, \quad
f^{(-)}_{lk} = \bar{f}_{l} \circ \bar{f}^{-1}_{k}, \quad
( f_{l} \circ f^{-1}_{k})( f_{c} \circ f^{-1}_{d})=
( f_{l} \, f_{c})\circ ( f^{-1}_{k} \, f^{-1}_{d}),
\end{equation}
where $l,k,c,d = 1,2,3,$ the transformation functions 
$ f_{k}\equiv f_{(+)}(\theta_{+k})$ and 
$\bar{f}_{l}\equiv f_{(-)}(\theta_{-k})$ are the operators in the space of 
internal degrees of freedom labeled by the subcolour index $(\pm k)$ such that 
the rotation through the angle $\theta_{\pm k}$ yields the subquark 
(antisubquark) field
\begin{equation}
\label {eq: R4.3}
q_{k} = f_{k} \, q_{0}, \quad
\bar{q}_{k} = \bar{q}_{0}\, \bar{f}_{k}. 
\end{equation}
The incompatibility commutation relations eq.(2.5.2) with the composition 
rule eq.(4.2) lead to the following commutation relations
\begin{equation}
\label {eq: R4.4}
\left[ f_{l},f_{k}\right] =\epsilon _{lkm}\,\bar{f}_{m}.
\end{equation}
Whence, the subquarks imply
\begin{equation}
\label {eq: R4.5}
\left[ q_{l},q_{k}\right] =Q_{0}\,\epsilon _{lkm}\,\bar{q}_{m}, 
\quad Q_{0}\equiv q_{0}^{2}\left/ \bar{q}_{0}\right..
\end{equation}
Following to [2] the symmetries of the $C-$ $(C\equiv s,c,b,t)$ and $Q-$ 
worlds are assumed to be respectively local and global unitary 
$diag(SU(3))$ symmetries (see subsec.12.1 and 12.2 ), for which the 
eq.(4.5) reduced to
\begin{equation}
\label {eq: R4.6}
q_{l}\,q_{k}=Q'_{0}\,\epsilon _{lkm}\,\bar{q}_{m}, 
\end{equation}
while for the $W-$world with the unified symmetry $SU(2)_{L}\otimes U(1)$ 
(subsec.12.8) the eq.(4.5) reduced to
\begin{equation}
\label {eq: R4.7}
\left[ q_{1L},q_{2L}\right] =Q_{0}\,\bar{q}_{2R}, \quad
\left[ q_{2L},q_{2R}\right] =Q_{0}\,\bar{q}_{1L}, \quad
\left[ q_{2R},q_{1L}\right] =Q_{0}\,\bar{q}_{2L}, 
\end{equation}
where the subcolour singlets are 
$Q_{2R}, \left[ q_{1L},q_{2L}\right]$ and 
$(q\,\bar{q}).$
Hence, for the electron and corresponding neutrino one gets (subsec.12.4)
\begin{equation}
\label {eq: R4.8}
\left[ \nu_{L}, e_{L} \right] =Q_{0} \, \bar{e}_{R}, \quad
\left[ e_{L}, e_{R}\right] =Q_{0} \,\bar{\nu}_{L}, \quad
\left[ e_{R}, \nu_{L}\right] =Q_{0} \,\bar{e}_{L}. 
\end{equation}
The eq.(4.5) yields the important relation
between the fermionic $(F)$ and bosonic $(B)$ subcolour singlets
\begin{equation}
\label {eq: R4.9}
(qqq)\equiv \FFr{1}{\sqrt{6}}\epsilon _{lkm}\,q_{l}q_{k}q_{m}=\Q_{u}\, 
(q\,\bar{q})\equiv \Q_{u}\,\FFr{1}{\sqrt{3}}
(q_{k}\,\bar{q}_{k}), \quad F\rightarrow \Q_{u}\,B, 
\end{equation}
and vice versa, where $F\equiv (qqq), \quad B\equiv (q\,\bar{q}),
\quad \Q_{u}\equiv\FFr{1}{\sqrt{2}}Q_{0} $.
It means that considered physical system must respect the invariance under 
a symmetry group of the fermion-boson transformations 
occurred in the internal worlds. The latter is known as a ``supersymmetry''. 
It is why the basis vectors in the 
Hilbert space $\cal{H}$ have taken to be in the form $\mid n_{B}n_{F}>$, 
where the boson and fermion occupation numbers respectively are 
$n_{B}=1,2,...,\infty $ and $n_{F}=0,1$. 
It is convenient, then, to describe such a 
quantum mechanical system as the spin-$1/2$ like supersymmetric particle 
with mass $m=\left( \FFr{\hbar}{\Q_{u}}\right)^{2}$ moving along the 
one-dimensional Euclidean line $\cal{R}$. Therefore, one introduces a 
generalized bosonic operator $b$ and a fermionic operator $f$ acting on the 
Hilbert space ${\cal H}= L^{2} ({\cal R})\otimes {\cal C}^{2}$: 
\begin{equation}
\label {eq: R4.10}
\begin{array}{l}
b:\,  L^{2} ({\cal R})\rightarrow  L^{2} ({\cal R}), \quad 
b=\FFr{1}{2}\left( \FFr{\partial}{\partial \, u}+ W(u) \right) \\
f:\, {\cal C}^{2} \rightarrow {\cal C}^{2}, \quad 
f=\left( \begin{array}{ll}
0 \,\, 0 \\
1 \,\, 0
\end{array}
\right),
\end{array}
\end{equation}
where the supersymmetric potential $W(u):\,{\cal R}\rightarrow {\cal R} $ 
defined on the given $(i)$ internal world is assumed to be piecewise 
continuously differentiable function. 
The commutation and anticommutation relations for these operators read
\begin{equation}
\label {eq: R4.11}
[b,\,b^{+}]=W'(u), \quad \{f,\,f^{+}\}=1.
\end{equation}
Employing standard technique, in a line with eq.(4.9), next we define the 
nilpotent supercharge operators 
\begin{equation}
\label {eq: R4.12}
\1Q_{u}=\Q_{u}\,b\otimes f^{+}=\Q_{u}\,\left( \begin{array}{ll}
0 \,\, b \\
0 \,\, 0
\end{array}
\right), \quad 
{\1Q_{u}}^{+}=\Q_{u}\,b^{+} \otimes f=\Q_{u}\,\left( \begin{array}{ll}
0 \,\, 0 \\
b^{+} \,\, 0
\end{array}
\right),
\end{equation}
which obey the anticommutation relations
\begin{equation}
\label {eq: R4.13}
\{\1Q_{u},\,\1Q_{u}\}=\{{\1Q_{u}}^{+},\,{\1Q_{u}}^{+}\}=0,
\end{equation}
and according to eq.(4.9) act as follows: 
\begin{equation}
\label {eq: R4.14}
\begin{array}{l}
\1Q_{u}\mid n_{B}, n_{F}>\,\, \propto \,\,\mid n_{B}-1, n_{F}+1>, \quad
{\1Q_{u}}^{+}\mid n_{B}, n_{F}>\,\, \propto \,\,\mid n_{B}+1, n_{F}-1>. 
\end{array}
\end{equation}

\section {SUSY quantum mechanics on the given internal world}
\label {qmech}
The quantum mechanical system of previous section is described by 
the self-adjoint Hamiltonian
\begin{equation}
\label {eq: R5.1}
\begin{array}{l}
H=\{\1Q_{u},\1Q_{u}\}={\Q_{u}}^{2}\,
\left( \begin{array}{ll}
b\,b^{+} \,\, 0 \\
0 \,\, \,\,b^{+}b
\end{array}
\right)=\left( \begin{array}{ll}
{\H1_{u}}_{+} \,\, 0 \\
0 \,\,\,\,{\H1_{u}}_{-} 
\end{array}
\right),
\end{array}
\end{equation}
provided by
\begin{equation}
\label {eq: R5.2}
{\H1_{u}}_{\pm}=\FFr{1}{2}{\Q_{u}}^{2}\,
\left[ -
\FFr{\partial^{2}}{\partial \, u^{2}}+ W^{2}(u) \pm W'(u)
\right]\ge 0,
\end{equation}
which are a standard Schr\"odinger operators of the conventional SUSY 
quantum mechanics (see e.g. [64-79] and references therein) acting on 
$ L^{2} ({\cal R})$. The eq.(4.13) and eq.(5.1) provide the conservation 
of the supercharge and the non-negativity of the Hamiltonian
\begin{equation}
\label {eq: R5.3}
\left[ 
\H1_{u}, \1Q_{u}
\right]=\left[ 
{\H1_{u}}, {\1Q_{u}}^{+}
\right] 
=0, \quad {\H1_{u}}\ge 0.
\end{equation}
Replacing the operators eq.(5.3) with a new ones
\begin{equation}
\label {eq: R5.4}
{\1Q_{u}}_{1}=\1Q_{u} + {\1Q_{u}}^{+}, \quad
{\1Q_{u}}_{2}=\1Q_{u} - {\1Q_{u}}^{+},
\end{equation}
one gets the superalgebra
\begin{equation}
\label {eq: R5.5}
\left
\{{\1Q_{u}}_{i}, {\1Q_{u}}_{k}
\right\}=2 \delta_{ik}\,\H1_{u},\quad
\left[
\1Q_{u}, \H1_{u}
\right]=0.
\end{equation}
A multiplicity of degeneracy of the levels of Hamiltonian $\H1_{u}$ with the 
energy $\E1_{u}$ equals to a dimension of invariant subspace with respect to 
the action of all the $\1Q_{u}$. If $\E1_{u}=0$, then the corresponding 
subspace is one-dimensional - a level of zero point energy. In general, the 
superalgebra eq.(5.5) with an arbitrary number of the supercharge operators  
${\1Q_{u}}_{i} (i=1,...,N)$ defines the Clifford algebra with the basis of 
${\q1_{u}}_{i}= {\1Q_{u}}_{i}\left/\sqrt{\E1_{u}}\right.$ for nonzero energy 
levels of $\H1_{u}$, which remains a central point in the SUSY theories. 
Due to it a definition of the multiplicity of degeneracy of the levels 
reduced to a definition of a dimension of the representations of the 
Clifford algebra, which is well known. For the even and odd number $N$ 
a dimension of the representation of Clifford algebra is given by the 
formula
\begin{equation}
\label {eq: R5.6}
\nu = 2^{n} = 2^{\left[N/2\right]},
\end{equation}
where $[\cdots]$ means the integer part, namely the $\nu$ defines a number of 
states in given supermultiplet. Thus, incompatibility relations eq.(2.6.2) 
yields the major law for the supermultiplets that each of them contains an 
equal number of fermion and boson degrees of freedom
\begin{equation}
\label {eq: R5.7}
n_{B}=n_{F}.
\end{equation}
Certainly, considering the operator $(-1)^{2S}$, where $S$ is the spin angular 
momentum having eigenvalue $+1$ acting on a bosonic state and eigenvalue $-1$ 
acting on a fermionic state, by means of eq.(5.1) one gets
\begin{equation}
\label {eq: R5.8}
\begin{array}{l}
\S_{i}<i\mid (-1)^{2S}  \H1_{u} \mid i>=
\S_{i}<i\mid (-1)^{2S}   \1Q_{u} \,{\1Q_{u}}^{+} \mid i>+
\S_{i}<i\mid (-1)^{2S}   {\1Q_{u}}^{+} \1Q_{u} \mid i>=\\
\S_{i}<i\mid (-1)^{2S}   \1Q_{u} \,{\1Q_{u}}^{+} \mid i>-
\S_{j}<j\mid (-1)^{2S}   \1Q_{u} \,{\1Q_{u}}^{+} \mid j>=0.
\end{array}
\end{equation}
Here one has used the relation of completeness
$\S_{i}\mid i>\, < i \mid =1$ within the subspace of states invariant 
with respect to the action of $\1Q_{u}$ and $ {\1Q_{u}}^{+}$, and the fact 
that the operator $(-1)^{2S}$ must anticommute with  $\1Q_{u}$. The 
$\S_{i}<i\mid (-1)^{2S}  \H1_{u} \mid i>=\E1_{u}Tr\left[ (-1)^{2S}\right]$   
is proportional to the number of bosonic degrees of freedom $n_{B}$ minus the 
number of fermionic degrees of freedom $n_{F}$ in the trace. Hence, the 
eq.(5.7) holds for any  $\E1_{u} \neq 0$ in each supermultiplet. \\ 
The concept of shape invariance [66] in SUSY quantum mechanics has proven to be 
very useful because it leads immediately to exactly solvable potentials, 
namely a subset of the potentials for which the Schr\"odinger equations are 
exactly solvable share an integrability condition, while the partner 
potentials
\begin{equation}
\label {eq: R5.9}
V_{1}(u)=W^{2}(u)-\FFr{\Q_{u}}{\sqrt{2}}W'(u),\quad
V_{2}(u)=W^{2}(u)+\FFr{\Q_{u}}{\sqrt{2}}W'(u),
\end{equation}
satisfy the condition of shape invariance
\begin{equation}
\label {eq: R5.10}
V_{2}(u; a_{1}) = V_{1}(u; a_{2})+R(a_{1}),
\end{equation}
where $a_{1,2}$ are a set of parameters specifying space-independent 
properties of the potentials. Therefore, the Hamiltonian eq.(5.2) can be 
readily factorized $ \H1_{u}-{\E1_{u}}_{0}={\hat{A}}^{+}\hat{A},$ where 
${\E1_{u}}_{0}$ is the ground state energy, and 
\begin{equation}
\label {eq: R5.11}
\hat{A}=W(u)+\FFr{i\Q_{u}}{\hbar\,\sqrt{2}}\hat{P}_{u},\quad
{\hat{A}}^{+}=W(u)-\FFr{i\Q_{u}}{\hbar\,\sqrt{2}}\hat{P}_{u}.
\end{equation}
The ground state wavefunction of  $ \H1_{u}$ either belongs to 
${\H1_{u}}_{+}$ or ${\H1_{u}}_{-}$
\begin{equation}
\label {eq: R5.12}
\psi^{\pm}_{0}(u)=\psi^{\pm}_{0}(\theta)\exp 
\left(
\pm \IIn^{u}_{0}W(u)\, du
\right)
\end{equation}
satisfies the condition $\hat{A}\mid \psi_{0}> = 0.$
The shape invariance has an underlying algebraic structure [75]. Depending on 
the asymptotic behavior of the superpotential $W(u)$ one of the two functions 
$\psi^{\pm}_{0}$ will be normalizable (good SUSY) or both are not be 
normalizable (broken SUSY). Clearing up this situation the Witten index [65]
is turned out to be one of the useful tool 
which, according to the Atiyah-Singer index theorem [79,80], associates with 
the operator index. The latter is a topological characteristic and does not 
vary with the variation of the parameters of theory. 
Thus, the Witten index
\begin{equation}
\label {eq: R5.13}
\Delta (\beta) =Tr
\left(
P\, e^{-\beta  \H1_{u}}
\right), \quad \beta > 0,
\end{equation}
depends only on the asymptotic values of $W(u).$ Here a self-adjoint operator 
$P=P^{+}$, called Witten parity, which anticommutes with the supercharges, and therefore commutes with the Hamiltonian, i.e. 
$$
\{ \1Q_{u}{}_{i},\, P \}=0, \quad [H,\, P]=0, \quad P^{2}=1,
$$
explicitly can be written 
$P=I\otimes \sigma_{3}=
\left( \begin{array}{ll}
1 \,\, \,\,\,0 \\
0 \, \,-1 
\end{array}
\right).$
Then, if $\Delta (\beta) \neq 0$ one has a good SUSY.

\section{ $n \geq 1$ MW-SUSY on the internal worlds}
\label{susy}
Hereinafter we shall use the four-dimensional Weyl spinor notation for the 
$n$ supersymmetry generators $Q^{A}_{\alpha} (A=1,...,n; \,\, \alpha =1,2))$. 
The relativistic generalization of eq.(5.1)-eq.(5.5) for the $n \ge 1$ MW-SUSY 
algebra with the complex scalar central charges $X^{AB}$ defined on the given 
$i$-th (i=1,..., N) internal manifold $\G1_{u_{i}}$ reads
\begin{equation}
\label {eq: R6.1}
\begin{array}{l}
\{ 
{\1Q_{u}}{}^{A}_{\alpha}, \, {\bar{\1Q_{u}}}{}_{\dot{\beta} \,B}
\}=2\,\,{}^{i}{\sig1_{u}}{}^{(\lambda, \delta)}_{\alpha \,\dot{\beta}}
\,{\P_{u_{i}}}{}_{(\lambda, \delta)}\,\delta_{B}{}^{A},\quad
\{
{\1Q_{u}}{}^{A}_{\alpha}, \, {\1Q_{u}}{}_{\beta}^{B}
\}=\epsilon_{\alpha\beta}\, X^{AB},\\ \\
\left[ 
{\1Q_{u}}{}^{A}_{\alpha}, \,{\P_{u_{i}}}{}_{(\lambda, \delta)}
\right]=
\left[ 
{\bar{\1Q_{u}}}{}_{\dot{\alpha}\, A}, \,{\P_{u_{i}}}{}_{(\lambda, \delta)}
\right]=
\left[ 
{\P_{u_{i}}}{}_{(\lambda, \delta)}, \, {\P_{u_{i}}}{}_{(\tau, \rho)}
\right]=0.
\end{array}
\end{equation}
Here we have used the MW-notation, namely, $\lambda = \pm, \delta = 1,2,3,$ 
$\epsilon_{\alpha\beta}$ is the 2-dimensional Levi-Civita symbol, and
imposed the convention 
${\1Q_{u_{1}}}{}^{A}_{\alpha}=\cdots={\1Q_{u_{N}}}{}^{A}_{\alpha}\equiv 
{\1Q_{u}}{}^{A}_{\alpha}$. The matrices 
${}^{i}{\sig1_{u}}^{(\lambda, \delta)}$ read
\begin{equation}
\label {eq: R6.2}
{}^{i}{\sig1_{u}}{}^{(\pm \delta)} = \FFr{1}{\sqrt{2}}
\left(
\kappa^{\delta}\,\sigma^{0}_{i} \pm \sigma^{\delta}_{i}
\right),
\end{equation}
where $<\kappa^{\delta},\, \kappa^{\rho}>=\delta^{\delta\rho},\,\,
\sigma^{m}_{i}\equiv (\sigma^{0}_{i},\, \vec{\sigma}_{i})$ 
are the $i$-th sample of the Pauli 
matrices such that $\bar{\sigma}^{m}_{i}\equiv 
(\sigma^{0}_{i},\, -\vec{\sigma}_{i})$, and
$$
\{
\sigma^{m}_{i}, \, \sigma^{n}_{j}
\}
=
\{
\sigma^{m}_{i}, \, \bar{\sigma}^{n}_{j}
\}
=
\{
\bar{\sigma}^{m}_{i}, \, \bar{\sigma}^{n}_{j}
\}
=0,
\quad \mbox{if}
\,\, i \neq j.
$$
To trace a maximal resemblance in outward appearance to the conventional SUSY 
theories [24-59] within the next two sections we omit the subscript 
$(u_{i})$ and the MW-index (i) to the left in eq.(6.1), denote the 
6-dimensional vector index by the $m \equiv (\lambda, \delta)$. 
But it goes without saying that all 
the results obtained refer to the given 6-dimensional internal world 
$\G1_{u_{i}}$. The resulting mathematical structure of eq.(6.1) is closely 
similar to those of conventional SUSY algebra, although not identical. Thus, 
adopting a standard technique in the following it is worth briefly recording 
some questions but we shall forbear to write them out in details as they are 
so standard. We start with a discussion of the representations of SUSY on 
asymptotic single particle states. The Casimirs for $n=1$ MW-SUSY 
(the extension to $n > 1$ is straightforward) are $P^{2}=P_{m}P^{m}$ and 
$C^{2}=C_{mn}C^{mn}$, where $C_{mn}=B_{M}P_{n}-B_{n}P_{m}, \,\, B_{m}= W_{m}-
\FFr{1}{4}\bar{Q}_{\dot{\alpha}}\,\bar{\sigma}^{\dot{\alpha}\, \beta}_{m}\,
Q_{\beta},\,\, W_{m}$ is the Pauli-Ljubanski vector, which has eigenvalues 
$-m^{2}\,s(s+1),\,\, s=0,\FFr{1}{2},1,...$ for massive states, and 
$W_{m}=\lambda P_{m}$ for massless states with the helicity $\lambda$.
For the $n=1$ MW-SUSY, no central charges, massive states (from the rest 
frame) the Clifford vacuum state $\mid \Omega >$ is actually an eigenstate 
of spin angular momentum  $\mid \Omega >=m,s,s_{3}$. Hence all the states 
in the MW-SUSY irrep are characterized only by the mass and spin. The 
normalized creation and annihilation operators are defined as follows: 
\begin{equation}
\label {eq: R6.3}
a_{1,2}=\FFr{1}{2\,m}Q_{1,2}, \quad 
a^{+}_{1,2}=\FFr{1}{2\,m}\bar{Q}_{1,2},  
\end{equation}
while for a given $\mid \Omega >$ the full massive MW-SUSY irrep contains the 
states $\mid \Omega >,\\ a^{+}_{1}\mid \Omega >,\,\, 
a^{+}_{2}\mid \Omega >$ and 
$\FFr{1}{\sqrt{2}} a^{+}_{1}a^{+}_{2}\mid \Omega >=
-\FFr{1}{\sqrt{2}} a^{+}_{2}a^{+}_{1}\mid \Omega >$. There 
are a total of 4(2j +1) states in the massive irrep, where $j$ equal integers 
or half-integers. In the fundamental $n=1$ MW-SUSY massive irrep (j = 0) 
there are four states with $S_{3}=0,-\FFr{1}{2}, \FFr{1}{2}$ and $0$, 
respectively. 
Taking into account that the parity operation interchanges $a^{+}_{1}$ and 
$a^{+}_{2}$, then the fundamental massive irrep contains one massive Weyl 
fermion, one real scalar and one real pseudoscalar. In the same manner one 
gets for the massless states of the $n=1$ MW-SUSY, no central charges, that 
the $\mid \Omega >$ is nondegenerate and has definite helicity 
$\lambda$- there is only one pair of creation and annihilation operators 
$(\bar{Q}_{2}\equiv 0)$ and the massless $n=1$ MW-SUSY irrep contains two 
states $\mid \Omega >$ with helicity  $\lambda$ and $\lambda +\FFr{1}{2}$. 
Such an analysis can be readily extended to the massive and massless states 
of the $ n > 1$ MW-SUSY, no central charges [26,35,41]. For example, there are 
$2^{2n}(2j +1)$ states in a massive irrep, while the helicity in the 
massless irrep are 
$\lambda, \lambda=\FFr{1}{2}, ... , \lambda + \FFr{n}{2}$.\\
In the presence of the central charges with the Wess and Bagger convention 
$(X^{A\,B}=-X_{A\,B})$ the basis to describe them must be rediagonalized 
using the Zumino's decomposition for the massive state in the rest frame, 
where the eigenvalues of the central charges $Z_{M}$ are real and 
non-negative. As far as the $2n$ pairs of creation and annihilation operators 
$\{a,\, a^{+}\}$ and $\{b,\, b^{+}\}$ are positive definite operators and 
$Z_{M}$ are non-negative, then: 1) in any MW-SUSY irrep $Z_{M}\leq 2m$; 2) if 
$ Z_{M}\leq 2m$ then the multiplicities of the massive irreps are the same 
as the case of no central charges; 3) if one saturates the bound, i.e.  
$Z_{M} = 2m$ for some or all $Z_{M}$, there are the reduced massive 
multiplets called short multiplets and the states are referred to as BPS-
saturated states after the BPS monopoles in SUSY gauge theories. 
The MW-SUSY irreps, like the conventional SUSY, on asymptotic single 
particle state will automatically carry a representation of the 
automorphism group
($\1Q_{u}{}^{A}_{\alpha}\rightarrow U^{A}{}_{B}\,\1Q_{u}{}^{B}_{\alpha}, 
\quad
\bar{\1Q_{u}}{}_{\dot{\alpha}\, A}\rightarrow \bar{\1Q_{u}}{}
_{\dot{\alpha}\, B}\,U^{+}{}^{B}{}_{A}$). For massless irreps $U(N)$ is 
the largest automorphism symmetry which respects helicity, while the same for 
massive irreps which respects spin is the unitary symplectic group of rank $N$ 
- $USp(2N)$. In the presence of central charges, if none of them saturates 
the BPS bound, the automorphism group still remains $USp(2N)$. At last, if 
one central charge saturates the BPS bound, the automorphism group reduces to 
$USp(N)$ for $N$ even, or $USp(N+1)$ for $N$ odd. The automorphism symmetries 
impose some constraints on the internal symmetry group.\\
When the MW-SUSY is represented on quantum fields, instead of asymptotic 
states, the Clifford vacuum condition reads
\begin{equation}
\label {eq: R6.4}
\left[
\bar{Q}_{\dot{\alpha}},\, \mid \Omega >
\right]=0,
\end{equation}
where $\mid \Omega >$ must be a complete scalar field.

\section{ Coset construction and $n=1$ rigid internal manifold 
$\SG1_{u}$}
\label{coset}
A universal geometric description of field systems respecting the 
invariance under a symmetry group $G$ is provided by the method of nonlinear 
or the coset space realizations. One considers $G$ as a group of 
transformations acting in some coset space  with an 
properly chosen stability group $H$ and identifies the coset parameters 
with fields. In conventional SUSY theories is widely used the superspace-
superfield formalism, i.e. the manifestly supersymmetric technique for 
constructing superfields carried out by a generalization of the coset 
construction [81-93]. Introducing constant Grassmann spinors $\theta^{\alpha},\,
\bar{\theta}_{\dot{\alpha}}$, one rewrites the $n=1$ MW-SUSY algebra eq.(6.1) 
as a Lie algebra of supergroup $G$, where the Maurer-Cartan form is valued. 
One constructs the coset $G \left/ H \right.\,\,\Omega \sim \Omega \, h, 
\,\, \Omega \in G,\,\, h \in H$, where $H$ is the stability group closing 
the generators of unbroken 6-dimensional rotations in $\G1_{u}$ and 
unbroken internal symmetries. Given a Lie algebra one can exponentiate to 
yield the general group element
\begin{equation}
\label {eq: R7.1}
\Omega(u,\theta,\bar{\theta},\omega)=e^{i(-u^{m}P_{m}+\theta Q +
\bar{\theta}\bar{Q})}\, e^{-\FFr{1}{2}\omega^{mn}M_{mn}},
\end{equation}
where one chooses to keep all of G unbroken $(h=1)$.
Whence it is clear that $Z^{A}_{u}\equiv (u,\theta,\bar{\theta})$ 
parametrizes a $(6+4)$-dimensional coset space: $n=1$ rigid (global SUSY) 
internal supermanifold $\SG1_{u}$ (internal super world).
A SUSY transformation is specified by the group element
\begin{equation}
\label {eq: R7.2}
g= e^{i( \xi ^{\alpha} Q_{\alpha} +
\bar{\xi}_{\dot{\alpha}}\bar{Q}^{\dot{\alpha}})}
\end{equation}
provided by conventional Grassmann parameters $(\xi\, \bar{\xi})$. 
The differential operators $Q$ and $\bar{Q}$ closed into the MW-SUSY algebra
and explicitly  read in standard manner.  
The Maurer-Cartan form $\Omega^{-1}\,\Omega$ transforms under a rigid group 
$G$ 
\begin{equation}
\label {eq: R7.3}
\Omega \rightarrow g\,\Omega\,h^{-1}, \quad 
\Omega^{-1} \,d\,\Omega\,\rightarrow h\,(\Omega^{-1} \,d\,\Omega\,)\,h^{-1}-
d\,h\,\,h^{-1},
\end{equation}
which induces the motion $(n=1)$
\begin{equation}
\label {eq: R7.4}
\begin{array}{l}
u^{m}\rightarrow u^{m} + i\theta \sigma^{m}\bar{\xi}-i\xi 
\sigma^{m}\bar{\theta},\\
\theta^{\alpha}\rightarrow \theta^{\alpha}+ \xi^{\alpha}, \quad 
\bar{\theta}_{\dot{\alpha}}\rightarrow \bar{\theta}_{\dot{\alpha}}+ 
\bar{\xi}_{\dot{\alpha}}.
\end{array}
\end{equation}
Computing the Maurer-Cartan form one can extract the covariant vielbein and 
the $H$ connection $\omega^{i}=0$. The $(6+4)$-dimensional vielbein 
$E_{M}{}^{A}$ and spin connection $W_{A}^{mn}$ enable to write the general 
form of covariant derivative in a such a space
\begin{equation}
\label {eq: R7.5}
D_{M}=E_{M}{}^{A}(\partial_{A}+\FFr{1}{2}\,W_{A}^{mn}M_{mn}),
\end{equation}
where $M$ is to denote an superspace index, $A$ is the super-tangent space 
index. With these building blocks it is readily to treat in fairy conventional 
manner a whole formalism of $n=1$ superfields and total supersymmetric 
actions invariant under the group $G$.

\section{ Broken super operator multimanifold $\hat{\rlap/S G}_{N}$}
\label{bsomm}
Following [2] (see App.A) the frame field of the particles (left-handed 
leptons and quarks) defined on the MW geometry  
$G_{N}=\G1_{\eta}\oplus
\G1_{u_{1}}\oplus \cdots\oplus\G1_{u_{N}}$
can be written
\begin{equation}
\label{eq: R8.1}
\chi_{p}={\ch1_{\eta}}{}_{p}(\eta){\ch1_{u}}{}_{p}(u),\quad
{\ch1_{u}}{}_{p}(u)\equiv\ch1_{u_{1}}(u_{1})\cdots \ch1_{u_{N}}(u_{N}).
\end{equation}
Here $\ch1_{u_{i}}$ is the left-handed Weyl spinor field defined 
on the manifold where the subquarks of corresponding 
species are involved, the suffix $p$ refers to particle. To develop 
some feeling for our approach and to avoid irrelevant complications 
here temporarily we skeletonize the microscopic structure of the particles 
presented it in complete generality, but in the subsec.12.4 and 12.5 we 
shall specialize to the case of leptons and quarks. To reproduce a 
supersymmetric generalization of this model the idea is now to promote 
$\G1_{u_{i}}$ into the $n=1$ rigid internal supermanifold 
$\SG1_{u_{i}}$. Accordingly, each left-handed fermionic 
$\ch1_{u_{i}}(u_{i})$ component must be promoted to chiral multiplets by 
adding superpartner as required by SUSY.  Corresponding to 
$\ch1_{u_{i}}(u_{i})$ the chiral superfield 
$\widetilde{\PH1_{u_{i}}}{}_{ch}(u_{i})$ contains off-shell four real 
fermionic degrees of freedom $(n_{F}=4)$ and four real bosonic degrees of 
freedom $(n_{B}=4)$, namely, it contains a left-handed Weyl fermion 
$\chi_{p\,\alpha}(u_{i})$ of spin $\FFr{1}{2}$ and mass dimension    
$\FFr{3}{2}$, a complex scalar field $A_{sp}(u_{i})$ of spin $0$ and 
mass dimension $1$, and an auxiliary complex scalar field $F(u_{i})$ of 
spin $0$ and mass dimension $2$ (a suffix (s) denotes a sparticle):
$\widetilde{\PH1_{u_{i}}}(u_{i})=
\left(
\chi_{p\,\alpha}(u_{i}), A_{sp}(u_{i}), F(u_{i}) 
\right).$  
Since the chain rule holds, i.e., any product of chiral superfields is also 
a chiral superfield, then the $u$-component in eq.(8.1) can be rewritten as 
follows: 
\begin{equation}
\label{eq: R8.2}
\ch1_{u}(u)\rightarrow \widetilde{\Phi}_{ch}(u)=
\widetilde{\PH1_{u_{1}}}{}_{ch}(u_{1})
\cdots
\widetilde{\PH1_{u_{N}}}{}_{ch}(u_{N}),
\end{equation}
with the component fields
\begin{equation}
\label{eq: R8.3}
\widetilde{\PH1_{u}}{}_{ch}(u)=\left(\ch1_{u}{}_{p\,\alpha}(u), 
\A1_{u}{}_{sp}(u), F(u)\right),
\end{equation}
provided by the following MW representations
\begin{equation}
\label{eq: R8.4}
\begin{array}{l}
\ch1_{u}{}_{p\,\alpha}(u)=
\left[ 
\ch1_{u_{1}}(u_{1})\cdots\ch1_{u_{N}}(u_{N})
\right]_{\alpha}, \quad
\A1_{u}{}_{sp}(u)=\A1_{u_{1}}{}_{sp}(u_{1})\cdots 
\A1_{u_{N}}{}_{sp}(u_{N}), \\
F(u))=F(u_{1})\cdots F(u_{N}).
\end{array}
\end{equation}
Accordingly, the particle field $\chi_{p}(\zeta)$ in eq.(8.1) should be 
promoted to the $\widetilde{\Phi}_{\rlap/S ch}$ field defined on the broken 
super MM (\rlap/SMM)
\begin{equation}
\label{eq: R8.5}
\rlap/S G_{N}=\G1_{\eta}\oplus \SG1_{u}{}_{N}\equiv\G1_{\eta}\oplus 
\SG1_{u_{1}}\oplus \cdots\oplus \SG1_{u_{N}},
\end{equation}
where the suffix $(\rlap/S )$ denotes a broken SUSY. The 
$\eta$-component of the particle breaks the whole symmetry 
between the frame superfields of particle and sparticle, since the 
$\eta$-component of the sparticle is absent:   
\begin{equation}
\label{eq: R8.6}
\chi_{p}(\zeta)\rightarrow \widetilde{\Phi}_{\rlap/S ch}=
\left({\ch1_{\eta}}{}_{p}(\eta){\ch1_{u}}{}_{p}(u),
\, A_{sp}(u), F(u) \right).
\end{equation}
Certainly, note that due to subquark algebra eq.(4.5) the MW-SUSY has 
arisen only on the internal worlds where the subquarks are defined. 
By this we arrive to the major point for an understanding the principle 
fact why there is no any direct experimental evidence for the existence 
of any of the numerous sparticles predicted by SUSY. Our view point is as 
follows: 

{\em 
$\bullet$ the sparticles merely could not survive on the manifold 
$\G1_{\eta}$, which clearly means 
that they always would be absent in the Minkowski spacetime continuum 
(sec.2.1)}.

To see the nature of the supersymmetric generalization involved, it prompt us 
further to introduce a new mathematical framework of the \rlap/SOMM:\, 
$\hat{\rlap/S G}_{N}$. The first step towards it is to promote the given OM:\,
$\hG_{u_{i}}$ into the internal super operator manifold (SOM) 
$\hSG1_{u_{i}}$. According to general scheme given in sec.2, it 
implies that the operator vectors  
$\hps_{u}= {}^{i}{\hgam_{u}}{}_{m}\,\,
{}^{i}{\ps1_{u}}{}^{m},\,\, (m=(\lambda,\delta), \lambda=\pm, \delta=1,2,3)$ 
of a tangent section of principle bundle with the 
base $\hG_{u}=\hG_{u_{1}}\oplus\cdots\oplus\hG_{u_{N}}$ are promoted to 
the MW super operator vectors  
$\hat{\widetilde{\PH1_{u}}}(Z_{u})$ 
given by the 
expansion 
\begin{equation}
\label{eq: R8.7}
\hps_{u} \rightarrow \hat{\widetilde{\PH1_{u}}}(Z_{u})=
{}^{i}{\hgam_{u}}{}_{M}\,\,{}^{i}{\widetilde{\PH1_{u}}}{}^{M}(Z_{u})
\left(\in 
\hSG1_{u}=\hSG1_{u_{1}}\oplus \cdots\oplus \hSG1_{u_{N}}\right),
\end{equation} 
which has arisen by the corresponding substitution
\begin{equation}
\label{eq: R8.8}
{}^{i}{\hgam_{u}}{}_{m}\rightarrow {}^{i}{\hgam_{u}}{}_{M}=
\left(
{}^{i}{\hgam_{u}}{}_{m}, a_{\alpha}, a_{\alpha}^{+}
\right),\quad
{}^{i}{\ps1_{u}}{}^{m}\rightarrow {}^{i}{\widetilde{\PH1_{u}}}{}^{M}(Z_{u})=
\left(
{}^{i}{\ps1_{u}}{}^{m}, \psi^{\mu}, \bar{\psi}{}^{\dot{\mu}}
\right),
\end{equation} 
where $M$ is the superspace index, the annihilation and creation operators 
$a, a^{+}$ are given in eq.(6.3), also
we take into account that in MW-SUSY algebra eq.(6.1) the convention 
${}^{1}\1Q_{u}=\cdots= {}^{N}\1Q_{u}\equiv Q$ is imposed, and introduce 
required spinor components
\begin{equation}
\label{eq: R8.9} 
\psi^{\mu}=\theta^{\mu}\psi(\theta),
\quad
 \bar{\psi}{}^{\dot{\mu}}= \bar{\theta}{}^{\dot{\mu}} 
\bar{\psi}(\bar{\theta}).
\end{equation} 
Hence
\begin{equation}
\label{eq: R8.10} 
\hat{\widetilde{\PH1_{u}}}(Z_{u})=
\hps_{u} + \hat{\psi}+\hat{\bar{\psi}},
\end{equation} 
where 
\begin{equation}
\label{eq: R8.11} 
\hat{\psi}=\hat{\theta}\psi(\theta),
\quad \hat{\bar{\psi}}= \hat{\bar{\theta}} \bar{\psi}(\bar{\theta}),  
\quad \hat{\theta}=a\,\theta,
\quad \hat{\bar{\theta}}=a^{+}\,\bar{\theta}.
\end{equation} 
Given the SOM:\, $\hSG1_{u}$ we can extend it to 
\rlap/SOMM:\,  $\hat{\rlap/S G}_{N}$: 
\begin{equation}
\label{eq: R8.12} 
\hat{\rlap/S G}_{N}=\hat{\G1_{\eta}}\oplus \hSG1_{u}{}_{N}=
\hat{\G1_{\eta}}\oplus \hSG1_{u_{1}}\oplus \cdots\oplus \hSG1_{u_{N}}.
\end{equation} 
The super operator vectors read 
\begin{equation}
\label{eq: R8.13}
\begin{array}{l}
\hat{\widetilde{\Phi}}_{\rlap/S}(Z)=
\left\{
\begin{array}{l}
\hps_{\eta}{}_{p}(\eta) + \hat{\widetilde{\PH1_{u}}}(Z_{u}) \quad 
\mbox{for particle},\\
\hat{\widetilde{\PH1_{u}}}(Z_{u}) \quad 
\mbox{for sparticle},
\end{array}
\right. =\\
\hat{\Phi} + \hat{\psi}+\hat{\bar{\psi}}=\left\{
\begin{array}{l}
\hat{\Phi}_{p}(\zeta)+ \hat{\psi}+\hat{\bar{\psi}}\quad \mbox{for particle},
\\
\hat{A}_{sp}(u)+ \hat{\psi}+\hat{\bar{\psi}}\quad \mbox{for sparticle},
\end{array}
\right.
\end{array}
\end{equation} 
where  
\begin{equation}
\label{eq: R8.14}
\begin{array}{l}
\hat{\Phi}_{p}(\zeta)=\hps_{\eta}{}_{p}(\eta) + \hps_{u}{}_{p}(u)\in 
\hat{G_{N}},\quad 
\hat{A}_{sp}(u) \equiv \Asp_{u}{}_{sp}(u) \in 
\hat{G_{N}}, \\
\hps_{\eta}{}_{p}(\eta) = \hgam_{\eta}{}_{m} \ps1_{\eta}{}^{m}(\eta).
\end{array}
\end{equation} 
It is remarkable that such a MW-SUSY generalization does not upset a major 
condition of the MW geometry $G_{N}$ realization on which the MW frame 
fields of particles are defined [2] (subsec.12.6). Certainly, taking into 
account that $\theta$ and $\bar{\theta}$ are Grassmann variables, 
a straightforward computation gives the important MW-SUSY generalization of 
matrix element of the bilinear form eq.(3.1.1)
\begin{equation}
\label{eq: R8.15}
\begin{array}{l}
\S_{\lambda=\pm}<\widetilde{\chi}_{\lambda}\mid
\hat{\widetilde{\Phi}}_{\rlap/S}(Z)\,
\bar{\hat{\widetilde{\Phi}}}_{\rlap/S}(Z)
\mid
\widetilde{\chi}_{\lambda}>= 
\S_{\lambda=\pm}<\widetilde{\chi}_{\lambda}\mid
\bar{\hat{\widetilde{\Phi}}}_{\rlap/S}(Z)\,
\hat{\widetilde{\Phi}}_{\rlap/S}(Z)
\mid \widetilde{\chi}_{\lambda}>= \\ 
\S_{\lambda=\pm}<\chi_{\lambda}\mid\hat{\Phi}(\zeta)\,
\bar{\hat{\Phi}}(\zeta)\mid
\chi_{\lambda}>= 
-i\Lm_{\zeta\rightarrow\zeta'}(\zeta\zeta')\G1_{\zeta}(\zeta-\zeta')
=\\
\left\{
\begin{array}{l}
-i\,\left(\eta^{2}{\G1_{\eta}}(0)-
\S^{N}_{i=1}{u_{i}}^{2}{\G1_{u_{i}}}(0)\right)
\quad 
\mbox{for particle},  \\
i\,\S^{N}_{i=1}{u_{i}}^{2}{\G1_{u_{i}}}(0),
\quad 
\mbox{for sparticle},
\end{array}
\right. 
\end{array}
\end{equation} 
provided by the corresponding Green's functions 
and generalized state vectors \\ 
$\widetilde{\chi}_{\lambda}=(\chi_{\lambda},\, \mid \Omega >)$, 
where $\mid \Omega >$ is the 
Cliford vacuum. The $G_{N}$-realization condition readily stems from the 
requirement that the bilinear form eq.(8.15) as a geometric object should 
be finite
\begin{equation}
\label{eq: R8.16}
\left|\S_{\lambda=\pm}<\widetilde{\chi}_{\lambda}\mid
\hat{\widetilde{\Phi}}_{\rlap/S}(Z)\,
\bar{\hat{\widetilde{\Phi}}}_{\rlap/S}(Z)
\mid
\widetilde{\chi}_{\lambda}> \right| =
\left| \S_{\lambda=\pm}<\chi_{\lambda}\mid\hat{\Phi}(\zeta)\,
\bar{\hat{\Phi}}(\zeta)\mid
\chi_{\lambda}>\right|
< \infty.
\end{equation}
As it can be seen form eq.(8.15), the latter may be satisfied only for 
the particles but not for sparticles.
For particles it reduced to
\begin{equation}
\label{eq: R8.17}
\zeta^{2}{\G1_{\zeta}}{}_{F}(0) = \eta^{2}{\G1_{\eta}}{}_{F}(0)-
\S^{N}_{i=1}{u_{i}}^{2}{\G1_{u_{i}}}{}_{F}(0) < \infty.
\end{equation}
The particle should be on the mass shell if
\begin{equation}
\label{eq: R8.18}
m\equiv  |p_{u}|\equiv \left(\S^{N}_{i=1}p^{2}_{u_{i}}\right)^{1/2}=
|p_{\eta}|,
\end{equation}
i.e. the MW geometry $G_{N}$ comes into being (sec.3) if
\begin{equation}
\label{eq: R8.18}
{\G1_{\zeta}}{}_{F}^{p}(0) = {\G1_{\eta}}{}_{F}^{p}(0) = 
{\G1_{u}}{}_{F}^{p}(0), \quad u^{2}{\G1_{p}}{}_{F}^{u}(0)\equiv 
\S^{N}_{i=1}{u_{i}}^{2}{\G1_{u_{i}}}{}_{F}(0).  
\end{equation}

\section{Realistic realization of the $n\geq 1$ MW-SUSY}
\label {real}
Our major goal is to develop a realistic MW-SUSY extension of the 
MSM [2], which is not convenient to carry out directly within considered 
scheme, where the broken MW-SUSY is defined on the \rlap/SMM:\,  
$\rlap/S G_{N}$ eq.(8.5).
However, much easier to handle it at first within the exact MW-SUSY defined
on the whole MW coset manifold 
\begin{equation}
\label{eq: R9.1} 
SG_{N}=\SG1_{\eta}\oplus  \SG1_{u}{}_{N}=\SG1_{\eta}\oplus 
\SG1_{u_{1}}\oplus \cdots\oplus \SG1_{u_{N}},
\end{equation} 
and afterwards to get back to the broken MW-SUSY. 
The MW coset manifold eq.(9.1) can be restored by {\em lifting up} the 
manifold $\G1_{\eta}$ to supermanifold  $ \SG1_{\eta}$. Then, we introduce 
at once the corresponding supercharge operators 
$\1Q_{\eta}=\Q_{\eta}\, b\otimes f^{+}$
and $\1Q_{\eta}{}^{+}=\Q_{\eta}\, b^{+}\otimes f$ (eq.(4.12)). 
According to eq.(4.5) one has 
$\Q_{\eta}=\FFr{1}{\sqrt{2}}\FFr{q^{2}_{0}}{\bar{q_{0}}}=i\,\Q_{u_{1}}=\cdots
i\,\Q_{u_{N}}$. In the same time one must {\em lift up} also each sparticle 
to corresponding particle state by assigning to it 
the $\A1_{\eta}{}_{sp}(\eta)$ component defined on $\G1_{\eta}$ in 
order to provide the mass for sparticle equaled exactly to the mass of 
corresponding particle. It will enable the sparticle to be included in the 
same supermultiplet with corresponding particle. Then, the MW frame field
of sparticle eq.(8.4) can be rewritten as follows:
\begin{equation}
\label{eq: R9.2} 
\A1_{u}{}_{sp}(u)\rightarrow A_{sp}(\zeta)=
\A1_{\eta}{}_{sp}(\eta)\A1_{u}{}_{sp}(u),
\end{equation}
while according to eq.(8.18) one gets 
\begin{equation}
\label{eq: R9.3}
\p1_{\eta}{}_{sp}=\p1_{\eta}{}_{p}= m =\p1_{u}{}_{p}=\p1_{u}{}_{sp}.
\end{equation}
Given eq.(9.3) now the sparticles like particles are free to emerge in 
$\G1_{\eta}$ because of a validity of the condition imposed upon 
the Green's function 
$\G1_{\eta}{}_{sp}(0)=\G1_{u}{}_{sp}(0)$. Hence, we can write down the 
first of the anticommutation relations given in eq.(6.1) for the exact
$n\geq 1$ MW SUSY algebra 
\begin{equation}
\label{eq: R9.4}
\begin{array}{l}
\{ 
{\1Q_{\eta}}{}^{A}_{\alpha}, \, {\bar{\1Q_{\eta}}}{}_{\dot{\beta} \,B}
\}=2\,\,{}^{i}{\sig1_{\eta}}{}^{(\lambda, \delta)}_{\alpha \,\dot{\beta}}
\,{\P_{\eta}}{}_{(\lambda, \delta)}\,\delta_{B}{}^{A},\\
\{ 
{\1Q_{u_{1}}}{}^{A}_{\alpha}, \, {\bar{\1Q_{u_{1}}}}{}_{\dot{\beta} \,B}
\}=2\,\,{}^{i}{\sig1_{u_{1}}}{}^{(\lambda, \delta)}_{\alpha \,\dot{\beta}}
\,{\P_{u_{1}}}{}_{(\lambda, \delta)}\,\delta_{B}{}^{A},\\
.....................................................\\
.....................................................\\
\{ 
{\1Q_{u_{N}}}{}^{A}_{\alpha}, \, {\bar{\1Q_{u_{N}}}}{}_{\dot{\beta} \,B}
\}=2\,\,{}^{i}{\sig1_{u_{N}}}{}^{(\lambda, \delta)}_{\alpha \,\dot{\beta}}
\,{\P_{u_{N}}}{}_{(\lambda, \delta)}\,\delta_{B}{}^{A}.
\end{array}
\end{equation}
Summing over all them we get the first anticommutation relation of the exact
$n \geq 1$ MW SUSY algebra defined on the SMM:\, $SG_{N}$:
\begin{equation}
\label{eq: R9.5}
\{ 
Q^{A}_{\alpha}, \, \bar{Q}_{\dot{\beta} \,B}
\}
=2\,\,{}^{i}\sigma^{m}_{\alpha \,\dot{\beta}}
\,\,{}^{i}P_{m}\,\delta_{B}{}^{A}.
\end{equation}
Here $Q\equiv \sqrt{N}\1Q_{\eta}=i\,\sqrt{N}\1Q_{u}, \,\, Q_{0}\equiv
\sqrt{N}\Q_{\eta}=i\,\sqrt{N}\Q_{u}$, and having used the MW- notations, 
i.e., 
now $m\equiv (\lambda, \mu, \delta)$ is the $12$-dimensional vector index 
where we let the first two subscripts in the parentheses
to denote the pseudovector components while the third refers to the
ordinary vector components, ${}^{i}P_{m}$ is the $12$-dimensional 
momentum defined on the given internal world $\G1_{u_{i}}$. The matrices 
${}^{i}\sigma^{m}$ read
\begin{equation}
\label{eq: R9.6}
\begin{array}{l}
{}^{i}\sigma^{(1,1,\delta)} = \FFr{1}{\sqrt{2}}
\left(
\nu_{i}\,\sig1_{\eta}{}^{(+\delta)} + {}^{i}\sig1_{u}{}^{(+\delta)} 
\right),\quad 
{}^{i}\sigma^{(2,1,\delta)} = \FFr{1}{\sqrt{2}}
\left(
\nu_{i}\,\sig1_{\eta}{}^{(+\delta)} - {}^{i}\sig1_{u}{}^{(+\delta)} 
\right),\\
{}^{i}\sigma^{(1,2,\delta)} = \FFr{1}{\sqrt{2}}
\left(
\nu_{i}\,\sig1_{\eta}{}^{(-\delta)} + {}^{i}\sig1_{u}{}^{(-\delta)} 
\right),\quad 
{}^{i}\sigma^{(2,2,\delta)} = \FFr{1}{\sqrt{2}}
\left(
\nu_{i}\,\sig1_{\eta}{}^{(-\delta)} - {}^{i}\sig1_{u}{}^{(-\delta)} 
\right),
\end{array}
\end{equation}
where
\begin{equation}
\label{eq: R9.7}
{}^{i}{\sig1_{\eta}}{}^{(\pm \delta)} = \FFr{1}{\sqrt{2}}
\left(
\kappa^{\delta}\,\sigma^{0}_{\eta} \pm \sigma^{\delta}_{\eta}
\right),
\end{equation}
for $\nu_{i}$ see the eq.(3.2), the 
$\kappa^{\delta}$ and matrices $\sigma^{m}$ are given in eq.(6.2).
In the same manner we get the rest of the MW-SUSY relations. 
The resulting exact $n \geq 1$ MW-SUSY, central charges, algebra 
defined on the SMM:\, $SG_{N}$ reads
\begin{equation}
\label{eq: R9.8}
\begin{array}{l}
\left\{ 
Q^{A}_{\alpha}, \, \bar{Q}_{\dot{\beta} \,B}
\right\}
=2\,\,{}^{i}\sigma^{m}_{\alpha \,\dot{\beta}}
\,\,{}^{i}P_{m}\,\delta_{B}{}^{A}, \quad
\left\{ 
Q^{A}_{\alpha}, \, Q^{B}_{\beta}
\right\} 
= \epsilon_{\alpha\beta}\,X^{A\,B},\quad m\equiv (\lambda,\mu,\delta),\\
\left[ Q^{A}_{\alpha}, \,{}^{i}P_{}\right]=
\left[ \bar{Q}_{\dot{\alpha}\, A}, \,{}^{i}P_{m}\right]=
\left[ {}^{i}P_{m}{}^{j}P_{n}\right]=0,\\
\left[ Q^{A}_{\alpha}, \,{}^{i}M_{mn}\right]=
{}^{i}\sigma_{mn\,\,\alpha}{}^{\beta}\,Q^{B}_{\beta},\quad
\left[ \bar{Q}^{\dot{\alpha}}_{A}, \,{}^{i}M_{mn}\right]=
{}^{i}\sigma_{mn}^{\dot{\alpha}}{}_{\dot{\beta}}\,
\bar{Q}^{\dot{\beta}}_{A},
\end{array}
\end{equation} 
provided by $X^{AB}\equiv\sqrt{N}\X1_{\eta}{}^{AB}=
i\,\sqrt{N}\X1_{u}{}^{AB}$.
Due to different features of particles and sparticles in the 
physically realistic framework of eq.(8.15), 
one must have always to distinguish them,  
especially, for obtaining the realistic theory by passing back to the 
physical limit 
$A_{sp}(\zeta) \rightarrow \A1_{u}{}_{sp}(u)$ 
(see eq.(9.11)). 
Only natural way to do that is to formulate conservation law  
by introducing an
additional discrete internal symmetry, which assigns the conserved charges 
$+1$ to the particles and $-1$ to the sparticles. This is a major  
motivation of a new symmetry due to which the sparticles can only be produced 
in pairs. The Jacobi identity for $[[Q,\,B],\,B]$ ($B$ are the generators 
of internal symmetry group) implies that the structure constants vanish, 
i.e., the internal symmetry algebra is Abelian. Only one independent 
combination of the Abelian generators actually has a nonzero commutator 
with $Q$ and $\bar{Q}\,\,(n=1)$. This $U(1)$ generator denoted by $R$ 
\begin{equation}
\label{eq: R9.9}
[Q_{\alpha},\, R]=Q_{\alpha}, \quad [\bar{Q}_{\alpha},\, R]=-
\bar{Q}_{\dot{\alpha}}
\end{equation} 
is known in conventional SUSY theories as the $R$-parity. Thus, the 
$n=1$ MW SUSY algebra in general possesses an internal (global) $U(1)$ 
symmetry, for which SUSY generators have $R$-charge $+1$ and $-1$, 
respectively. This is a multiplicative $Z_{2}$ discrete symmetry defined as
\begin{equation}
\label{eq: R9.10}
R=(-1)^{3B+L+2S},
\end{equation} 
where $S,\,B$ and $L$ are the spin, the baryon and the lepton numbers 
(sec.12).
R-parity leaves invariant the fields of the SM, and flips the sign of their 
superpartners. Therefore, it implies that sparticles are pair produced and 
the lightest sparticle cannot decay. Reflecting upon said above as 
the first step towards constructing the realistic theory we shall start with 
unbroken MW-SUSY implemented on the SMM:\, $SG_{N}$ by lifting up all the 
sparticles to corresponding particle states provided by the $R$-parity 
conservation. Using then the conventional rules governing the MW-SUSY theory 
we may write the most generic renormalizable MW-SUSY action 
$S_{SG_{N}}$ (sec.14) involving gauge and matter frame fields, and, thus, the 
corresponding generating functional $Z_{SG_{N}}[{\cal J}]$. 
Of course, it cannot be the {\em exact action} describing the 
{\em exact symmetry} in nature. Along this line it has to be realized in its 
broken phase, and at the second step the corresponding generating functional 
$Z_{SG_{N}}[{\cal J}]$ should be defined on $ \rlap/S G_{N}$ by
passing back to the physical limit 
$A_{sp}(\zeta) \rightarrow \A1_{u}{}_{sp}(u)$
for all the sparticles involved which respects the $R$-parity:
\begin{equation}
\label{eq: R9.11}
Z_{real}[{\cal J}]\equiv Z_{\rlap/S G_{N}}[{\cal J}]= 
\Lm_{\left( R\,;\,\, A_{sp}(\zeta) \rightarrow \A1_{u}{}_{sp}(u)\right) }
Z_{SG_{N}}[{\cal J}].
\end{equation}  
Such a breaking of the MW-SUSY can be implemented on 
$\rlap/S G_{N}$ first of all by subtracting back all the explicit 
{\em soft} mass terms $S_{soft}$ formerly introduced for the sparticles 
eq.(9.3)
\begin{equation}
\label{eq: R9.12}
S_{\rlap/S G_{N}}=S_{SG_{N}} + S_{soft},
\end{equation} 
i.e., the mass terms of the scalar members of the chiral multiplets, 
apart from the Higgs chiral multiplet (sec.13), for which one must subtract 
the mass term of the Higgsino instead of scalar, and for the gaugino members 
of vector supermultiplets in the Lagrangian (sec.14). These soft terms do 
not reintroduce the quadratic diagrams [94], which motivated the introduction of SUSY. 
By this one has indeed returned to the $\rlap/S G_{N}$
\begin{equation}
\label{eq: R9.13}
\SG1_{\eta}
{}_{ \left( R;\,\, A_{sp}(\zeta) \rightarrow \A1_{u}{}_{sp}(u)\right)}
\,\,\rightarrow \G1_{\eta}.
\end{equation} 
Hence, {only the particles should be survived on $\G1_{\eta}$, but not the 
sparticles at all}.
The resulting formula eq.(9.11) can be regarded as the major recipe for 
a {\em realistic realization of any MW-SUSY} including the 
MW-supergravity and MW-superstrings. A remarkable feature of such an 
approach resides in the important fact that even if the MW-SUSY is broken, 
due to the splitting between the fermion-boson supersymmetric partners is 
itself of O(VEV), the naturalness problem 
in the MSM is resolved. Actually, the systems with the unbroken MW-SUSY 
defined on the $SG_{N}$ are very constrained in the sense that the 
observables vary 
holomorphically with the parameters of theory. According to Cauchy's 
theorem, such complex analytic functions are determined in 
terms of very little data: the singularities and the asymptotic behavior on 
the hypersurface $S G_{N}$. It is then as well true a vice versa that 
the boson-fermion cancellation in the hierarchy problem can be thought 
to be a consequence of a constraint stemming from holomorphy and it will be   
held on the $\rlap/S G_{N}$ as well at the limit eq.(9.11).

\section{MW super operator multimanifold $\hat{SG}_{N}$}
\label {SOMM}
The structure of the resulting algebra eq.(9.8) is similar to those of 
eq.(6.1). The sole difference is that in eq.(9.8) instead of 
$6$-dimensional vector indices  now we use the $12$-dimensional vector 
indices $m=(\lambda,\tau,\delta)$, where the first two in parenthesis 
denote the pseudovector components $(\lambda,\tau=1,2)$ while the third 
refers to the ordinary vector components $\delta=1,2,3$, and the double 
occurrence of dummy MW-indices to the left will be taken to denote a 
summation extended over all their values $i=1,..., N$.
Hence the Casimirs, the representations and the coset constructions of the 
exact MW-SUSY can be treated in similar manner of sec.6. For example, the 
Maurer-Cartan form is valued in a Lie algebra of supergroup $G$ has a 
standard expansion 
\begin{equation}
\label {eq: R10.1}
\Omega^{-1}\,d\,\Omega=i\left(
\omega^{A}\Gamma_{A}+\omega^{a}\Gamma_{a}+\omega^{i}\Gamma_{i}
\right),
\end{equation}
where constructing the coset $G\left/ H\right.$, the generators 
$\Gamma_{i}$ of unbroken $12$-dimensional rotations in $G_{N}$ and 
unbroken internal symmetries close into the stability group 
$H, \,\, \Gamma_{A}$ are the generators of unbroken translations in 
$G_{N},\,\, \Gamma_{i}$ are the generators of spontaneously broken internal 
symmetries and symmetries of $G_{N}$. The $\omega^{A},\, \omega^{a}$ and 
$\omega^{i}$ are the set of one-forms on the SMM $SG_{N}$ parametrized by 
the coordinates 
${}^{i}Z^{A}=\left({}^{i}\zeta, \theta, \bar{\theta}\right)$. 
The general group element can be written 
\begin{equation}
\label {eq: R10.2}
\Omega\left({}^{i}\zeta, \theta, \bar{\theta},\omega\right)=
e^{i\,\left(-{}^{i}\zeta^{m} \,\,{}^{i}P_{m}  +\theta Q + 
\bar{\theta}\bar{Q}\right)} \,
e^{-\FFr{1}{2}\,{}^{i}\omega^{mn}\,\,{}^{i}M_{mn}}.
\end{equation}
As usual, one chooses to keep all of $G$ unbroken. The ${}^{i}Z^{A}$ 
parametrizes a $N(6+4)$-dimensional coset space $SG_{N}$. The transformation
\begin{equation}
\label {eq: R10.3}
\Omega \rightarrow g\,\Omega\,h^{-1}
\end{equation} $(h \in H)$ induces the motion
\begin{equation}
\label {eq: R10.4}
{}^{i}\zeta^{m}\rightarrow {}^{i}\zeta^{m}+ 
i\,\theta \,\,{}^{i}\sigma^{m}\,\xi-i\,\,{}^{i}\xi\, \sigma^{m}\,
\bar{\theta},\quad
\theta^{\alpha}\rightarrow \theta^{\alpha}+\xi^{\alpha}, \quad 
\bar{\theta}_{\dot{\alpha}}\rightarrow \bar{\theta}_{\dot{\alpha}}+
\bar{\xi}_{\dot{\alpha}},
\end{equation} and $h=1.$ The differential operators $Q$ and $\bar{Q}$ close into the $n=1$,  
no central charges, MW-SUSY algebra
\begin{equation}
\label{eq: R10.5}
\left\{ Q_{\alpha}, \, \bar{Q}_{\dot{\alpha}}\right\}
=2\,\,{}^{i}\sigma^{m}_{\alpha \,\dot{\alpha}}
\,\,{}^{i}\partial_{m},\quad
\left\{ Q_{\alpha}, \, Q_{\beta}\right\}=
\left\{ \bar{Q}_{\dot{\alpha}}, \, \bar{Q}_{\dot{\beta}}\right\}=0,
\end{equation}
where
\begin{equation}
\label {eq: R10.6}
Q_{\alpha}= \FFr{\partial}{\partial\,\theta^{\alpha}}
-i\,\,{}^{i}\sigma^{m}_{\alpha \,\dot{\alpha}}\,
\bar{\theta}^{\dot{\alpha}}\,\,{}^{i}\partial_{m},\quad
\bar{Q}^{\dot{\alpha}}= \FFr{\partial}{\partial\,\theta_{\dot{\alpha}}}
-i\theta^{\alpha}\,\,{}^{i}\sigma^{m}_{\alpha \,\dot{\beta}}\,
\epsilon^{\dot{\alpha} \,\dot{\beta}}\,\,{}^{i}\partial_{m}.
\end{equation} The covariant vielbein and covariant derivative respectively are
\begin{equation}
\label {eq: R10.7}
E_{M}{}^{A}=\left(
\begin{array}{lll}
\delta_{m}{}^{n}\quad\quad\quad \quad\quad\,0 \quad\quad 0 \\
-i\,\,{}^{i}\sigma^{n}_{\mu \,\dot{\mu}}\,
\bar{\theta}^{\dot{\mu}}\quad \quad\,\delta_{\mu}{}^{\alpha} \quad \,0 \\
-i\theta^{\mu}\,\,{}^{i}\sigma^{n}_{\mu \,\dot{\nu}}\,
\epsilon^{\dot{\nu} \,\dot{\mu}}\quad 0 \quad\,\,
\delta_{\dot{\alpha}}{}^{\dot{\mu}}
\end{array}
\right)
\end{equation}
and
\begin{equation}
\label {eq: R10.8}
\begin{array}{l}
{}^{i}D_{m}={}^{i}\partial_{m},\quad
D_{\alpha}= \FFr{\partial}{\partial\,\theta^{\alpha}}
+i\,\,{}^{i}\sigma^{m}_{\alpha \,\dot{\alpha}}\,
\bar{\theta}^{\dot{\alpha}}\,\,{}^{i}\partial_{m},\quad
\bar{D}^{\dot{\alpha}}= \FFr{\partial}{\partial\,\theta_{\dot{\alpha}}}
+i\theta^{\alpha}\,\,{}^{i}\sigma^{m}_{\alpha \,\dot{\beta}}\,
\epsilon^{\dot{\alpha} \,\dot{\beta}}\,\,{}^{i}\partial_{m}.
\end{array}
\end{equation}
The rest of the whole formalism of $n=1$ MW-SUSY irreps and their 
supersymmetric actions invariant under the group $G$ can be treated by 
standard technique. The irreps contain an equal number of bosonic and 
fermionic degrees of freedom with degenerate mass, which is a generic feature 
of supersymmetric field theories stemming from the fact that $P^{2}$ is a 
Casimir operator of SUSY algebra. Since $Q$ has mass dimension $1/2$ and thus 
the  mass dimensions of the fields in the irrep differ by $3/2$. 
According to generic scheme  given in sec.8, the $\chi_{p}({}^{i}\zeta)$ 
field in eq.(8.1) should be promoted to the 
${}^{i}\widetilde{\Phi}({}^{i}Z^{A}))$ defined on the SMM:\, 
$SG_{N}$: ${}^{i}\widetilde{\Phi}^{M}({}^{i}Z^{A})=\left( {}^{i}\Phi^{m}
({}^{i}\zeta),\, \psi^{\mu}(\theta),\,
\bar{\psi}{}^{\dot{\mu}}(\bar{\theta})\right)$, where $\psi^{\mu}(\theta)$ 
and $\bar{\psi}{}^{\dot{\mu}}(\bar{\theta})$ are given in eq.(8.9). 
The operator vectors  
$\hat{\Phi}({}^{i}\zeta)= {}^{i}{\hgam_{u}}{}_{m}\,\,
{}^{i}\Psi^{m}({}^{i}\zeta),\,\, (i=\eta,\,1,...,N))$ 
of the tangent section of principle bundle with the 
base $\hat{G}_{N}=\hG_{\eta}\oplus\hG_{u_{1}}\oplus\cdots\oplus\hG_{u_{N}}$ 
should be promoted to the MW super operator vectors  
$\hat{\widetilde{\Phi}}_{N}({}^{i}Z^{A})$ defined on the SOMM 
\begin{equation}
\label{eq: R10.9}
\hat{SG}_{N}=\hat{\SG1_{\eta}}\oplus\hat{\SG1_{u_{1}}}\oplus\cdots\oplus
\hat{\SG1_{u_{N}}},
\end{equation} 
where the following expansion holds
\begin{equation}
\label{eq: R10.10}
\hat{\widetilde{\Phi}}_{N}({}^{i}Z^{A})={}^{i}{\hgam_{u}}{}_{M}\,\,
{}^{i}\widetilde{\Phi}^{M}({}^{i}Z^{A})=\hat{\Phi}({}^{i}\zeta)
+\hat{\psi}+\hat{\bar{\psi}}\in \hat{SG}_{N},
\end{equation} 
provided by the 
${}^{i}\hat{\gamma}_{M}=\left({}^{i}\hat{\gamma}_{m}\,a,\,a^{+}\right)$, 
and ${}^{i}\partial_{A}=\FFr{\partial}{\partial \,\,{}^{i}Z^{A}}
=\left({}^{i}\partial_{m},\,\partial_{\alpha},\,
\bar{\partial}_{\dot{\alpha}}\right)$. 
In the case at hand the particles and sparticles are in the massless states 
since they are on the mass shell eq.(9.3). Such states can be analyzed from 
the light-like reference frame $P_{m}=(\p1_{\eta}, 0,0,\p1_{u})$, where the 
Casimirs are 
\begin{equation}
\label{eq: R10.11}
P^{2}={}^{i}P_{m}\,\,{}^{i}P^{m}=0, \quad
C^{2}=-2\,p_{u}^{2}(B_{0}-B_{3})^{2}=-\FFr{1}{2}\,p_{u}^{2}\,
\bar{Q}_{\dot{2}}Q_{2}\bar{Q}_{\dot{2}}Q_{2}=0,
\end{equation} 
provided by 
$
\left\{ Q_{1},\,Q_{\dot{1}} \right\}=4\,p_{u},\quad 
\left\{ Q_{2},\,Q_{\dot{2}} \right\}=0.
$
Thus we can set $Q_{\dot{2}}=0$ in the operator sense, which yields only one 
pair of creation and annihilation operators 
\begin{equation}
\label{eq: R10.12}
a=\FFr{1}{2\sqrt{p_{u}}}Q_{1}, \quad a^{+}=\FFr{1}{2\sqrt{p_{u}}}Q_{\dot{1}}.
\end{equation} 
The corresponding MW-SUSY generalized bilinear form can be readily computed 
to be in the form
\begin{equation}
\label{eq: R10.13}
\begin{array}{l}
\S_{\lambda=\pm}<\widetilde{\chi}_{\lambda}\mid
\hat{\widetilde{\Phi}}_{N}\,
\bar{\hat{\widetilde{\Phi}}}_{N}
\mid
\widetilde{\chi}_{\lambda}> =
\S_{\lambda=\pm}<\chi_{\lambda}\mid\hat{\Phi}(\zeta)\,
\bar{\hat{\Phi}}(\zeta)\mid
\chi_{\lambda}>=\\ 
-i\,\zeta^{2}{\G1_{\zeta}}(0) = 
-i\,\left(\eta^{2}{\G1_{\eta}}(0)-\S^{N}_{i=1}{u_{i}}^{2}{\G1_{u_{i}}}(0)
\right),
\end{array}
\end{equation}
which is required to be finite. The latter yields the major condition of 
the MW geometry realization
\begin{equation}
\label{eq: R10.14}
{\G1_{\zeta}}{}_{F}(0) = {\G1_{\eta}}{}_{F}(0) = 
{\G1_{u}}{}_{F}(0), \quad u^{2}\G1_{u}{}_{F}(0)\equiv 
\S^{N}_{i=1}{u_{i}}^{2}{\G1_{u_{i}}}{}_{F}(0),
\end{equation}
or
\begin{equation}
\label{eq: R10.14}
m\equiv  |p_{u}|\equiv \left(\S^{N}_{i=1}p^{2}_{u_{i}}\right)^{1/2}=
|p_{\eta}|.  
\end{equation}
To produce the MW-SUSY generalization of the frame field defined on 
MW geometry, one must promote each left-handed fermionic components
$\ch1_{\eta}(\eta)$ and $\ch1_{u}(u)$  in eq.(8.1) to corresponding chiral 
multiplets. Then, the total frame field itself will be a chiral superfield. 
Thus, the two irreps are important for further development of MSMSM. 
They are the chiral and vector multiplets of the MW-SUSY eq.(9.8), which will 
be discussed in the next section.

\section{The supersymmetric frame field action defined on $SG_{N}$}
\label {act}
The MW-SUSY (eq.(9.8))  generalization of the skeletonized frame field eq.(8.1), 
which is a major ingredient of microscopic structure of leptons and quarks, 
defined on the $n=1$ rigid supermanifold $SG_{N}$ can be written in terms of 
chiral superfields
\begin{equation}
\label{eq: R11.1}
\widetilde{\Phi}_{ch}(Z)=\widetilde{\BB_{\eta}}(Z_{\eta})\,
\widetilde{\BB_{u_{1}}}(Z_{u_{1}})\cdots
\widetilde{\BB_{u_{N}}}(Z_{u_{N}}),
\end{equation}
where the generalized coordinates $Z=({}^{i}\zeta,\,\theta,\,\bar{\theta}),
\,\,Z_{\eta}=(\eta,\,\theta,\,\bar{\theta}),\,\, 
Z_{u_{1}}=(u_{1},\,\theta,\,\bar{\theta}), ...,$\\
$Z_{u_{N}}=(u_{N},\,\theta,\,\bar{\theta})$
parametrize the $(N+1)(6+4)$-dimensional coset space $SG_{N}$. According to 
the standard technique, the chiral irrep can be obtained from right 
covariant constrained general scalar superfield $\widetilde{\Phi}(Z)$, which 
is not fully reducible $\bar{D}_{\alpha} \widetilde{\Phi}_{ch}(Z)=0$. To 
solve this constraint the scalar superfield is regarded as a function of 
new bosonic coordinates  
\begin{equation}
\label{eq: R11.2}
{}^{i}y^{m}= {}^{i}\zeta^{m}+i\,\theta \,\,{}^{i}\sigma^{m}\bar{\theta}
\end{equation}
and $\theta$. Hence
\begin{equation}
\label{eq: R11.3}
\begin{array}{l}
\widetilde{\Phi}_{ch}({}^{i}y^{m},\, \theta)= 
A({}^{i}y^{m})+\sqrt{2}\,\theta\,\chi({}^{i}y^{m})+\theta\theta\, 
F({}^{i}y^{m})=
A({}^{i}\zeta)+
i\,\theta \,\,{}^{i}\sigma^{m}\bar{\theta}\,\,{}^{i}\partial_{m}\,A({}^{i}\zeta)+\\
\FFr{1}{4}\,\theta\theta\,\bar{\theta}\bar{\theta}\,\Box A({}^{i}\zeta)+
\sqrt{2}\,\theta\,\chi({}^{i}\zeta) -\FFr{i}{\sqrt{2}}\,\theta\theta\,
{}^{i}\partial_{m}\,\chi({}^{i}\zeta)\,\,{}^{i}\sigma^{m}\,\bar{\theta} +
\theta\theta\,F({}^{i}\zeta),
\end{array}
\end{equation}
where $\Box =-\,\,{}^{i}\partial_{m}\,\,{}^{i}\partial^{m}, \,\,
\chi_{\alpha}({}^{i}\zeta) $ is the left-handed Weyl fermion of spin $1/2$, 
$A({}^{i}\zeta)$ is the scalar superpartner of spin $0$, and 
$F({}^{i}\zeta)$ is the complex scalar auxiliary field of spin $0$. Thus, 
$\widetilde{\Phi}_{ch}$ has off-shell four real bosonic degrees of freedom 
$n_{B}=4$ and four real fermionic degrees of freedom $n_{F}=4$. An 
infinitesimal $n=1$ supersymmetric transformation of the component fields are 
as follows:
\begin{equation}
\label{eq: R11.4}
\delta_{\xi}\,A=\sqrt{2}\,\xi\,\chi,\quad
\delta_{\xi}\,\chi=i\,\sqrt{2} \,\,{}^{i}\sigma^{m}\,\bar{\xi}
\,\,{}^{i}\partial_{m}\,A+\sqrt{2} \,\xi\,F,\quad
\delta_{\xi}\,F=i\,\sqrt{2}\,\bar{\xi} \,\,{}^{i}\bar{\sigma}{}^{m}
\,\,{}^{i}\partial_{m}\,\chi.
\end{equation}
Due to chain rule, the chiral superfield eq.(11.1) can be rewritten 
in terms of standard ingredients of chiral supermultiplet
\begin{equation}
\label{eq: R11.5}
\widetilde{\Phi}_{ch}(Z)=
\left(
\chi({}^{i}\zeta),\,A({}^{i}\zeta)\, F({}^{i}\zeta)
\right),
\end{equation}
which have the following MW-component expansions:
\begin{equation}
\label{eq: R11.6}
\begin{array}{l}
\chi({}^{i}\zeta)=
\ch1_{\eta}(\eta)\ch1_{u_{1}}(u_{1})\cdots\ch1_{u_{N}}(u_{N}), \quad
A({}^{i}\zeta)= \A1_{\eta}(\eta)\A1_{u_{1}}(u_{1})\cdots\A1_{u_{N}}(u_{N}), \\
F({}^{i}\zeta)=\F1_{\eta}(\eta)\F1_{u_{1}}(u_{1})\cdots\F1_{u_{N}}(u_{N}), 
\end{array}
\end{equation}
The local gauge transformations of chiral superfields read
\begin{equation}
\label{eq: R11.7}
\widetilde{\Phi}_{ch}(Z)\rightarrow 
e^{\Lambda^{a}\,T^{a}}\,\widetilde{\Phi}_{ch}(Z),
\end{equation}
where, as usual,  $\Lambda$ is the chiral superfield satisfying the 
constraint $\bar{D} \, \Lambda =0. $ To make invariant the kinetic term 
$\widetilde{\Phi}^{+}_{ch}\,\widetilde{\Phi}_{ch}$ in the action of chiral 
superfield, one must introduce a vector superfield $V=V^{a}T^{a},\,\, 
V=V^{+}$ to exponentiate the finite transformation
\begin{equation}
\label{eq: R11.8}
e^{g\,V}\rightarrow e^{i\,\Lambda^{+}}\,e^{g\,V}\,e^{-i\,\Lambda},
\end{equation}
where $\Lambda\equiv \Lambda^{a}\,T^{a}$. The generators $T^{a}$ of a compact 
gauge group $G$, which will be specified in sec.12, have to commute with the 
supercharge operators, i.e., all members $(\chi,\,A,\,F)$ of chiral 
supermultiplet are in the same representation of the gauge group. 
Using a standard technique, without loss of generality one can decompose any 
vector superfield 
\begin{equation}
\label{eq: R11.9}
V\left({}^{i}\zeta, \,\theta,\,\bar{\theta}\right)=V_{WZ}+
\widetilde{\Phi}_{ch}+\widetilde{\Phi}^{+}_{ch},
\end{equation}
and some of unphysical degrees of freedom then can be gauged away like the 
unitary gauge in ordinary field theory, such that in the Wess-Zumino gauge the 
vector superfield $V_{WZ}$, similar to the chiral multiplet, contains only 
equal number of bosonic and fermionic degrees of freedom $n_{B}=n_{F}=4$: a 
gauge boson ${}^{i}\upsilon_{m}({}^{i}\zeta)$ of spin $1$ and mass dimension 
$1$, a Weyl fermion (gaugino) $\lambda$ of spin $1/2$ and mass dimension 
$3/2$, and a real scalar field $D$ of spin 0 and mass dimension $2$: 
$V_{WZ}=\left({}^{i}\upsilon_{m}({}^{i}\zeta),\, 
\lambda_{\alpha}({}^{i}\zeta),\, D({}^{i}\zeta),\, \right)$. They all are Lie
algebra valued fields, namely 
${}^{i}\upsilon_{m}={}^{i}\upsilon_{m}^{a}T^{a},$ etc. Hence,
\begin{equation}
\label{eq: R11.10}
V_{WZ}=-\theta\,\,{}^{i}\sigma^{m}\,\bar{\theta}\,\,{}^{i}\upsilon_{m}-
i\,\bar{\theta}\bar{\theta}\, \lambda + i\,\theta\theta \,\bar{\theta}
\bar{\lambda}+\FFr{1}{2}\, \theta\theta\,\bar{\theta}\bar{\theta}\, D.
\end{equation}
The SUSY transformations of the components read
\begin{equation}
\label{eq: R11.11}
\begin{array}{l}
\delta_{\xi}\,\,{}^{i}\upsilon_{m}^{a}=-i\,\bar{\lambda}{}^{a}
\,\,{}^{i}\bar{\sigma}{}^{m}\,\xi + i\xi\,\,{}^{i}\bar{\sigma}{}^{m}
\lambda^{a}. \quad 
\delta_{\xi}\,\lambda^{a}=i\,\xi\,D^{a}+\sigma^{mn}\,\xi\,F^{a}_{mn},\\
\delta_{\xi}\,D^{a}=-\xi\,\,{}^{i}\sigma^{m}\,\,{}^{i}D_{m}\,\bar{\lambda}^{a}-
{}^{i}D_{m}\,\lambda^{a}\,\,{}^{i}\sigma^{m}\,\bar{\xi},
\end{array}
\end{equation}
where $\sigma^{mn}=\FFr{1}{4}\left[{}^{i}\sigma^{m},
\,\,{}^{i}\sigma^{m}\right]$, The Yang-Mills field strength of the vector 
bosons $F^{a}_{mn}$ and covariant derivative are as follows:
\begin{equation}
\label{eq: R11.12}
\begin{array}{l}
F^{a}_{mn}={}^{i}\partial_{n}\,\,{}^{i}\upsilon_{m}^{a}-
{}^{i}\partial_{m}\,\,{}^{i}\upsilon_{n}^{a}-
g\,f^{abc}\,\,{}^{i}\upsilon_{m}^{b}\,\,{}^{i}\upsilon_{n}^{c},\\
{}^{i}D_{m}\,\lambda^{a} = {}^{i}\partial_{m}\,\lambda^{a}-
g\,f^{abc}\,\,{}^{i}\upsilon_{m}^{b}\,\lambda^{c},
\end{array}
\end{equation}
provided by the structure constants $f^{abc}$ of the Lie algebra.\\
A gauge invariant, renormalizable, supersymmetric action of the gauge vector 
superfield reads
\begin{equation}
\label{eq: R11.13}
S_{gauge}=\FFr{1}{4}\IIn \,d^{6}\zeta\,\IIn\,d\,\theta^{2}\,Tr\, 
W^{\alpha}\, W_{\alpha},
\end{equation}
where $W^{\alpha}$ and $ \bar{W}_{\alpha}=(W_{\alpha})^{+}$ are the 
left and right-handed spinor superfields. This is an off-shell irrep known as 
the curl multiplet or field strength multiplet containing $4+1+3=8$ real 
components $\lambda,D$ and $F_{mn}$. In the Wess-Zumino gauge the 
$W^{\alpha}$ and $ \bar{W}_{\alpha}$ are written
\begin{equation}
\label{eq: R11.14}
W_{\alpha}\equiv -\FFr{1}{4\,g}\,\bar{D}\,\bar{D}\,\,
e^{-g V_{WZ}}\,D_{\alpha}\,
e^{g\,V_{WZ}}, \quad
\bar{W}_{\alpha}\equiv \FFr{1}{4\,g}\,D\,D\,\,e^{g V_{WZ}}\,
\bar{D}_{\dot{\alpha}}\,e^{-g\,V_{WZ}}.
\end{equation}
A straightforward computation gives 
\begin{equation}
\label{eq: R11.15}
\begin{array}{l}
W^{\alpha}({}^{i}y^{m},\, \theta)=-
i\lambda^{\alpha}\,({}^{i}y^{m},\, \theta)+
\theta^{\alpha}\,D\,({}^{i}y^{m},\, \theta)-
i\,\,{}^{i}\sigma_{mn\,\alpha}^{\beta}\, F_{mn}\,\theta_{\beta}\\
+
\theta\theta\,\,{}^{i}\sigma_{\alpha\,\dot{\alpha}}^{m}\,\,{}^{i}D_{m}\,
\bar{\lambda}{}^{\alpha}\,({}^{i}y^{m}),
\end{array}
\end{equation}
and the action eq.(11.13) gives rise to
\begin{equation}
\label{eq: R11.16}
S_{gauge}=\IIn \,d^{6}\zeta\,\IIn\,d\,\theta^{2}\,
\left[
-\FFr{1}{4}\,F_{mn}^{a}\,F^{mn\, a}-
i\,\bar{\lambda}{}^{\alpha}\,\,{}^{i}\bar{\sigma}^{m}\,\,{}^{i}D_{m}\,
\lambda^{\alpha}+\FFr{1}{2}D^{\alpha}\,D^{\alpha}
\right].
\end{equation}
The $F$-component of a chiral superfield and the $D$-component of a vector 
superfield transform by a total derivative. Generic renormalizable action 
involving the chiral and vector gauge superfields [25-49] reads
\begin{equation}
\label{eq: R11.17}
\begin{array}{l}
S =\IIn \,d^{6}\zeta\,\IIn\,d^{4}\,\theta\,
\widetilde{\Phi}^{+}_{ch}\,e^{g\, V}\,\widetilde{\Phi}_{ch}\\
+
\left[\IIn \,d^{6}\zeta\,\IIn\,d^{2}\,\theta\,
\left(\FFr{1}{4}\,Tr\, W^{\alpha}\, W_{\alpha} 
+ P\left(\widetilde{\Phi}_{ch}\right)
\right)+\mbox{h.c.}
\right].
\end{array}
\end{equation}
where $P\left(\widetilde{\Phi}_{ch}\right)$ is the superpotential, which is 
the holomorphic function of $\widetilde{\Phi}_{ch}$ characterizing the 
interactions of chiral superfields of internal components
\begin{equation}
\label{eq: R11.18}
P\left(\widetilde{\Phi}_{ch}\right)=
\P_{u_{1}}\left(\widetilde{\BB_{u_{1}}}{}_{ch}(u_{1})\right)+\cdots +
\P_{u_{N}}\left(\widetilde{\BB_{u_{N}}}{}_{ch}(u_{N})\right).
\end{equation}
All the terms in eq.(11.17) fixed by symmetry expect for those in the 
superpotential. A gauge invariant, renormalizable Lagrangian containing the 
chiral multiplets $(A^{I},\,\chi^{I},\, F^{I})$ explicitly labeled by the 
indices $I,J,...,$ coupled to vector-multiplets stems from the eq.(11.17) 
\begin{equation}
\label{eq: R11.19}
\begin{array}{l}
L\left( A^{I},\,\chi^{I},\, F^{I},\,\,{}^{i}\upsilon_{m}^{a},\, 
\lambda^{a},\,D^{a}\right)=
-{}^{i}D_{m}\, A^{*}_{I}\,\,{}^{i}D^{m} \,A^{I}-i\,\bar{\chi}_{I}
\,\,{}^{i}\bar{\sigma}{}^{m}\,\,{}^{i}D_{m}\,\chi^{I}\\
-\FFr{1}{4}\,F^{a}_{mn}\, F^{mn\,a}-i\,\bar{\lambda}^{a}
\,\,{}^{i}\bar{\sigma}{}^{m}\,\,{}^{i}D_{m}\,\lambda^{a}-
i\sqrt{2}\,g\,\bar{\lambda}^{a}\,\bar{\chi}_{I}\,T^{a}{}_{J}^{I}\, A^{J}+
i\sqrt{2}\,g\, A^{*}_{J}\,T^{a}{}_{I}^{J}\,\chi^{I}\,\lambda^{a}\\
- \bar{F}^{I} \,F^{I}-\FFr{g^{2}}{2}\, D^{a}\,D^{a} -
g\,D^{a}\, A^{*}{}^{I}\,T^{a}_{IJ}\,A^{J}-
\FFr{1}{2}\,P_{IJ}\,\chi^{I}\,\chi^{J}-
\FFr{1}{2}\,\left(P_{IJ}\right)^{*}\,\bar{\chi}_{I}\,\bar{\chi}_{J}\\
-F^{I}\, P_{I}-\bar{F}{}^{I}\, P_{I}^{*},
\end{array}
\end{equation}
where 
\begin{equation}
\label{eq: R11.20}
{}^{i}D_{m}\,A^{I}={}^{i}\partial_{m}\,A^{I}+i\,\,{}^{i}\upsilon_{m}^{a}
T^{a}{}^{I}_{J}\,A^{J},\quad
{}^{i}D_{m}\,\chi^{I}={}^{i}\partial_{m}\,\chi^{I}+i\,\,{}^{i}\upsilon_{m}^{a}
T^{a}{}^{I}_{J}\,\chi^{J},
\end{equation}
$P_{I}$ and $P_{IJ}$ are the derivatives of a holomorphic function $P(A)$:
\begin{equation}
\label{eq: R11.21}
P_{IJ}=\FFr{\partial^{2}}{\partial\, A^{I}\,\,\partial\, A^{I}}\,P(A), \quad 
P_{I}=\FFr{\partial}{\partial\, A^{I}}\,P(A).
\end{equation}
The only freedom in constructing the supersymmetric Lagrangian eq.(11.19) 
after choosing of superfields and gauge symmetries is to specify a 
superpotential $P\left(\widetilde{\Phi}{}_{ch}\right)$, which is not allowed 
to contain their complex conjugates $\widetilde{\Phi}{}^{+}_{ch}.$ 
Since the terms in the superpotential with more that $3$ chiral superfields 
would yield non-renormalizable interactions in the Lagrangian, then it  
contains terms with $2$ and $3$ chiral superfields as in the Wess-Zumino 
simplest (sensible) $n=1$ SUSY toy model [25]. Furthermore, as an analytic 
function of the superfields the superpotential is not allowed to contain 
derivative interactions. From the superpotential can be found both the scalar 
potential and Yukawa interactions of the fermions with the scalars 
\begin{equation}
\label{eq: R11.22}
P\left(\widetilde{\Phi}{}_{ch}\right)=\FFr{1}{2}\,m\,
\widetilde{\Phi}{}_{ch}^{2}+\FFr{1}{3}\,Y\,\widetilde{\Phi}{}_{ch}^{3}.
\end{equation}
From the action eq.(11.17) one gets
\begin{equation}
\label{eq: R11.23}
\left
[\FFr{1}{2}\,m\,\widetilde{\Phi}{}_{ch}^{2}+
\FFr{1}{3}\,Y\,\widetilde{\Phi}{}_{ch}^{3}
\right]_{\theta\,\theta}=m\,A\,F +Y\,A^{2}\,F.
\end{equation}
Whence, the action eq.(11.17) contains no derivatives acting on the auxiliary 
field $F$, which can be eliminated by solving its equations of motion
\begin{equation}
\label{eq: R11.24}
\begin{array}{l}
\FFr{\delta\,L}{\delta\,F_{I}}=F^{*}_{I}-P_{I}=F^{*}_{I}-m_{IJ}\,A^{J}-
Y_{IJK}\,A^{J}\,A^{K}=0,\\\\
\FFr{\delta\,L}{\delta\,F^{*}_{I}}=F_{I}-P^{*}_{I}=F_{I}-m_{IJ}\,A^{*}{}^{J}-
Y_{IJK}\,A^{*}{}^{J}\,A^{*}{}^{K}=0.
\end{array}
\end{equation}
The algebraic equation of motion for the other auxiliary field $D^{a}$ can be 
obtained
\begin{equation}
\label{eq: R11.25}
\FFr{\delta\,L}{\delta\,D^{a}}=D^{a}+g\,A^{*}{}^{I}\,T^{a}_{IJ}\,A^{*}{}^{J}=0.
\end{equation}
The equations (11.24) and (11.25) can be used to eliminate the auxiliary 
fields $F^{I}$ and $D^{a}$ from the Lagrangian eq.(11.19)
\begin{equation}
\label{eq: R11.26}
\begin{array}{l}
L\left( A^{I},\,\chi^{I},\,\,{}^{i}\upsilon_{m}^{a},\, 
\lambda^{a},\, F_{I}=P^{*}_{I},\,D^{a}=
-g\,A^{*}{}^{I}\,T^{a}_{IJ}\,A^{*}{}^{J}\right)=
-\FFr{1}{4}\,F^{a}_{mn}\, F^{mn\,a}\\
-i\,\bar{\lambda}^{a}\,\,{}^{i}\bar{\sigma}{}^{m}\,\,{}^{i}D_{m}\,\lambda^{a}-
{}^{i}D_{m} \, A^{*}_{I}\,\,{}^{i}D^{m} \,A^{I}-
i\,\bar{\chi}_{I}\,\,{}^{i}\bar{\sigma}{}^{m}\,\,{}^{i}D_{m}\,\chi^{I}
+i\sqrt{2}\,g\, 
\left(
A^{*}_{J}\,T^{a}{}_{I}^{J}\,\chi^{I}\,\lambda^{a}\right.\\
\left.
-\bar{\lambda}^{a}\,\bar{\chi}_{I}\,T^{a}{}_{J}^{I}\, A^{J}
\right)-
\FFr{1}{2}\,P_{IJ}\,\chi^{I}\,\chi^{J}-
\FFr{1}{2}\,\left(P_{IJ}\right)^{*}\,\bar{\chi}_{I}\,\bar{\chi}_{J}-
V(A,\,A^{*}),
\end{array}
\end{equation}
where the positive semi-definite scalar potential $V(A,\,A^{*})$ is 
extracted:
\begin{equation}
\label{eq: R11.27}
\begin{array}{l}
V(A,\,A^{*})=
\left(
F^{I}\,F^{*}{}^{I}+\FFr{1}{2}\,D^{a}\,D^{a}
\right)_{\FFr{\delta\,L}{\delta\,F}=0,\, \FFr{\delta\,L}{\delta\,D^{a}}=0}
=\\
P_{I}\,P^{*}_{I}+\FFr{1}{2}\,g^{2}\,
\left(A^{*}{}^{I}\,T^{a}_{IJ}\,A^{J}\right)\,
\left(A^{*}{}^{I}\,T^{a}_{IJ}\,A^{J}\right)\geq 0.
\end{array}
\end{equation}
We purposely expose the explicit mass term and Yukawa coupling 
in the superpotential eq.(11.22) in order to emphasize the most remarkable 
feature of SUSY theory that such choices are fixed by supersymmetry and 
ensured that all quadratic divergences cancel between bosonic and fermionic 
loops. Certainly, from the eq.(11.26) can be seen that like the Wess-Zumino 
simplest toy model the fields interact via Yuakawa and scalar couplings, 
where the quadratic divergences exactly cancel due to the supersymmetric 
relations $Y_{F}=Y$, and $Y_{B}=Y^{*}\,Y$. But the values of the masses and 
Yukawa couplings in eq.(11.24) still are free parameters of the theory at 
hand, which are implemented as much as they needed to be further constrained 
by the appropriate choice of the concrete MW-superpotential eq.(11.18) 
inserting the nonlinear interactions of internal components of the MW frame 
field. The latter will be adopted to build up the SUSY extension of 
microscopic MW structure of all the particles in MSM. 
Here we shall attempt to amplify and substantiate the assertions made in the 
MSM and further expose via MSMSM. In pursuing this aim we are once again led 
to the principal point of drastic change of the standard SUSY scheme to 
specialize, hereinafter, the superpotential eq.(11.18) to be in the form
\begin{equation}
\label{eq: R11.28}
P(Q)=P_{Q}(A_{Q})+P_{W}(A_{W})
\end{equation}
such that 
\begin{equation}
\label{eq: R11.29}
\mid F \mid ^{2}=\mid F_{Q} \mid ^{2}+\mid F_{W} \mid ^{2}=
\left|\FFr{\partial\,P_{Q}}{\partial \, A_{Q}}\right|+
\left|\FFr{\partial\,P_{Q}}{\partial \, A_{W}}\right|,
\end{equation}
provided by the scalars
\begin{equation}
\label{eq: R11.30}
\left|\FFr{\partial\,P_{Q}}{\partial \, A_{Q}}\right|\equiv 
\L1_{Q}{}_{I}=\FFr{1}{4}\,\lambda_{Q}
\left(
\J1_{Q}{}_{L}\,\J1_{Q}{}_{R}^{+}+\J1_{Q}{}_{R}\,\J1_{Q}{}_{L}^{+}
\right),\quad
\left|\FFr{\partial\,P_{Q}}{\partial \, A_{W}}\right|\equiv
\L1_{W}{}_{I}=\FFr{1}{2}\,\lambda_{W}\,S_{W}\,S_{W}^{+}.
\end{equation}
Here we admit that the MW-index $(i)$ will be running through $i=Q,\,W$ 
specifying the internal worlds formally denoting $Q$-world of electric 
charge and the $W$-world of weak interactions [2](subsec.12.2 and 12.3). Also adopt the
conventions of [2]
\begin{equation}
\label{eq: R11.31}
\begin{array}{l}
{\J1_{Q}}_{L,R}=\V_{Q} \mp \A1_{Q},\quad
\V_{Q}{}^{m}=\bp_{Q}{}_{D}\,\gamma^{m}\,\ps1_{Q}{}_{D},
\quad 
(\V_{Q}{}^{m}){}^{+}=\V_{Q}{}_{m}=
\bp_{Q}{}_{D}\,\gamma_{m}\,\ps1_{Q}{}_{D},\\
\A1_{Q}{}^{m}=\bp_{Q}{}_{D}\,\gamma^{m}\,\gamma^{5}\,\ps1_{Q}{}_{D},
\quad 
(\A1_{Q}{}^{m}){}^{+}=
\A1_{Q}{}_{m}=\bp_{Q}{}_{D}\,\gamma_{m}\,\gamma_{5}\,\ps1_{Q}{}_{D},\quad
S_{W}=\bp_{W}{}_{D}\,\ps1_{W}{}_{D},\\
m\equiv (\lambda,\delta)\quad 
\gamma^{5}=i\gamma^{0}\gamma^{1}\gamma^{2}\gamma^{3}, \quad
\lambda=\pm,\,\,\delta=1,2,3 
\end{array}
\end{equation}
$\gamma^{m}$ are given in App.A, 
$\gamma^{\mu}$ and $\gamma^{5}$ are Dirac matrices, 
$\psi_{D}$ is the Dirac spinor in the Weyl basis 
containing a left-handed Weyl spinor $\chi_{\alpha}$ and a right-handed Weyl 
spinor 
$P_{L}\,\psi_{D}=
\left(
\begin{array}{c}
\chi_{\alpha}\\
0
\end{array}
\right), \quad
P_{R}\,\psi_{D}=
\left(
\begin{array}{c}
0\\
{\psi_{\dot{\alpha}}}
\end{array}
\right).
$
Using the Fiertz identities one gets
\begin{equation}
\label{eq: R11.32}
\L1_{Q}{}_{I}=\FFr{1}{2}\,\lambda_{Q}\,(\V_{Q}\,\V_{Q}{}^{+}-
\A1_{Q}\,\A1_{Q}{}^{+})=-\lambda_{Q}\,(S_{Q}\,S_{Q}^{+}-p_{Q}\,p_{Q}^{+}),
\end{equation}
where $S_{Q}=\bar{\psi}{}_{D}\,\psi_{D}$ and $p_{Q}=\bar{\psi}{}_{D}\,
\gamma^{5} \,\psi_{D}$. Thus, despite the fact that in conventional SUSY 
theories the superpotential gives rise the fermion masses and Yukawa 
couplings, in suggested framework, in accordance with eq.(11.30), such terms 
are substituted by a new ones using a following set of constraints 
\begin{equation}
\label{eq: R11.33}
\begin{array}{l}
\F1_{Q}{}^{*}_{I}=-\FFr{\partial \, P_{Q}}{\partial A^{I}_{Q}}\equiv
2\,i\,(\lambda_{Q})^{1/2}\,(\ch1_{Q}\,\ps1_{Q})=
\m1_{Q}{}_{IJ}\,A^{J}_{Q}+
\Y1_{Q}{}_{IJK}\,A^{J}_{Q}\,A^{K}_{Q},\\
\F1_{W}{}^{*}_{I}=-\FFr{\partial \, P_{W}}{\partial A^{I}_{W}}\equiv
i\,(\FFr{\lambda_{W}}{2})^{1/2}\,(\ch1_{W}\,\ps1_{W}+
\bar{\ch1_{W}}\,\bar{\ps1_{W}})=
\m1_{W}{}_{IJ}\,A^{J}_{W}+
\Y1_{W}{}_{IJK}\,A^{J}_{W}\,A^{K}_{W},\\
\P_{Q}{}_{IJ}\equiv
\FFr{\partial }{\partial A^{I}_{Q}}\,
\left[
-2\,i\,(\lambda_{Q})^{1/2}\,(\ch1_{Q}\,\ps1_{Q})
\right]=
\m1_{Q}{}_{IJ}+2\,\Y1_{Q}{}_{IJK}\,A^{K}_{Q},\\
\P_{W}{}_{IJ}\equiv
\FFr{\partial }{\partial A^{I}_{W}}\,
\left[
-i\,(\FFr{\lambda_{W}}{2})^{1/2}\,(\ch1_{W}\,\ps1_{W}+
\bar{\ch1_{W}}\,\bar{\ps1_{W}})
\right]=
\m1_{W}{}_{IJ}+2\, \Y1_{W}{}_{IJK}\,A^{K}_{W},
\end{array}
\end{equation}
where, as usual, one abbreviates expressions with two-spinor fields by 
suppressing undotted indices contracted like ${}^{\alpha}{}_{\alpha}$, and 
dotted indices contracted like ${}_{\alpha}{}^{\alpha}$.
While, the positive semidefinite scalar potential eq.(11.27) gives rise to
\begin{equation}
\label{eq: R11.34}
V(A,A^{*})=\L1_{Q}{}_{I}+\L1_{W}{}_{I}+
\FFr{1}{2}\,g^{2}\,
\left(A^{*}{}^{I}\,T^{a}_{IJ}\,A^{J}\right)^{2}\geq 0.
\end{equation}
Furthermore, putting it all together and inserting eq.(11.34) into eq.(11.26) 
with $P_{IJ}\equiv 0$ the Lagrangian of supersymmetric frame fields reduces to 
one contained only the MW internal components of particles and their SUSY 
partners such that all of them are massless:
\begin{equation}
\label{eq: R11.35}
\begin{array}{l}
L\left( A^{I},\,\chi^{I},\,\,{}^{i}\upsilon_{m}^{a},\, 
\lambda^{a}\right)=
-\FFr{1}{4}\,F^{a}_{mn}\, F^{mn\,a}
-i\,\bar{\lambda}^{a}\,\,{}^{i}\bar{\sigma}{}^{m}\,\,{}^{i}D_{m}\,\lambda^{a}
-{}^{i}D_{m} \, A^{*}_{I}\,\,{}^{i}D^{m} \,A^{I}\\
-
i\,\bar{\chi}_{I}\,\,{}^{i}\bar{\sigma}{}^{m}\,\,{}^{i}D_{m}\,\chi^{I}
+i\sqrt{2}\,g\, 
\left(
A^{*}_{J}\,T^{a}{}_{I}^{J}\,\chi^{I}\,\lambda^{a}
-\bar{\lambda}^{a}\,\bar{\chi}_{I}\,T^{a}{}_{J}^{I}\, A^{J}
\right)+
4\,\lambda_{Q}\,(\ch1_{Q}\,\ps1_{Q})(\bar{\ch1_{Q}}\,\bar{\ps1_{Q}})\\
-
\FFr{\lambda_{W}}{2}\,(\ch1_{W}\,\ps1_{W}+
\bar{\ch1_{W}}\,\bar{\ps1_{W}})^{2}-
\FFr{1}{2}\,g^{2}\,
\left(A^{*}{}^{I}\,T^{a}_{IJ}\,A^{J}\right)^{2},
\end{array}
\end{equation}
where all the fields $ A^{I},\,\chi^{I},\,\,{}^{i}\upsilon_{m}^{a},\, 
\lambda^{a}$ imply the MW-expansion. 

\renewcommand{\theequation}{\thesubsection.\arabic{equation}}
\section{The brief outline of theoretical issues in the MSM}
We now have all the tools we need to build up the MSMSM.
For our immediate purpose, however, to provide sufficient background let 
us recapitulate. The following  subsections contain  concisely some of the 
necessary knowledge of generic on MSM, but for more details we refer to [2].
Within the MSM the possible elementary particles are thought to be 
composite dynamical systems in analogy to quantum mechanical stationary 
states of compound atom, i.e., the systems arisen from the primary fundamental principle. But, now a dynamical treatment built up on the 
MW geometry is quite different and more amenable to qualitative 
understanding.

\subsection{The MW-structure of the particles}
\label {Struct}
We start by  considering the MW-structure of all the collection of matter 
fields $\Psi(\zeta)$ with nontrivial MW internal structure
$
\Psi(\zeta)=\ps1_{\eta}(\eta)\,\,{}^{1}\ps1_{u}(\theta_{1})
\cdots\,\,{}^{N}\ps1_{u}(\theta_{N}),
$
where the component ${}^{i}\ps1_{u}(\theta_{i})$ is made of product of 
some subquarks and antisubquarks
$
{}^{i}\ps1_{u}({}^{i}\theta)={}^{i}\ps1_{u}(\{ {}^{i}q \}
,\{ {}^{i}\bar{q} \})$. We admit that the MW index $(i)$ will be running
only through $i=Q,W,B,$ $s,c,b,t$ specifying the internal worlds formally
taken to denote in following nomenclature: Q-world of electric 
charge; W-world of weak interactions;
B-baryonic world of strong interactions; the s,c,b,t  are
the worlds of strangeness, charm, bottom and top. Also, we admit that the 
distortion rotations in the worlds Q,W and B are local depending only of 
the $\eta$-coordinates 
${}^{i}\theta_{\pm k}(\eta)$ and they are global in the flavour worlds 
s,c,b,t.  
We note that due to concrete symmetries of internal worlds, the 
MW- structure of particles will come into being if the following major 
conditions hold, namely, besides the MW geometry realization 
requirement eq.(3.1.4) a condition of MW connections must be held 
too, which will be discussed in the next subsection.
We assign to  each distortion rotation mode in the three dimensional spaces
${\RR_{u_{i}}}{}_{+}^{3}$ and ${\RR_{u_{i}}}{}_{-}^{3}$ 
a scale $1/3$, namely
each of the subquarks associated with the rotations 
around the axes of given world carries the corresponding charge in the 
scale $1/3$ ; antisubquark carries respectively the $(-1/3)$ charge.
In the case of the worlds C=s,c,b,t, where distortion rotations are
global and diagonal with respect to axes 1,2,3, the physical system of
corresponding subquarks is invariant under the global transformations
$f^{(3)}_{C}(\theta^{c})$ of the global unitary group $diag(SU^{c}_{3})$ 
parametrized by the three angles $\theta^{c}_{1},\,\theta^{c}_{2}$ and 
$\theta^{c}_{3}$.
While
$
f^{(3)}_{C}\left( f^{(3)}_{C}\right)^{+}=1, \quad 
\|f^{(3)}_{C}\| = 1
$.
That is 
$
\theta^{c}_{1}+\theta^{c}_{2}+\theta^{c}_{3}=0.
$
We explore the simplest possibility 
$\theta^{c}\equiv \theta^{c}_{1}=\theta^{c}_{2}$, then one gets
\begin{equation}
\label{eq: R12.1.4}
f^{(3)}_{C}=
\exp\left( 
-i\FFr{\lambda_{8}}{\sqrt{3}}\theta^{c}
\right)=
e^{-iY^{c}\theta^{c}},
\end{equation}
provided by the operator of hypercharge $Y^{c}$ of diagonal group 
$diag(SU^{c}_{i})$. 
If all the internal worlds are involved, then
$
Y^{c} = s + c + b + t.
$
The conservation of each rotation mode in 
Q- and B- worlds, where the distortion rotations are local, means that
corresponding subquarks carry respectively the conserved charges
Q and B in the scale $1/3$, and antisubquarks - $(-1/3)$ charges.
It can be provided by including the matrix $\lambda_{8}$ as
the generator  with the others in the symmetries of corresponding worlds
(Q, B), and expressed in the invariance of the system of corresponding 
subquarks under the transformations of these symmetries.
The incompatibility relations eq.(2.5.2) for global 
distortion rotations in the worlds C=s,c,b,t reduced to
$$
f^{c}_{11}f^{c}_{22}=\bar{f}^{c}_{33},\quad
f^{c}_{22}f^{c}_{33}=\bar{f}^{c}_{11},\quad
f^{c}_{33}f^{c}_{11}=\bar{f}^{c}_{22},
$$
where 
$
\|f^{(3)}_{C}\| = f^{c}_{11} f^{c}_{22} f^{c}_{33}=1, \quad
f^{c}_{ii}\bar{f}^{c}_{ii}=1 \quad \mbox{for} \quad i=1,2,3.
$
It means that two subcolour singlets are available:
$
\left( q\bar{q}\right)^{c}_{i}=inv, 
\quad
\left(q_{1}q_{2}q_{3} \right)^{c}=inv,
$
carrying respectively the charges
$
C_{\left( q\bar{q}\right)^{c}_{i}}=0 \quad 
C_{\left(q_{1}q_{2}q_{3} \right)^{c}}=1.
$
We make use of convention
$
\left( q\bar{q}\right)^{c}_{i}\equiv {}^{c}q_{i}{}^{c}\bar{q}_{i}, 
\quad \left(q_{1}q_{2}q_{3} \right)^{c}\equiv
{}^{c}q_{1}{}^{c}q_{2}{}^{c}q_{3} .
$
Including the baryonic charge into {strong hypercharge} 
\begin{equation}
\label{eq: R12.1.5}
Y = B + s + c + b + t,
\end{equation}
we conclude that the hypercharge Y is a
sum of all conserved rotation modes in the internal worlds B,s,c,b,t involved
in the MW geometry realization condition eq.(3.1.4).

\subsection {Realization of Q-world and Gell-Mann-Nishijima relation}
\label {Rel}
The symmetry of Q-world of electric charge, assumed to be a 
local unitary symmetry $diag\left( SU^{loc}(3)\right)$, is diagonal with 
respect to axes 1,2,3. The unitary unimodular matrix 
$f^{(3)}_{Q}$ of local distortion rotations takes the form
$$
f^{(3)}_{Q}=e^{-i\lambda_{Q}\theta_{Q}},
$$
where $f^{(3)}_{Q}\left( f^{(3)}_{Q}\right)^{+}=1, \quad 
\|f^{(3)}_{Q}\| = 1$,
provided
$
\theta_{1}+\theta_{2}+\theta_{3}=0.
$
Taking into account the scale of rotation mode,
in other than eq.(12.1.1) simple case 
$
\theta_{2}=\theta_{3}=-\FFr{1}{3}\theta_{Q}
$
it follows that
$
\theta_{1}=\FFr{2}{3}\theta_{Q} .
$
The matrix $\lambda_{Q}$ may be written 
$\lambda_{Q}=\FFr{1}{2}\lambda_{3}+\FFr{1}{2\sqrt{3}}\lambda_{8}.
$
Making use of the corresponding operators of the group $SU(3)$
we arrive at Gell-Mann-Nishijima relation
\begin{equation}
\label{eq: R12.2.1}
Q= T_{3} + \FFr{1}{2}Y,
\end{equation}
where $Q=\lambda_{Q}$ is the generator of electric charge,
$T_{3}=\FFr{1}{2}\lambda_{3}$ is the third component of isospin 
$\vec{T}$, and $Y=\FFr{1}{\sqrt{3}}\lambda_{8}$ is the hypercharge.
The eigenvalues of 
these operators will be defined later on by considering the
symmetries and microscopic structures of fundamental fields. We 
think of operators $T_{3}$ and  $Y$ as the MW connection
charges and of relation eq.(12.2.1) as the condition of
the realization of the MW connections. Thus, during realization of 
MW- structure the symmetries of corresponding internal
worlds must be unified into more higher symmetry including also the 
$\lambda_{3}$ and $\lambda_{8}$. Meanwhile, the realization conditions of
the MW- structure are embodied in eq.(3.1.4) and eq.(12.2.1), 
provided by the conservation law of each rotational mode in the 
corresponding internal worlds involved.
For example, in the case of quarks (see eq.(12.5.1)) the eq.(3.1.4) reads
\begin{equation}
\label{eq: R12.2.2}
\S_{i=B,s,c,b,t}\omega_{i}{\G1_{i}}^{\theta}_{F}(0)={\G1_{\eta}}_{F}(0),
\end{equation}
and according to eq.(12.1.5), the Gell-Mann-Nishijima relation is 
written down
\begin{equation}
\label{eq: R12.2.3}
Q= T_{3} + \FFr{1}{2}(B+s+c+b+t).
\end{equation}
In the case of leptons (see eq.(12.4.1)), the realization conditions 
reduced to the following:
\begin{equation}
\label{eq: R12.2.4}
{\G1_{Q}}^{\theta}_{F}(0)={\G1_{\eta}}_{F}(0), \quad u\equiv u_{Q},
\quad (i\equiv Q)
\end{equation}
and
\begin{equation}
\label{eq: R12.2.4}
Q= T_{3}^{w} + \FFr{1}{2}Y^{w},
\end{equation}
where $T_{3}^{w}$ and $Y^{w}$ are respectively the operators of third 
component of weak isospin $\vec{T}^{w}$ and weak hypercharge
(subsec.12.3, 12.8).
The incompatibility relations eq.(2.5.2) lead to 
$$
f^{Q}_{11}f^{Q}_{22}=\bar{f}^{Q}_{33},\quad
f^{Q}_{22}f^{Q}_{33}=\bar{f}^{Q}_{11},\quad
f^{Q}_{33}f^{Q}_{11}=\bar{f}^{Q}_{22},
$$
where 
$
f^{Q}_{ii}\bar{f}^{Q}_{ii}=1, \quad \mbox{for} \quad i=1,2,3,\quad
\|f^{(3)}_{Q}\| = f^{Q}_{11} f^{Q}_{22} f^{Q}_{33}=1.
$
This in turn suggests two subcolour singlets
$
\left( q\bar{q}\right)^{Q}_{i}=inv,
\quad
\left(q_{1}q_{2}q_{3} \right)^{Q}=inv,
$
with the electric charges 
$
Q_{\left( q\bar{q}\right)^{Q}_{i}}=0, \quad 
Q_{\left(q_{1}q_{2}q_{3} \right)^{Q}}=1,
$
respectively.
The singlets $\left( q\bar{q}\right)^{Q}_{i}$ for given $i$ allow us to think 
of the $\left( q\bar{q}\right)^{Q}$ system as the mixed ensemble, such that a 
fraction of the members with relative population $L_{1}$ are characterized by 
the $(q_{1})^{Q}$, some other fraction with relative population  $L_{2}$, by 
$(q_{2})^{Q}$, and so on. Namely, the $\left( q\bar{q}\right)^{Q}$ ensemble 
can be regarded as a mixture of pure ensembles. The fractional populations are 
constrained to satisfy the normalization condition
\begin{equation}
\label{eq: R4.6}
\S_{i} L_{i}=1, \quad L_{i}\equiv \left( q\bar{q}\right)^{Q}_{i}\left.\right/
\left( q\bar{q}\right)^{Q}.
\end{equation}
The $ L_{i}$ also imply the orthogonality condition ensued from the symmetry 
of the $Q$-world
\begin{equation}
\label{eq: R4.7}
<L_{i}, L_{j}>=0 \quad \mbox{if}\quad i\neq j.
\end{equation}
This prompts us to define the usual quantum mechanical density operator
\begin{equation}
\label{eq: R4.8}
\rho_{1}^{Q}=\S_{i}L_{i}\,(q_{i})^{Q})\,(q_{i})^{Q\,+}, \quad 
tr(\rho_{1}^{Q})=1.
\end{equation}
The eq.(12.2.6) suggests another singlets as well
\begin{equation}
\label{eq: R4.9}
\left(q_{1}q_{2}q_{3} \right)^{Q}_{i}\equiv
L_{i}\,\left(q_{1}q_{2}q_{3} \right)^{Q},
\end{equation}
which will be used to build up the MW-structures of the leptons.

\subsection{The symmetries of the W-,B-, and global worlds}
\label {Wworld}
$\bullet$ The W-world\\
It will be seen in subsec.12.8 that the symmetry of 
W-world of weak interactions is $SU^{loc}(2)_{L}\otimes U^{loc}(1)_{Y}$
invoking local group of weak hypercharge $Y^{w}$ ($U^{loc}(1)_{Y}$).
However, for the present it is worthwhile to restrict oneself 
by admitting that the symmetry of W-world is simply expressed by the group
of weak isospin  $SU^{loc}(2)$, i.e., from the very first we consider the 
case of two dimensional distortion transformations through the angles 
$\theta_{\pm}$ around two arbitrary axes in the W-world. 
In accordance with the results of subsec.3.2, the fields of subquarks and 
antisubquarks will come in doublets, which form the basis for fundamental 
representation of weak isospin group $SU^{loc}(2)$ often called a 
``custodial'' symmetry [5,20,126]. The doublet states
are complex linear combinations of up and down states of weak isotopic 
spin. Three possible doublets of six subquark states are
$
\left( \begin{array}{c} q_{1}\\ q_{2}
\end{array} \right)^{W},\quad
\left( \begin{array}{c} q_{2}\\ q_{3}
\end{array} \right)^{W},\quad
\left( \begin{array}{c} q_{3}\\ q_{1}
\end{array} \right)^{W}.
$
\\
\\
$\bullet$ The B-world\\
The B-world is responsible for strong interactions. The internal 
symmetry group is $SU^{loc}_{c}(3)$ enabling to introduce
gauge theory in subcolour space with subcolour charges as exactly 
conserved quantities (sec.3 in [1]).
The local distortion transformations implemented
on the subquarks $(q_{i})^{B},\quad i=1,2,3$ through a
$SU^{loc}_{c}(3)$ rotation matrix $U$ in the fundamental 
representation. Taking into account a conservation of rotation 
mode, each subquark carries $(1/3)$ baryonic  
charge, while the antisubquark carries the $(-1/3)$ baryonic charge.
\\
$\bullet$ The global worlds\\
We adopt a simplified view-point on the field
component $(q_{f}^{c})$ (f=u,d,s,c,b,t) associated with the 
global distortion rotations in the given worlds s,c,b,t, such that they have 
following microscopic structure with corresponding global charges:
\begin{equation}
\label{eq: R13.1}
\begin{array}{l} 
q^{c}_{u}=q^{c}_{d}=1, \quad 
\bar{q}^{c}_{s}=\left( \overline{q_{1}^{c}q_{2}^{c}q_{3}^{c}} \right)^{s}, 
\quad s =-1;\quad
q^{c}_{c}=\left( q_{1}^{c}q_{2}^{c}q_{3}^{c} \right)^{c}, 
\quad c = 1;\\
\bar{q}_{b}^{c}=\left( \overline{q_{1}^{c}q_{2}^{c}q_{3}^{c}} \right)^{b}, 
\quad b = -1;\quad
q_{t}^{c}=\left( q_{1}^{c}q_{2}^{c}q_{3}^{c} \right)^{t}, 
\quad t = 1.
\end{array} 
\end{equation}
To realize the MW-structure the global symmetries
of internal worlds have unified into more higher symmetry including
the generators $\lambda_{3}$ and $\lambda_{8}$ (subsec.12.2).
This global group is the flavour group $SU_{f}(6)$ 
unifying all the symmetries $SU^{c}_{i}$ of the worlds Q,B,s,c,b,t:
$
SU_{f}(6)\supset
SU_{f}(2)\otimes SU^{c}_{B}\otimes SU^{c}_{s}\otimes 
SU^{c}_{c}\otimes SU^{c}_{b}\otimes SU^{c}_{t}.
$

\subsection {The microscopic structure of leptons}
\label {Lept}
After a quantitative discussion of the properties of symmetries of internal
worlds, below we will attempt to show how the known fermion fields of 
leptons and quarks fit into this scheme. 
In this section we start with the leptons.
Taking into account the eq.(12.1.1) and eq(12.2.4), we may
consider six possible lepton fields forming three doublets of 
lepton generations
$
\left( \begin{array}{c} \nu_{e}\\ e
\end{array} \right),\quad
\left( \begin{array}{c} \nu_{\mu}\\ \mu
\end{array} \right),\quad
\left( \begin{array}{c} \nu_{\tau}\\ \tau
\end{array} \right),
$
where
\begin{equation}
\label{eq: R11.2}
\begin{array}{l}
\left\{ \begin{array}{l}
\nu_{e}\equiv {\ps1_{\eta}}{}_{\nu_{e}}(\eta)\, 
(q_{1}\bar{q_{1}})^{Q}(q_{1})^{w}=
L_{e}{\ps1_{\eta}}{}_{\nu_{e}}(\eta)\, (q\bar{q})^{Q}(q_{1})^{w},\\
e \equiv {\ps1_{\eta}}{}_{e}(\eta)\,
(\overline{q_{1}q_{2}q_{3}})^{Q}_{1}(q_{2})^{w}=
L_{e}{\ps1_{\eta}}{}_{e}(\eta)\, (\overline{q_{1}q_{2}q_{3}})^{Q}(q_{2})^{w},
\end{array} \right. 
\\
\left\{ \begin{array}{l}
\nu_{\mu}\equiv {\ps1_{\eta}}{}_{\nu_{\mu}}(\eta)\, 
(q_{2}\bar{q_{2}})^{Q}(q_{2})^{w}=
L_{\mu}{\ps1_{\eta}}{}_{\nu_{\mu}}(\eta)\, (q\bar{q})^{Q}(q_{2})^{w},\\
\mu \equiv {\ps1_{\eta}}{}_{\mu}(\eta)\,
(\overline{q_{1}q_{2}q_{3}})^{Q}_{2}(q_{3})^{w}=
L_{\mu}{\ps1_{\eta}}{}_{\mu}(\eta)\, (\overline{q_{1}q_{2}q_{3}})^{Q}(q_{3})^{w},
\end{array} \right. 
\\ 
\left\{ \begin{array}{l}
\nu_{\tau}\equiv  {\ps1_{\eta}}{}_{\nu_{\tau}}(\eta)\,
(q_{3}\bar{q_{3}})^{Q}(q_{3})^{w}=
L_{\tau}{\ps1_{\eta}}{}_{\nu_{\tau}}(\eta)\, (q\bar{q})^{Q}(q_{3})^{w},\\
\tau \equiv  {\ps1_{\eta}}{}_{\tau}(\eta)\,
(\overline{q_{1}q_{2}q_{3}})^{Q}_{3}(q_{1})^{w}=
L_{\tau}{\ps1_{\eta}}{}_{\tau}(\eta)\, (\overline{q_{1}q_{2}q_{3}})^{Q}(q_{1})^{w}
\end{array} \right. .
\end{array} 
\end{equation}
Here $e,\mu, \tau$ are the electron, the muon and the tau meson, 
$\nu_{e},\nu_{\mu},\nu_{\tau}$ are corresponding neutrinos,
$L_{e}\equiv L_{1},L_{\mu}\equiv L_{2},L_{\tau}\equiv L_{3},$
are leptonic charges. The leptons carry leptonic charges 
as follows: $L_{e}: \,(e,\nu_{e}),\,\,$
$ L_{\mu}:\,(\mu,\nu_{\mu})$ and
$ L_{\tau}:\,(\tau,\nu_{\tau}),$ which are conserved in all interactions. 
The leptons carry also the weak isospins:
$T^{w}_{3}=\FFr{1}{2}$ for  $\nu_{e},\nu_{\mu},\nu_{\tau}$; and
$T^{w}_{3}=-\FFr{1}{2}$ for $e,\mu, \tau$, respectively,
and following electric charges:
$
Q_{\nu_{e}}=Q_{\nu_{\mu}}=Q_{\nu_{\tau}}=0, \quad
Q_{e}=Q_{\mu}=Q_{\tau}=-1.
$
The Q-components $\ps1_{Q}(u_{Q})$ of lepton fields eq.(12.4.1) are made
of singlet combinations of subquarks in Q-world. They imply subcolour
confinement eq.(12.2.4). Then, the MW geometry realization condition 
is already satisfied and leptons may emerge in geometry in free combinations 
without any constraint. 
Thus, in suggested scheme there are only three possible
generations of six leptons with integer electric and leptonic charges
being free of confinement.

\subsection {The microscopic structure of quarks}
\label {Quark}
The only possible MW- structures of 18 quark 
fields read
\begin{equation}
\label{eq: R12.1}
\begin{array}{l}
\left\{ \begin{array}{l}
u_{i}\equiv {\ps1_{\eta}}{}_{u}(\eta)\,
(q_{2}q_{3})^{Q}(q_{1})^{w}(q_{i}^{B}),\\
d_{i} \equiv  {\ps1_{\eta}}{}_{d}(\eta)\,
(\bar{q}_{1})^{Q}(q_{2})^{w}(q_{i}^{B}),
\end{array} \right. 
\quad
\left\{ \begin{array}{l}
c_{i}\equiv {\ps1_{\eta}}{}_{c}(\eta)\,
(q_{3}q_{1})^{Q}(q_{2})^{w}(q_{i}^{B})(q_{c}^{c}),\\
s_{i} \equiv {\ps1_{\eta}}{}_{s}(\eta)\,
(\bar{q}_{2})^{Q}(q_{3})^{w}(q_{i}^{B})(\bar{q}_{s}^{c}),
\end{array} \right. 
\\
\left\{ \begin{array}{l}
t_{i}\equiv {\ps1_{\eta}}{}_{t}(\eta)\,
(q_{1}q_{2})^{Q}(q_{3})^{w}(q_{i}^{B})(q_{t}^{c}),\\
b_{i} \equiv {\ps1_{\eta}}{}_{b}(\eta)\,
(\bar{q}_{3})^{Q}(q_{1})^{w}(q_{i}^{B})(\bar{q}_{b}^{c}),
\end{array} \right. ,
\end{array} 
\end{equation}
where the subcolour index $(i)$ runs through $i=1,2,3$, the 
$(q_{f}^{c})$ are given in eq.(12.1.3).
Henceforth the subcolour index will be left implicit, but always a 
summation must be extended over all subcolours in B-world.
These fields form three possible doublets of weak isospin in the W-world
$
\left( \begin{array}{c} u \\ d
\end{array} \right),\quad
\left( \begin{array}{c} c \\ s
\end{array} \right),\quad
\left( \begin{array}{c} t \\ b
\end{array} \right).
$
The quark flavour mixing and similar issues are left for discussion  
in subsec.17.1.
The corresponding electric charges of quarks
read
$
Q_{u}=Q_{c}=Q_{t}=\FFr{2}{3},\quad
Q_{d}=Q_{s}=Q_{b}=-\FFr{1}{3},
$
in agreement with the rules governing the MW connections 
eq.(12.2.3), where the electric charge difference of up and down quarks
implies
$
\Delta Q=\Delta T^{w}_{3}=1.
$
The explicit form of structure of 
$(q_{f}^{c})$ are given in the next section. Note 
that all components of $(q_{f}^{c})$ are made of singlet combinations 
(eq.(12.1.1))of global subquarks in corresponding internal worlds, i.e., they 
obey the subcolour confinement condition. According to eq.(12.2.2), this
condition for B-world still remains to be satisfied. Therefore the
total quark fields obey the confinement.
Thus, three quark generations of six possible quark 
fields exist. They carry fractional electric and baryonic charges
and imply a confinement.
Although within considered schemes the subquarks are 
defined on the internal worlds, however the resulting 
$\eta$-components , which we are going to deal with to describe the leptons 
and quarks defined on the spacetime continuum, are affected by 
them. Actually, as it is seen in subsec.2.4
the rotation through the angle $\theta_{+k}$ yields
a total subquark field 
$$
{q}_{k}(\theta)=\Psi(\theta_{+k})={\ps1_{\eta}}^{0}
\ps1_{u}(\theta_{+k})
$$
where ${\ps1_{\eta}}^{0}$ is the plane wave defined on $\G1_{\eta}$.
Hence, one gets
$$
{q}_{k}(\theta(\eta))={\ps1_{\eta}}^{0}\,{\q1_{u}}{}_{k}(\theta(\eta))=
{\q1_{\eta}}{}_{k}(\theta(\eta))\,{\ps1_{u}}^{0},
\quad {\q1_{\eta}}{}_{k}(\theta(\eta))\equiv f_{(+)}(\theta_{+k}(\eta))
\,{\ps1_{\eta}}^{0}, 
$$
where ${\ps1_{u}}^{0}$ is a plane wave defined on $\G1_{u}$. The
${\q1_{\eta}}{}_{k}(\theta(\eta))$ can be considered as the subquark field
defined on the flat manifold $\G1_{\eta}$ with the same quantum numbers of
${\q1_{u}}{}_{k}(\theta(\eta))$.
Thus, instead of the eq.(12.4.1) and eq.(12.5.1) we may consider on 
equal footing only the resulting $\eta$-components of leptons and quarks 
implying the given same structures. This
enables to pass back to the Minkowski spacetime continuum 
$\G1_{\eta}\rightarrow M_{4}$ (subsec.2.1).

\subsection{The particle frame field}
\label{Fund}
All the fields including the leptons eq.(12.4.1) and
quarks eq.(12.5.1), along with the spacetime components have also MW 
components made of the various subquarks defined on the corresponding 
internal worlds:
\begin{equation}
\label{eq: R12.7.1}
\Psi(\theta)=\ps1_{\eta}(\eta)\ps1_{Q}(\theta_{Q})\ps1_{W}(\theta_{W})
\ps1_{B}(\theta_{B})\ps1_{C}(\theta^{c}).
\end{equation}
We assume that this field has arisen from the frame field with the same 
components defined on $G_{N}$ in the lowest state $(s_{0})$ 
\begin{equation}
\label{eq: R12.7.2}
\Psi(0)=\ps1_{\eta}(\eta)\ps1_{Q}(0)\ps1_{W}(0)
\ps1_{B}(0)\ps1_{C}(0)
\end{equation}
which serves as the ready made frame into which the distorted
ordinary structures of the same species should be involved.
The components $\ps1_{Q}(\theta_{Q}), \ps1_{W}(\theta_{W}),
\ps1_{B}(\theta_{B})$ are primary massless bare Fermi fields,
Let us now extract from the MW-SUSY Lagrangian eq.(11.35)  
the piece containing 
only $i=\eta,\,Q,\,W$ fermionic components of the particle frame fields. To keep 
resemblance with [2] it is convenient to rewrite them in Dirac 
spinor terms suppressing the subscript $D$ and superscript $I$ at 
$\psi^{I}_{D}$. The resulting Lagrangian of the frame field with nonlinear 
fermion interactions of the internal components looks like 
Heisenberg theory and reads
\begin{equation}
\label{eq: R12.7.3}
L(D)=\L1_{\eta}(\D1_{\eta})-
\L1_{Q}(\D1_{Q})-
\L1_{W}(\D1_{W}),
\end{equation}
where
\begin{equation}
\label{eq: R12.7.4}
\begin{array}{l}
\L1_{\eta}(\D1_{\eta})=
{\L1_{\eta}}'\,{}^{(0)}_{0}(\D1_{\eta})-
\FFr{1}{2}Tr ({\bf \B1_{\eta}}\bar{\bf \B1_{\eta}}),
\quad
\L1_{Q}(\D1_{Q})=
{\L1_{Q}}'{}^{(0)}_{0}(\D1_{Q})-
{\L1_{Q}}{}_{I}-
\FFr{1}{2}Tr ({\bf \B1_{Q}}\bar{\bf \B1_{Q}}),
\\
\L1_{W}(\D1_{W})=
{\L1_{W}}'{}^{(0)}_{0}(\D1_{W})-
{\L1_{W}}{}_{I}-
\FFr{1}{2}Tr ({\bf \B1_{W}}\bar{\bf \B1_{W}}).
\end{array}
\end{equation}
Here
\begin{equation}
\label{eq: R12.7.5}
\begin{array}{l}
{\L1_{\eta}}'{}^{(0)}_{0}=
\FFr{i}{2} \{ 
\bar{\Psi}\gamma\D1_{\eta}\Psi-
\bar{\Psi}\gamma\lD1_{\eta}\Psi
\}= {\ps1_{u}}^{+}
{\L1_{\eta}}{}_{0}^{(0)}
\ps1_{u},\\
{\L1_{u}}'{}^{(0)}_{0}=
\FFr{i}{2} \{ 
\bar{\Psi}\,{\gamma\D1_{u}}\Psi-
\bar{\Psi}\,{\gamma\lD1_{u}}\Psi
\}= {\ps1_{\eta}}^{+}
{\L1_{u}}{}_{0}^{(0)}
\ps1_{\eta},
\end{array}
\end{equation}
and 
\begin{equation}
\label{eq: R12.7.6}
{\L1_{\eta}}{}_{0}^{(0)}=
\FFr{i}{2} \{ 
\bp_{\eta}\,{\gamma\D1_{\eta}}\ps1_{\eta}-
\bp_{\eta}\,{\gamma\lD1_{\eta}}\ps1_{\eta}
\},\quad
{\L1_{u}}{}_{0}^{(0)}=
\FFr{i}{2} \{ 
\bp_{u}\,{\gamma\D1_{u}}\ps1_{u}-
\bp_{u}\,{\gamma\lD1_{u}}\ps1_{u}
\}.
\end{equation}
The Lagrangian eq.(12.6.3) has the global 
$\gamma_{5}$ and local gauge symmetries. We consider 
only $\gamma_{5}$ symmetry in Q-world, namely  
${\bf \B1_{Q}}\equiv 0$.
According to the OMM formalism,
it is the most important to fix the mass shell of 
the stable MW- structure eq.(3.5). Thus,
in the first we must take the variation of the Lagrangian eq.(12.6.3) with 
respect to frame field eq.(12.6.2), then 
switch on nonlinear fermion interactions of the components.
In other words we shall take the variation of the Lagrangian eq.(12.7.3) with
respect to these components only on the fixed mass shell.
The equations of free field (${\bf B}=0$) of the MW- structure follow 
at once, which can be written in terms of separate equations for the massless 
bare components $\ps1_{\eta}$, $\ps1_{Q}$ and $\ps1_{W}$:
\begin{equation}
\label{eq: R12.7.8}
\gamma\,\p1_{\eta}\,\ps1_{\eta}=
i\,\gamma\,{\pr_{\eta}}\,\ps1_{\eta}=0, \quad
\gamma\,{\p1_{Q}}\,\ps1_{Q}=
i\,\gamma\,{\pr_{Q}}\,\ps1_{Q}=0, \quad
\gamma\,{\p1_{W}}\,\ps1_{W}=
i\,\gamma\,{\pr_{W}}\,\ps1_{W}=0. 
\end{equation}
The important feature is that these equations respect
the simultaneous substitution
${\ps1_{Q}}{}^{(0)}\rightarrow {\ps1_{Q}}{}^{(m)}$
and
${\ps1_{\eta}}{}^{(0)}\rightarrow {\ps1_{\eta}}{}^{(m)}$, 
where
${\ps1_{Q,\,\eta}}{}^{(0)}$ and ${\ps1_{Q, \,\eta}}{}^{(m)}$
are respectively the massless and massive $Q,\,\,\eta$-component fields.
Then, in free state the massless field components
$\ps1_{\eta}$, $\ps1_{Q}$ and $\ps1_{W}$ are independent the Lagrangian 
\begin{equation}
\label{eq: R12.7.9}
L_{0}'{}^{(0)}= 
{\ps1_{u}}{}^{+}{\L1_{\eta}}{}_{0}^{(0)}{\ps1_{u}}-
{\ps1_{\eta}}{}^{+}{\L1_{u}}{}_{0}^{(0)}\ps1_{\eta}=
{\ps1_{u}}{}^{+}{\L1_{\eta}}{}_{0}^{(0)}{\ps1_{u}}-
{\ps1_{\eta}}{}^{+}({\L1_{Q}}{}_{0}^{(0)}+{\L1_{W}}{}_{0}^{(0)})\ps1_{\eta}
\end{equation}
reduces to the following:
\begin{equation}
\label{eq: R12.2.10}
L_{0}'{}^{(0)}= 
{\L1_{\eta}}{}_{0}^{(0)}-{\L1_{u}}{}_{0}^{(0)}=
{\L1_{\eta}}{}_{0}^{(0)}-{\L1_{Q}}{}_{0}^{(0)}-{\L1_{W}}{}_{0}^{(0)}.
\end{equation}
Thus, we shall implement our scheme as follows:
starting with the reduced Lagrangian $L_{0}'{}^{(0)}$ of free field we shall 
switch on nonlinear fermion interactions of the components.
After a generation of nonzero mass of $\ps1_{Q}$ component
in Q-world (next subsec.) shall look for the corresponding corrections via 
the eq.(12.6.5) to the reduced Lagrangian eq.(12.6.9).
These corrections mean only the interaction between the components, and do 
not imply at all the mass acquiring process for the $\eta$-component.

\subsection{Generation of mass of fermions in Q-world}
\label{Qrear}
We apply now a well known Nambu-Jona-Lasinio model [95] to 
generate a fermion mass in the Q-world and
start from the chirality invariant total Lagrangian of the field $\ps1_{Q}:
$ 
$
\L1_{Q}={\L1_{Q}}_{0}^{(0)}-{\L1_{Q}}_{I},
$
where a primary field $\ps1_{Q}$ is the massless bare spinor implying 
$\gamma_{5}$ invariance. However,
due to interaction the rearrangement of vacuum state has caused a 
generation of 
nonzero mass of fermion such like to appearance of energy gap in 
superconductor [96-98] and [99, 100]. After a vacuum 
rearrangement the total Lagrangian of initial massless bare field
${\ps1_{Q}}^{0}$ gives rise to corresponding Lagrangian ${\L1_{Q}}^{(m)}$ 
of massive field ${\ps1_{Q}}^{(m)}:\quad$
$
\L1_{Q}={\L1_{Q}}_{0}^{(0)}-{\L1_{Q}}_{I}={\L1_{Q}}^{(m)}
$
describing Dirac particle 
$
(\gamma p_{Q}-\Sigma_{Q}){\ps1_{Q}}^{(m)}=0$, where $\Sigma_{Q})$ is the 
self-energy operator.
In lowest order 
$
\Sigma_{Q}=m_{Q}\ll \lambda^{-1/2}.
$
Within the refined theory of superconductivity, the collective 
excitations of quasi-particle pairs arise in addition to the individual
quasi-particle excitations when a quasi-particle accelerated in the
medium [98, 102-106]. This leads to the conclusion given in [95, 101] that, in
general, the Dirac quasi-particle is only an approximate description
of an entire system with the collective excitations as the stable
or unstable bound quasi-particle pairs. In a simple approximation 
there arise CP-odd excitations of zero mass as well as CP-even
massive bound states of nucleon number zero and two.
Along the same line we must substitute 
in eq.(12.6.5) the massless field
$\Psi^{(0)}\equiv \ps1_{\eta}{\ps1_{Q}}^{(0)}\ps1_{W}$ by massive
field $\Psi^{(m)}\equiv \ps1_{\eta}{\ps1_{Q}}^{(m)}\ps1_{W}$. 
We obtain
\begin{equation}
\label{eq: R12.8.2}
\gamma p_{Q}{\ps1_{Q}}^{(m)}=
\Sigma_{Q}{\ps1_{Q}}^{(m)}, \quad
\gamma p_{W}{\Psi}^{(m)}=0,\quad
\gamma p_{\eta}{\Psi}^{(m)}=
(\gamma p_{Q}+\gamma p_{W}){\Psi}^{(m)}=
\Sigma_{Q}{\Psi}^{(m)}. 
\end{equation}
Such redefinition 
${\ps1_{Q}}^{(0)}\rightarrow {\ps1_{Q}}^{(m)}$
leaves the structure of that piece of Lagrangian eq.(12.6.3) involving
only the fields $\ps1_{\eta}$ and
$\ps1_{W}$ unchanged 
\begin{equation}
\label{eq: R12.8.4}
\begin{array}{l}
L_{0}={\L1_{\eta}}_{0}^{(0)}-{\L1_{W}}_{0}^{(0)}=
\left( {\L1_{\eta}}_{0}^{(0)}-
\Sigma_{Q}\bar{\Psi}\Psi \right)-
\left( 
{\L1_{W}}_{0}^{(0)}-
\Sigma_{Q}\bar{\Psi}\Psi
\right)=
{\L1_{\eta}}_{0}^{(m)}-{\L1_{W}}_{0}^{(m)},
\end{array}
\end{equation}
where the component $\ps1_{Q}$ is left implicit.
Upon combining and rearranging relevant terms we separate the following
pieces in the resulting gauge invariant Lagrangian eq.(12.6.3): 
\begin{equation}
\label{eq: R12.8.6}
\L1_{\eta}(\D1_{\eta})=
\FFr{i}{2} \{ 
\bp_{\eta}\gamma\D1_{\eta}\ps1_{\eta}-
\bp_{\eta}\gamma\lD1_{\eta}\ps1_{\eta}
\}-f_{Q}\bar{\Psi}\Psi -\FFr{1}{2}Tr({\bf \B1_{\eta}}\bar{\bf \B1_{\eta}})
\end{equation}
\begin{equation}
\label{eq: R12.8.7}
\L1_{W}(\D1_{W})=
\FFr{i}{2} \{ 
\bp_{W}\gamma\D1_{W}\ps1_{W}-
\bp_{W}\gamma\lD1_{W}\ps1_{W}
\}-\Sigma_{Q}\bar{\Psi}\Psi-
\FFr{\lambda}{2}S_{W}{S_{W}}^{+}-
\FFr{1}{2}Tr({\bf \B1_{W}}\bar{\bf \B1_{W}}),
\end{equation}
provided by
$
f_{Q}\equiv \Sigma_{Q}(p_{Q},m_{Q},\lambda,\Lambda),\quad 
\Psi=\ps1_{\eta}\ps1_{W}. 
$
The eq.(12.7.3) and eq.(12.7.4) are the Lagrangians that we shall be 
concerned within the following.

\subsection {The electroweak symmetry; the P-violation and the Weinberg mixing 
angle}
\label{Symm}
The microscopic approach creates a particular incentive for 
the pertinent concepts and ideas of the unified electroweak interactions. 
If, according to the subsec.12.3, one admit at very beginning that the local 
rotations in the W-world are occurred  only around two arbitrary axes, 
then one immediately concludes that under such circumstances  
the weak interacting particles could not be realized, because of the
condition of MW connections eq.(12.2.5), which is not satisfied yet. 
That is, the Q- and W-worlds cannot be realized separately.
Following the [2], a simple way of effecting a reconciliation is to assume 
that during a realization of weak interacting charged fermions, under the 
action of the Q-world, instead of initial symmetry the spanning of 
the $W-$ world into the 
world of unified electroweak interaction $W^{loc}_{(3)}$ took place, where
the local rotations always occur around all three axes. 
Taking into account that at the very beginning all subquark fields 
in W-world are massless,
we cannot rule out the possibility that they are transformed independently.
On the other hand, when this situation prevails the spanning
$
W^{loc}_{(2)}\rightarrow W^{loc}_{(3)}
$
must be occurred compulsory in order to provide a necessary background for 
the condition eq.(12.2.5) to be satisfied.
The most likely attitude here is that doing away
this shortage the subquark fields $q_{L_{1}},q_{L_{2}},q_{R_{1}},$
and $q_{R_{2}}$ tend to give rise to triplet.
The three dimensional effective space 
$W^{loc}_{(3)}$ will then arise
\begin{equation}
\label{eq: R12.9.1}
W^{loc}_{(2)} \ni  q^{w}_{(2)}\,
({\vec{T}}^{w}=\FFr{1}{2})
\rightarrow q^{w}_{(3)}=
\left(\matrix{
q_{R}({\vec{T}}^{w}=0)\cr
\cr
q_{L}({\vec{T}}^{w}=\FFr{1}{2})\cr}
\right)= 
\left(\matrix{
q^{w}_{3}\cr
q^{w}_{1}\cr
q^{w}_{2}\cr}
\right) \equiv
\left(\matrix{
q_{R_{2}}\cr
q_{L_{1}}\cr
q_{L_{2}}\cr}
\right)\in W^{loc}_{(3)}.
\end{equation}
The latter holds if violating initial P-symmetry 
the components $q_{R_{1}},q_{R_{2}}$ still remain in isosinglet 
states, namely the components $q_{L}$ form isodoublet while
$q_{R}$ is a isosinglet:
$
q_{L}\,({\vec{T}}^{w}=\FFr{1}{2}),\quad
q_{R}\,({\vec{T}}^{w}=0),
$
i.e., the mirror symmetry is broken.
Corresponding local transformations are implemented upon triplet
$
{q^{w}}'_{(3)}=f_{W}^{(3)}q^{w}_{(3)},
$
where making use of incompatibility relations eq.(2.5.2) one gets
the expanded group of local rotations in W-world (see [2]):
\begin{equation}
\label{eq: R12.9.3}
f^{(3)}_{exp}=\left( \matrix{
e^{-i\beta}  & 0  & 0 \cr
0  & f_{11}e^{-i\frac{\beta}{2}}  & f_{12}e^{-i\frac{\beta}{2}} \cr
0  & f_{21}e^{-i\frac{\beta}{2}}  & f_{22}e^{-i\frac{\beta}{2}} \cr
} \right)\in SU^{loc}(2)_{L}\otimes U^{loc}(1),
\end{equation}
where
$
U=e^{-i\vec{T}^{w}\vec{\theta^{w}}}\in SU^{loc}(2)_{L}, \quad
U_{1}=e^{-iY^{w}\theta_{1}}\in U^{loc}(1).
$
Here $U^{loc}(1)$ is the group of weak hypercharge $Y^{w}$ taking
the following values for left- and right-handed subquark fields:
$
q_{R}:Y^{w}=0,-2,\quad q_{L}:Y^{w}=-1.
$
Whence 
$
q'^{w}_{(3)}=f_{exp}^{(3)}q^{w}_{(3)}
$
or
$$
q'_{L}= {\displaystyle e^{-i\vec{T}^{w}\vec{\theta}^{w}-iY^{w}_{L}\theta_{1}}}
q_{L}, \quad
q'_{R}={\displaystyle e^{-iY^{w}_{R}\theta_{1}}}q_{R}.
$$
\\
The spanning eq.(12.8.1) implies the P-violation in the W-world expressed in 
the reduction of initial symmetry group of local transformations of
right-handed components $q_{R}$: 
\begin{equation}
\label{eq: R12.10.1}
\left[ SU(2)\right]_{R}\rightarrow \left[ U(1)\right]_{R},
\end{equation}
where subscript $(R)$ specified the transformations implemented upon
right-handed components. The invariance of physical system of the fields
$q_{R}$ under initial group $\left[ SU(2)\right]_{R}$ may be
realized as well by introducing non-Abelian massless vector gauge fields
${\bf A}=\vec{T}^{w}\vec{A}$ with the values in Lie algebra 
of the group $\left[ SU(2)\right]_{R}$. 
Under a reduction eq.(12.8.3) the coupling 
constant $(g)$ changed into $(g')$ specifying the interaction strength
between $q_{R}$ and the Abelian gauge field $B$ associated 
with the local group $\left[ U(1)\right]_{R}$. Thereto
$
g = g'\tan \theta_{w},
$
where $\theta_{w}$ is the Weinberg mixing angle. To define it we consider 
the interaction vertices corresponding to the groups 
$\left[ SU(2)\right]_{R}:$
$
g {\bf A}\bar{q}_{R}\gamma\FFr{\bf \tau}{2}q_{R}
\quad$
and $\quad \left[ U(1)\right]_{R}:\quad $
$
g' B\bar{q}_{R}\gamma\FFr{Y^{w}}{2}q_{R}.
$
Note that $\FFr{\lambda_{8}}{2}$ is in the same 
normalization scale as each of the matrices $\FFr{\lambda_{i}}{2}\quad 
(i=1,2,3):$
$
Tr\left( \FFr{\lambda_{8}}{2}\right)^{2}=
Tr\left( \FFr{\lambda_{i}}{2}\right)^{2}=\FFr{1}{2}.
$
Hence, the vertex scale reads
$
(\mbox{Scale})_{SU(2)}=g \FFr{\lambda_{3}}{2},
$
which is equivalent to $g \FFr{\lambda_{8}}{2}$.
It is obvious that per generator scale could not be changed under the
reduction eq.(12.10.1), i.e.
$
\FFr{(\mbox{Scale})_{SU(2)}}{N_{SU(2)}}=
\FFr{(\mbox{Scale})_{U(1)}}{N_{U(1)}},
$
where $N_{SU(2)}$ and $N_{U(1)}$ are the numbers of generators 
respectively in the groups $SU(2)$ and $U(1)$. 
Thus,
$
(Scale)_{U(1)}=\FFr{1}{3}(Scale)_{SU(2)}.
$
Stated somewhat differently, the normalized vertex
for the group $\left[ U(1)\right]_{R}$  reads
$
\FFr{1}{3}g B\bar{q}_{R}\gamma\FFr{\lambda_{8}}{2}q_{R}.
$
In comparing the coefficients can then be equated 
$
\FFr{g'}{g} = \tan \theta_{w}=\FFr{1}{\sqrt{3}}.
$
We may draw a statement that during the realization of MW- structure
the spanning eq.(12.8.1) compulsory occurred, which is the source of
P-violation in W-world incorporated with the reduction eq.(12.8.3). The
latter is characterized by the Weinberg mixing angle 
with the value fixed at $30^{0}$.

\subsection {Emergence of composite isospinor-scalar 
mesons}
\label{Mes}
The field $q^{w}_{(2)}$ is the W-component of total field
$
q_{(2)}=\q1_{\eta}{}_{(2)}\,{\q1_{W}}{}_{(2)},
$
where the field component $\q1_{Q}$ is left implicit.
Instead of it, below we introduce the additional suffix 
$(Q=0,\pm)$ specifying electric charge of the field.
At the very outset there is an absolute symmetry between the components
$
q_{1}={\q1_{\eta}}{}_{1}\,{\q1_{W}}{}_{1}$ and
$
q_{2}={\q1_{\eta}}{}_{2}\,{\q1_{W}}{}_{2}.$
Hence, left-and right-handed components of fields may be written
\begin{equation}
\label{eq: R12.11.1}
q_{1L}={\q1_{\eta}}{}_{1L}^{(0)}\,{\q1_{W}}{}_{1L}^{(-)}, \quad
q_{2L}={\q1_{\eta}}{}_{2L}^{(-)}\,{\q1_{W}}{}_{2L}^{(0)},\quad
q_{1R}={\q1_{\eta}}{}_{1R}^{(0)}\,{\q1_{W}}{}_{1R}^{(-)}, \quad
q_{2R}={\q1_{\eta}}{}_{2R}^{(-)}\,{\q1_{W}}{}_{2R}^{(0)}.
\end{equation}
On the example of one lepton generation $e$ and $\nu$, without loss of 
generality, in [2] we have exploited the properties of these fields. A
further implication of other fermion generations is straightforward.
We shown that the term $-f_{Q}\bar{\Psi}\Psi$ arisen in the Lagrangian
eq.(12.6.3) accommodates the Yukawa 
couplings between the fermions and corresponding isospinor-scalar mesons 
in fairy conventional form
\begin{equation}
\label{eq: R12.11.2}
-f_{Q}\bar{\Psi}\Psi=
-f_{e}\left( \bar{L}\,H \,e_{R}+\bar{e_{R}}\,H^{+} \,L\right)-
f_{\nu}\left( \bar{L}\,H_{c} \,\nu_{R}+\bar{\nu_{R}}\,H^{+}_{c} 
\,L\right),
\end{equation}
where the charge conjugated field $H_{c}$ is defined 
$
\left( H_{c}\right)_{i}=H^{*\,k}\varepsilon_{ik},
$
and the isospinor-scalar meson field $H$ reads
$$
H \equiv  
\gamma^{0}\,{\q1_{W}}{}_{L}^{+}\,{\q1_{W}}{}_{R},\quad
H^{+}\equiv 
\gamma^{0}\,{\q1_{W}}{}_{R}^{+}\,{\q1_{W}}{}_{L}.
$$
Then, the two possible composite isospinor-scalar mesons are as follows:
$$
H_{u}=
\left(\matrix{
h_{u}^{(+)}\cr
\cr
h_{u}^{(0)}\cr}
\right), \quad
H_{d}=
\left(\matrix{
h_{d}^{(0)}\cr
\cr
h_{d}^{(-)}\cr}
\right),
$$
where
$$
h_{u}^{+}\equiv \left( {\q1_{W}}{}_{1L}^{(-)}\right)^{+}
{\q1_{W}}{}_{2R}^{(0)},\quad
h_{u}^{0}\equiv \left( {\q1_{W}}{}_{2L}^{(0)}\right)^{+}
{\q1_{W}}{}_{2R}^{(0)},
$$
$$
h_{d}^{0}\equiv \left( {\q1_{W}}{}_{1L}^{(-)}\right)^{+}
{\q1_{W}}{}_{1R}^{(-)},\quad
h_{d}^{-}\equiv \left( {\q1_{W}}{}_{2L}^{(0)}\right)^{+}
{\q1_{W}}{}_{1R}^{(-)},
$$
In accordance with eq.(12.2.5), the isospinor-scalar
meson carries following weak hypercharge
$
H:Y^{w}=1.
$
To compute the coupling constants $f_{e}$ and $f_{\nu}$ for the leptons one 
must retrieve their implicit field-components $\ps1_{Q}$. Hence
\begin{equation}
\label{eq: R12.11.2}
f_{i}=tr(\rho_{i}^{Q}\,\Sigma_{Q}), \quad
f_{i}^{\nu}=tr(\rho^{Q\,\nu}_{i}\,\Sigma_{Q}), \quad
\end{equation}
where the density operators $\rho_{i}^{Q}$ and 
$\rho^{Q\,\nu}_{i}$ for given $i$ of the pure ensembles are used
\begin{equation}
\label{eq: R12.11.2}
\begin{array}{l}
\rho_{i}^{Q}=\left(q_{1}q_{2}q_{3}\right)^{Q\,+}_{i}\,
\left(q_{1}q_{2}q_{3}\right)^{Q}_{i}, \quad
\rho^{Q\,\nu}_{i}=\left(q_{i}\bar{q}_{i}\right)^{Q\,+}\,
\left(q_{i}\bar{q}_{i}\right)^{Q},\\
tr(\rho_{i}^{Q})^{2}=tr(\rho_{i}^{Q})=1, \quad
tr(\rho_{i}^{Q\,\nu})^{2}=tr(\rho_{i}^{Q\,\nu})=1.
\end{array}
\end{equation}
According to eq.(12.2.6), one gets
\begin{equation}
\label{eq: R12.11.2}
f_{i}=L_{i}^{2}\,\bar{\Sigma}_{Q},\quad 
f_{i}^{\nu}=L_{i}^{2}\,\bar{\Sigma}_{Q}^{\nu},\quad 
\bar{\Sigma}_{Q}\equiv\Sigma_{Q}(\lambda, L)\, \rho^{Q},
\quad
\bar{\Sigma}_{Q}{}^{\nu}\equiv\Sigma_{Q}(\lambda, L)\, \rho^{Q\,\nu},
\end{equation}
where
\begin{equation}
\label{eq: R12.11.2}
\rho^{Q}=\left(q_{1}q_{2}q_{3}\right)^{Q\,+}\,
\left(q_{1}q_{2}q_{3}\right)^{Q}, \quad
\rho^{Q\,\nu}=\left(q\,\bar{q}\right)^{Q\,+}\,
\left(q\,\bar{q}\right)^{Q}.
\end{equation}
An implication of the quarks into this scheme is straightforward if one 
retrieves their implicit field-components 
$\ps1_{Q}, \ps1_{B},\ps1_{C},\quad (C=s,c,b,t)$ (see subsec.12.6). On the 
analogy of previous case the coupling constants read
\begin{equation}
\label{eq: R12.11.2}
f_{i}=tr(\rho_{i}\,\Sigma_{Q}),
\end{equation}
where $i=u,d,s,c,b,t.$ Taking into account the MW-  structure of the 
quarks eq.(12.5.1), we may write down the corresponding density operators 
\begin{equation}
\label{eq: R12.11.2}
\rho_{i}=\rho_{i}^{Q}\rho_{i}^{B}\rho_{i}^{C}
\end{equation}
given in a convention
\begin{equation}
\label{eq: R12.11.2}
\rho_{i}^{A}=\ps1_{A}{}_{i}^{+}\ps1_{A}{}_{i},
\end{equation}
where $\rho_{u}^{C}=\rho_{d}^{C}=1.$

\subsection{The Higgs boson}
\label{Higgs}
Within our approach the 
self-interacting isospinor-scalar Higgs bosons arise 
as the collective modes of excitations of bound quasi-particle iso-pairs. 
Pursuing the analogy with [96-109] in outlined here approach a key 
problem is to find out the eligible mechanism 
leading to the formation of pairs, somewhat like Cooper mechanism, but
generalized for relativistic fermions, of course, in absence of any lattice.
We suggested this mechanism in the framework of gauge invariance 
incorporated with the P-violation phenomenon in W-world [2].
To trace a maximum resemblance to the superconductivity theory, in this 
section it will be advantageous to describe our approach in terms of four 
dimensional Minkowski space ${\M1_{W}}{}_{4}$ corresponding
to internal W-world:
$\G1_{W}\rightarrow {\M1_{W}}{}_{4}$(subsec.2.1). 
Although we shall leave the suffix $(W)$  implicit, but it goes without 
saying that all results obtained within this section refer to W-world.
Following the previous section, we consider the 
isospinor-scalar $H$-meson arisen in W-world
$$
H(x)=\gamma^{0}\,{\Psi_{L}}^{+}(x)\,\Psi_{R}(x),
$$
where $x \in M_{4}$ is a point of W-world. The standard notational
conventions will be employed throughout
$$
{\q1_{W}}{}_{L}\equiv {\ps1_{W}}{}_{L}(\x1_{W})\rightarrow \Psi_{L}(x),
\quad {\M1_{W}}{}_{4}\rightarrow M_{4},\quad
{\q1_{W}}{}_{R}\equiv {\ps1_{W}}{}_{R}(\x1_{W})\rightarrow \Psi_{R}(x),
$$
where 
$
\Psi_{R}(x)=\gamma(1+\vec{\sigma}\vec{\beta})\Psi_{L}(x), \quad 
\vec{\beta}=\FFr{\vec{v}}{c},\quad
\Psi_{L}(x)=\gamma(1-\vec{\sigma}\vec{\beta})\Psi_{R}(x), \quad
\gamma =\FFr{E}{m},
$
provided by the spin $\vec{\sigma}$, energy $E$ and velocity
$\vec{v}$ of particle. In terms of Fourier integrals
\begin{equation}
\label{eq: R12.12.1}
\Psi_{L}(x)=\FFr{1}{(2\pi)^{4}}\IIn\Psi_{L}(p_{L})
{\displaystyle e^{ip_{L}x}} d^{4}p_{L},
\quad
\Psi_{R}(x)=\FFr{1}{(2\pi)^{4}}\IIn\Psi_{R}(p_{R})
{\displaystyle e^{ip_{R}x}} d^{4}p_{R},
\end{equation}
it is readily to get
\begin{equation}
\label{eq: R12.12.2}
H(k)=\IIn H(x)\,
{\displaystyle e^{-ikx}}d^{4}x=
\gamma^{0}\IIn\FFr{d^{4}p_{L}}{(2\pi)^{4}}
{\Psi_{L}}^{+}(p_{L})\Psi_{R}(p_{L}+k)=
\gamma^{0}\IIn\FFr{d^{4}p_{R}}{(2\pi)^{4}}
{\Psi_{L}}^{+}(p_{R}-k)\Psi_{R}(p_{R})
\end{equation}
provided by conservation law of fourmomentum
$
k = p_{R}-p_{L},
$
where $k=k(\omega, \vec{k})$, $\,p_{L,R}=p_{L,R}(E_{L,R}, \vec{p}_{L,R})$.
Our arguments on Bose-condensation are based on the local gauge invariance 
of the theory incorporated with the P-violation in weak interactions.
The rationale for this approach is readily forthcoming from the 
consideration of gauge transformations of the fields eq.(12.10.1) under
the P-violation in W-world
$$
\Psi'_{L}(x)=U_{L}(x)\Psi_{L}(x), \quad
\Psi'_{R}(x)=U_{R}(x)\Psi_{R}(x),
$$
where the Fourier expansions have carried out over corresponding 
{gauge quanta} with wave fourvectors $q_{L}$ and $q_{R}$
\begin{equation}
\label{eq: R12.12.3}
U_{L}(x)=\IIn\FFr{d^{4}q_{L}}{(2\pi)^{4}}
{\displaystyle e^{iq_{L}x}} U_{L}(q_{L}), \quad
U_{R}(x)=\IIn\FFr{d^{4}q_{R}}{(2\pi)^{4}}
{\displaystyle e^{iq_{R}x}} U_{L}(q_{R}),
\end{equation}
and
$
U_{L}(x)\neq U_{R}(x).
$
They induce the gauge transformations implemented upon $H$-field
$
H'(x) =U(x)\,H(x).
$
The matrix of {\em induced gauge transformations} may be written down
in terms of {\em induced gauge quanta}
\begin{equation}
\label{eq: R12.12.4}
U(x)\equiv U^{+}_{L}(x)U_{R}(x)=
\IIn\FFr{d^{4}q}{(2\pi)^{4}}
{\displaystyle e^{iq x}} U(q), 
\end{equation}
where $q=-q_{L}+q_{R}, \quad q(q^{0},\vec{q})$. In momentum space
one gets
\begin{equation}
\label{eq: R12.12.5}
\begin{array}{l}
H'(k')=\IIn\FFr{d^{4}q}{(2\pi)^{4}}\,
U(q) \, H(k'-q)=
\IIn\FFr{d^{4}k}{(2\pi)^{4}}\,
U(k'-k) \, H(k).
\end{array}
\end{equation}
Conservation of fourmomentum requires that
$
k'=k + q.
$
According to eq.(12.10.2) and eq.(12.10.5), we have
$$
-p'_{L}+p'_{R}=-p_{L}+p_{R} + q =-p''_{L}+p_{R}=
-p_{L}+p''_{R},
$$
where 
$
p''_{L}=p_{L}-q,\quad
p''_{R}=p_{R}+q.
$
As to the wave vectors of fermions, they imply the conservation law
$
\vec{p}_{L}+\vec{p}_{R}=\vec{p}''_{L}+\vec{p}''_{R}
$
characterizing the scattering process of two fermions with {\em effective
interaction caused by the mediating induced gauge quanta}.  We suggest the 
mechanism for the effective attraction between the fermions 
in the following manner: among all induced gauge 
transformations with miscellaneous gauge quanta we distinguish 
a subset with the induced gauge quanta of the frequencies characterized 
by the maximum frequency
$\FFr{\widetilde{q}}{\hbar}\quad (\widetilde{q}=max\{q^{0}\})$
greater than the frequency of inducing oscillations fermion force
$
\FFr{\bar{E_{L}}-\bar{E''_{L}}}{\hbar}< \FFr{\widetilde{q}}{\hbar}.
$
To the extent that this is a general phenomenon, we can expect under 
this condition the effective attraction 
(negative interaction) arisen between the fermions caused by exchange of
virtual induced gauge quanta if only the forced oscillations of
these quanta occur in the same phase with the oscillations of
inducing force (the oscillations of fermion density).
In view of this we may think of isospinor $\Psi_{L}$ and isoscalar
$\Psi_{R}$ fields as the fermion fields composing the iso-pairs with the 
same conserving net momentum $\vec{p}={\vec{p}}_{L}+{\vec{p}}_{R}$
and opposite spin, for which the maximum number of negative matrix
elements of operators composed by corresponding creation and annihilation
operators 
$
a^{+}_{\vec{p}\,''_{R}}\, a_{\vec{p}_{R}}\,
a^{+}_{\vec{p}\,''_{L}}\, a_{\vec{p}_{L}}
$
(designated by the pair wave vector $\vec{p}$)
may be obtained for coherent ground state with
$
\vec{p}=\vec{p}_{L}+\vec{p}_{R}=0.
$
The fermions filled up the Fermi sea block
the levels below Fermi surface. Hence, the fermions are in 
superconductive or normal state described
by Bloch individual particle model. Thus, the Bose-condensate arises
in the W-world as the collective mode of excitations of bound
quasi-particle iso-pairs described by the same wave function in the
superconducting phase
$
\Psi=<\Psi_{L}\Psi_{R}>,
$
where $<\cdots>$ is taken to denote the vacuum averaging. The vacuum of the 
W-world filled up by such iso-pairs at absolute zero $T=0$.\\
We make a final observation that 
$\Psi_{R}\Psi_{R}^{+}=n_{R}$ is a scalar density number of
right-handed particles. Then it readily follows:
\begin{equation}
\label{eq: R12.12.8}
(\Psi_{L}\Psi_{R})^{+}(\Psi_{L}\Psi_{R})=H H^{+},
\end{equation}
where
$
\mid \Psi\mid^{2}=
<HH^{+}>=\mid <H>\mid^{2}\equiv 
\mid H \mid^{2}.
$
It is convenient to abbreviate the $<H>$ by the 
symbol $H $. Thus,
the $H$-meson actually arises as the collective mode of 
excitations of bound quasi-particle iso-pairs.

\subsection{The energy gap function}
\label{Rel}
We start with total Lagrangian eq.(12.7.4) of self-interacting
fermion field in W-world
\begin{equation}
\label{eq: R12.14.1}
\begin{array}{l}
\L1_{W}(x)=
\FFr{i}{2} \{ 
\bp_{W}(x)\gamma^{\mu}{\pr_{W}}{}_{\mu}\ps1_{W}(x)-
\bp_{W}(x)\gamma^{\mu}{\lpr_{W}}{}_{\mu}\ps1_{W}(x)\}-
m\bp_{W}(x)\ps1_{W}(x)-\\
-\FFr{\lambda}{2}\bp_{W}(x)\left(\bp_{W}(x)\ps1_{W}(x)\right)\ps1_{W}(x),
\end{array}
\end{equation}
where, $m =\Sigma_{Q}$ is the self-energy operator of the fermion field 
component in Q-world, the suffix $(W)$ just was put forth for instance in 
illustration of a point at issue. For the sake of simplicity, we also admit 
${\bf\B1_{W}}(x)=0$, but of course one is free to restore the gauge field 
${\bf \B1_{W}}(x)$ whenever it should be needed. 
In lowest order the relation
$m\equiv m_{Q}\ll \lambda^{-1/2}$ holds.
At non-relativistic limit the function $\Psi$ reads
$
\Psi \rightarrow e^{imc^{2}t}\Psi,
$
and Lagrangian eq.(12.11.1) leads to Hamiltonian used in [99]. 
We make use of the Gor'kov's technique and evaluate the field equations 
ensued from the eq.(12.11.1) in following manner: 
The spirit of the calculation will be to treat interaction between 
the particles as being absent everywhere except the thin spherical shell
$2\widetilde{q}$ near the Fermi surface. The Bose
condensate of bound particle iso-pairs occurred at zero momentum.
The scattering processes between the particles are absent. 
We consider the matrix elements 
$
<T \left(\Psi_{\alpha}(x_{1})\Psi_{\beta}(x_{2})
\bar{\Psi}_{\gamma}(x_{3})\bar{\Psi}_{\delta}(x_{4})
\right)>
$
and introduce the functions
\begin{equation}
\label{eq: R12.14.3}
\begin{array}{l}
<N \mid
T \left( \gamma^{0} \Psi(x)\Psi(x')\right)
\mid N+2>=e^{-2i\mu't}F(x-x'), \\ 
<N +2\mid
T \left( \Psi^{+}(x)\gamma^{0}\Psi^{+}(x')\right)
\mid N>=e^{2i\mu't}F^{+}(x-x').
\end{array}
\end{equation}
Here, $\mu'=\mu + m$, $\mu$ is the chemical potential. 
We omit a prime over $\mu$, but should
understand  $\mu + m$ under it. 
Making use of Fourier integrals, it
renders the field equations easier to handle in momentum space [2]
\begin{equation}
\label{eq: R12.14.4}
\begin{array}{l}
(\gamma p-m)G(p)-i\lambda \gamma^{0}F(0+)\bar{F}(p)=
1, \\ 
\bar{F}(p)(\gamma p+m-2\mu \gamma^{0})+
i\lambda \bar{F}(0+)G(p)=0,
\end{array}
\end{equation}
where $G(p)$ is the thermodynamic Green's function, 
$
F_{\alpha\beta}(0+)=e^{2i\mu t}
<T \left( \gamma^{0} \Psi_{\alpha}(x)\Psi_{\beta}(x)\right)>=
\Lm_{x \rightarrow x'(t \rightarrow t')}F_{\alpha\beta}(x-x').
$
Next we substitute 
$$
(\gamma p-m)=(\omega' -\xi_{p})\gamma^{0},\quad
(\gamma p+m-2\mu \gamma^{0})=\gamma^{0}(\omega' +\xi_{p}^{+}),
$$
where
$$
\begin{array}{l}
\omega' =\omega -\mu'=\omega -m -\mu ,\quad
\xi_{p}=(\vec{\gamma}\vec{p}+m)\gamma^{0}-\mu'=
(\vec{\gamma}\vec{p}+m)\gamma^{0}-m-\mu, \\
\xi_{p}^{+}=\gamma^{0}({\vec{\gamma}}^{+}\vec{p}+m) -m-\mu,
\end{array}
$$
and omit a prime over $\omega'$ for the rest of this section.
We apply
$$
F(0+)=-JI, \quad 
I=\left( \matrix{
\hspace{0.3cm} 0 \quad 1\cr
-1 \quad 0 \cr
}\right),\quad
F^{+}(0+)F(0+)=-J^{2}I^{2}=J^{2}, 
$$
and
$\widehat{\omega} +\widehat{\xi}_{p}=
\gamma^{0}(\omega +\xi_{p})$.
The gap function $\Delta$ reads
$
\Delta^{2}=\lambda^{2} J^{2},
$
where
$
J=\IIn\FFr{d\omega\,d\vec{k}}{(2\pi)^{4}}F^{+}(p).
$
Using the standard rules [108], one may pass over the poles. 
This method allows oneself to extend the study up to limit of
temperatures, such that $T_{c}-T\ll T_{c}$ in terms of thermodynamic 
Green's function.
So
\begin{equation}
\label{eq: R12.14.5}
\begin{array}{l}
F^{+}(p)=-i\lambda J (\omega-\xi_{p}+i\delta)^{-1}
(\omega+\xi_{p}-i\delta)^{-1} 
-\FFr{\pi\Delta}{\varepsilon_{p}}n(\varepsilon_{p})
\{ \delta(\omega-\varepsilon_{p})+
\delta(\omega+\varepsilon_{p})\}, 
\\
G(p)=\gamma^{0}\{u_{p}^{2}(\omega-\xi_{p}+i\delta)^{-1}+
v_{p}^{2}(\omega+\xi_{p}-i\delta)^{-1}+ 
2\pi i n(\varepsilon_{p})
[ u_{p}^{2}\delta(\omega-\varepsilon_{p})-\\
v_{p}^{2}\delta(\omega+\varepsilon_{p})]\},
\end{array}
\end{equation}
where 
$u_{p}^{2}=\FFr{1}{2}\left( 1+ \FFr{\xi_{p}}{\varepsilon_{p}}\right),
\quad
v_{p}^{2}=\FFr{1}{2}\left( 1- \FFr{\xi_{p}}{\varepsilon_{p}}\right),\quad
\varepsilon_{p}=(\xi_{p}^{2}+\Delta^{2}(T))^{1/2}.
$
and
$n(\varepsilon_{p})$ is the usual Fermi function
$
n(\varepsilon_{p})=\left( \exp\FFr{\varepsilon_{p}}{T} +1\right)^{-1}.
$
Then 
\begin{equation}
\label{eq: R12.14.6}
1=\FFr{\mid \lambda \mid}{2(2\pi)^{3}}
\IIn d\vec{k}\, \FFr{1-2n(\varepsilon_{k})}{\varepsilon_{k}(T)}
\quad \left( \mid \xi_{p}\mid
< \widetilde{q} \right),
\end{equation}
determining the energy gap $\Delta$ as a function of $T$. According to
eq.(12.11.5), the $\Delta(T)\rightarrow 0$ at $T\rightarrow T_{c}\sim
\Delta(0)$ [96].

\subsection{Self-interacting potential of Bose-condensate}
\label{Pot}
To go any further in exploring the form and significance of obtained
results it is entirely feasible to include 
the generalization of the equations (12.11.3) in presence of spatially 
varying magnetic field with vector potential $\vec{A}(\vec{r})$, which is
straightforward $(t\rightarrow \tau=it)$
\begin{equation}
\label{eq: R12.15.1}
\begin{array}{l}
\left\{ 
-\gamma^{0}\FFr{\partial}{\partial \tau}-
i\vec{\gamma}\left(\FFr{\partial}{\partial \vec{r}}-ie\vec{A}(\vec{r}) 
\right)-m +\gamma^{0}\mu 
\right\}
G(x,x')+\gamma^{0}\Delta(\vec{r})\bar{F}(x,x')=
\delta (x-x'), \\ 
\bar{F}(x,x')
\left\{ 
\gamma^{0}\FFr{\partial}{\partial \tau}+
i\vec{\gamma}\left(\FFr{\partial}{\partial \vec{r}}+ie\vec{A}(\vec{r}) 
\right)-m +\gamma^{0}\mu 
\right\}-
\Delta^{*}(\vec{r})\gamma^{0}G(x,x')=0,
\end{array}
\end{equation}
where the thermodynamic Green's function [116,117] is used, 
the energy gap function is in the form
$
\Delta^{*}(\vec{r})=\lambda F^{+}(\tau,\vec{r};\,\tau,\vec{r}).
$
This function is logarithmically divergent, but with a cutoff of
energy of interacting fermions at the spatial distances in order
of $\FFr{\hbar v}{\widetilde{\omega}}$ can be made finite, 
where $\widetilde{\omega} \equiv \FFr{\widetilde{q}}{\hbar}$.
To handle the eq.(12.12.1) one uses the Fourier components of functions 
$G(x,x')$ and $F(x,x')$
\begin{equation}
\label{eq: R12.15.2}
G(\vec{r},\vec{r}';u)=T\S_{n}e^{-i\omega u}
G_{\omega}(\vec{r},\vec{r}'), \quad
G_{\omega}(\vec{r},\vec{r}')=\FFr{1}{2}\IIn^{1/T}_{-1/T}e^{i\omega u}
G(\vec{r},\vec{r}';u)d\,u,
\end{equation}
where $u=\tau - \tau'$, $\omega$ is the discrete index $\omega =
\pi T(2n + 1), \quad n=0,\pm 1, \ldots$.\, the gap function is defined by
$
\Delta^{*}(\vec{r})=\lambda T\S_{n}
F^{+}_{\omega}(\vec{r},\vec{r}').
$
Then (see [2])
\begin{equation}
\label{eq: R12.15.6}
G_{\omega}(\vec{r},\vec{r}')=\widetilde{G}_{\omega}(\vec{r},\vec{r}')-
\IIn\widetilde{G}_{\omega}(\vec{r},\vec{s})
\gamma^{0}\Delta(\vec{s})
\bar{F_{\omega}}(\vec{s},\vec{r}')d^{3}s,
\end{equation}
and
\begin{equation}
\label{eq: R12.15.7}
\bar{F_{\omega}}(\vec{r},\vec{r}')=
\IIn\widetilde{G}_{\omega}(\vec{s},\vec{r}')
\Delta^{*}(\vec{s})\gamma^{0}
\widetilde{G}_{-\omega}(\vec{s},\vec{r})d^{3}s.
\end{equation}
where $\widetilde{G}_{\omega}(\vec{r},\vec{r}')$ is the Bloch individual 
particle Green's function of the fermion in normal mode.  
The gap function $\Delta(\vec{r})$ as well as 
$\bar{F_{\omega}}(\vec{r},\vec{r}')$ are small ones at close neighbourhood
of transition temperature $T_{c}$ and varied slowly over a coherence
distance. This approximation, which went into the derivation of 
equations,  meets our interest in eq..(12.12.3), eq.(12.12.4).
Using standard procedure one may readily express them in power series of 
$\Delta$ and $\Delta^{*}$ by keeping
only the terms in $\bar{F_{\omega}}(\vec{r},\vec{r}')$ up to the cubic
and in $G_{\omega}(\vec{r},\vec{r}')$ - quadratic order in 
$\Delta$. The technique now is to average over the polarization of particles 
and expand the obtained equation up to the terms quadratic in 
$(\vec{r}-\vec{r}')$.
Keeping in mind aforesaid, after calculations the resulting equation can be 
written [2] 
\begin{equation}
\label{eq: R12.15.13}
\begin{array}{l}
\left\{
\left(i\hbar\vec{\nabla}+\FFr{e^{*}}{c}\vec{A} \right)^{2}+
\FFr{2m}{\nu}\left[ 
\FFr{2\pi^{2}}{\lambda m p_{0}}\left(\FFr{\mu}{m} \right)^{2}
\left( \FFr{\mu}{m}-1\right)+
\left(\FFr{\mu}{m} \right)^{2}
\left( \FFr{T}{T_{c\mu}}-1\right)+
\right. \right.\\ 
\left. \left.
\FFr{2}{N}\mid \Psi(\vec{r})\mid^{2}
\right]
\right\}\Psi(\vec{r})=0,
\end{array}
\end{equation}
where $\nu=\FFr{7\zeta(3)mv_{F}^{2}}{24(\pi k_{B}T_{c})^{2}}$ and 
$T_{c\mu}= \FFr{m}{\mu}T_{c}$.
Succinctly 
\begin{equation}
\label{eq: R12.15.14}
\left\{
\vec{p}_{A}^{2}-\FFr{1}{2}m_{\Psi}^{2}+\FFr{1}{4}\lambda_{\Psi}^{2}
\mid \Psi(\vec{r})\mid^{2}
\right\}\Psi(\vec{r})=0,
\end{equation}
provided by
\begin{equation}
\label{eq: R12.15.15}
\begin{array}{l}
m_{\Psi}^{2}(\lambda,T,T_{c\mu})=
\FFr{24}{7\zeta(3)}\left(\FFr{\hbar}{\xi_{0}} \right)^{2}
\left(\FFr{\mu}{m} \right)^{2}
\left[
1-\FFr{T}{T_{c\mu}}-\left( \FFr{\mu}{m}-1\right)
\ln\FFr{2\widetilde{\omega}}{\Delta_{0}}
\right],\\ 
\lambda_{\Psi}^{2}(\lambda,T_{c})=
\FFr{96}{7\zeta(3)}\left(\FFr{\hbar}{\xi_{0}} \right)^{2}
\FFr{1}{N}, \quad 
\Psi(\vec{r})=\Delta(\vec{r})\FFr{\left(7\zeta(3)N\right)^{1/2}}
{4\pi k_{B}T_{c}}, \quad 
\vec{p}_{A}=i\hbar\vec{\nabla}+\FFr{e^{*}}{c}\vec{A}.
\end{array}
\end{equation}
Whence, the transition temperature decreases inversely by relativistic 
factor $\FFr{\mu}{m}$.
A spontaneous breaking of symmetry of ground state occurs at
$
\eta_{\Psi}^{2}(\lambda,T < T_{c\mu}) > 0,
$
where
$
\eta_{\Psi}^{2}(\lambda,T,T_{c\mu})=\FFr{m_{\Psi}^{2}}{\lambda_{\Psi}^{2}}.
$
\\
The eq.(12.12.6) splits into the couple of equations for
$\Psi_{L}$ and $\Psi_{R}$. Thus, a Lagrangian of the
$H$ boson will be arisen  with the corresponding values of mass 
$m_{\Psi}^{2}\equiv m_{H}^{2}$
and coupling constant $\lambda_{\Psi}^{2}\equiv\lambda_{H}^{2}$.

\subsection{Four-component Bose-condensate in magnetic field}
\label{Four}
In [2] we have derived the equation of four-component 
bispinor field of Bose-condensate, wherein   
due to self-interaction the spin part of it is vanished. 
Actually, we start with the nonsymmetric state $\Delta_{L}\neq \Delta_{R}$, 
where $\Psi_{L}$ and $\Psi_{R}$ are two eigenstates of chirality 
operator $\gamma_{5}$. Then, the eq.(12.12.5) enables to postulate the 
equation of four-component Bose-condensate in equilibrium state at presence 
of the magnetic field 
\begin{equation}
\label{eq: R12.16.2}
i\hbar \FFr{\partial \Psi}{\partial t}=
\left\{
c\vec{\alpha}\left(\vec{p}+ \FFr{e^{*}}{c}\vec{A} \right)+\beta mc^{2}+
M(F)+L(F)\mid \Psi\mid^{2}
\right\}\Psi=0,
\end{equation}
where the functions $M(F)$ and $L(F)$ can be determined under the requirement 
that the second-order equations ensued from the eq.(12.13.1) must match onto 
eq.(12.12.6). Also, taking into account an approximation fitting our interest 
that the gap function is small at close neighbourhood of transition 
temperature, we get: 
$$
\begin{array}{l}
M(F)=\left(M^{2}_{0}+ \FFr{i}{2}e^{*}F \right)^{1/2}, \quad
M_{0}=\left( m^{2}+\FFr{1}{2}m^{2}_{H} \right)^{1/2},\quad
L(F)=-\FFr{\lambda^{2}_{H}}{8M(F)}, \\ 
L_{0}=-\FFr{\lambda^{2}_{H}}{8M_{0}},
M(F)L(F)=M_{0}L_{0}=-\FFr{1}{8}\lambda^{2}_{H},
\end{array}
$$
such that
\begin{equation}
\label{eq: R12.16.4}
\left\{ \vec{p}_{A}^{2} +m^{2}-\left(M_{0}+L_{0}\mid \Psi\mid^{2} 
\right)^{2} \right\}\Psi(\vec{r})\equiv
\left\{ \vec{p}_{A}^{2}-\FFr{1}{2}m^{2}_{H}+
\FFr{1}{4}\lambda^{2}_{H}\mid \Psi\mid^{2} 
 \right\}\Psi(\vec{r})=0.
\end{equation}
This has yet another 
important consequence. At $\Delta_{L}\neq 0$ and
imposed constraint\\
$
\left( m+M(F) + L(F)\mid \Psi\mid^{2}\right)_{F\rightarrow 0}
\rightarrow 0
$
we have
\begin{equation}
\label{eq: R12.16.5}
\Delta_{2}=
\FFr{1}{\sqrt{2}}(\Delta_{L}-\Delta_{R})=0, \quad
\Psi_{2}=0.
\end{equation}
Thus, the $\mid \Psi_{0}\mid $ is the gap function symmetry-restoring 
value
$$
\Delta_{2}\left( \mid \Psi_{0}\mid^{2}=\FFr{m+M_{0}}{-L_{0}}\right)
=0,
\quad
\Delta_{L}\left( \mid \Psi_{0}\mid^{2}\right)=
\Delta_{R}\left( \mid \Psi_{0}\mid^{2}\right),
$$
where, according to eq.(12.13.2), one has
\begin{equation}
\label{eq: R12.16.6}
V\equiv \left[m^{2}-\left( M_{0}+L_{0}\mid \Psi\mid^{2}\right)^{2}\right]
\Psi^{2} =\left[-\FFr{1}{2}m^{2}_{H}+
\FFr{1}{4}\lambda^{2}_{H}\mid \Psi\mid^{2}\right]
\Psi^{2},
\end{equation}
and 
\begin{equation}
\label{eq: R12.16.7}
V\left( \mid \Psi_{0}\mid^{2}=\FFr{m+M_{0}}{-L_{0}}=
\FFr{1}{2}\eta^{2}_{H}(\lambda,T,T_{c\mu})
\right)=0.
\end{equation}
We conclude that the field
of symmetry-breaking Higgs boson must be counted off from the
$\Delta_{L}=\Delta_{R}$ symmetry-restoring value of Bose-condensate
$
\mid \Psi_{0}\mid=\FFr{1}{\sqrt{2}}\eta_{H}(\lambda,T,T_{c\mu})
$
as the point of origin describing the excitation in the neighbourhood
of stable vacuum eq.(12.13.5).
The gauge invariant Lagrangian eq.(12.7.4) takes the form
\begin{equation}
\label{eq: R12.16.8}
\L1_{W}(\D1_{W})=\FFr{i}{2}
\left\{ 
\bp_{W} \gamma \D1_{W} \ps1_{W} -
\bp_{W} \gamma \overleftarrow{\D1_{W}}\ps1_{W}
\right\}- 
\bp_{W}\left\{ 
m +\gamma^{0}\left[M(F) + L(F)\mid \Psi\mid^{2}\right] 
\right\}\ps1_{W}.
\end{equation}
At the symmetry-restoring point, this Lagrangian 
can be replaced by
$$
\L1_{W}(\D1_{W})\rightarrow {\L1_{W}}_{1}(\D1_{W})=
\FFr{1}{2}\left( \D1_{W}{\ps1_{W}}_{1} \right)^{2} -
\V_{W}\left(\mid {\ps1_{W}}_{1} \mid^{2} \right),
$$
provided 
$$
\V_{W}\left(\mid {\ps1_{W}}_{1} \mid^{2} \right)=
-\FFr{1}{2}m^{2}_{H}{\ps1_{W}}_{1}^{2}+
\FFr{1}{4}\lambda^{2}_{H}{\ps1_{W}}_{1}^{4}.
$$
Taking into account the eq.(12.10.6), in which
$
\mid {\ps1_{W}}_{1}\mid^{2}=\mid H \mid^{2}=
\FFr{1}{2}\mid \eta_{H} +\chi\mid^{2},
$
one gets
\begin{equation}
\label{eq: R12.16.9}
{\L1_{W}}_{H}(\D1_{W})=
\FFr{1}{2}\left( \D1_{W} H \right)^{2} -
\V_{W}\left(\mid H \mid^{2} \right),
\quad
\V_{W}\left(\mid H \mid^{2} \right)=
-\FFr{1}{2}m^{2}_{H}\varphi^{2} +
\FFr{1}{4}\lambda^{2}_{H}H^{4}.
\end{equation}
Finally, recording the question of whether or not it is 
possible to extend the ideas of former approach to lower temperatures as 
it was investigated in the case of Gor'kov's theory by others [112-114], 
in [2], as usual, we admit that the order parameter and vector potential 
vary slowly over distances of the order of the coherence length. 
We restrict ourselves to the London limit and the derivation of equations
was proceeded by iterating to a low order giving only the leading 
terms. Taking into account the eq.(12.12.1), eq.(12.12.3) and  eq.(12.12.4), 
it is straightforward to derive the separate integral
equations for $G$ and $F^{+}$ in terms of $\Delta$, $\Delta^{*}$
and $\widetilde{G}$.
The mathematical structure of obtained equations is closely similar to
that studied by [112,119,120] in somewhat different context. Then, adopting 
their technique in [2] we introduce sum and difference coordinates, and
Fourier transform with respect to the difference coordinates.
To obtain resulting expressions we have proceeded with further 
standard calculations. There is only one thing to be noticed about the 
integration. Due to the angular integration in momentum space, the terms 
linear in the vector $\vec{p}$ will be vanished , as well as the integration 
over the energies removes the linear terms in 
$
\epsilon (\vec{p},\vec{R})\equiv \FFr{1}{2m}\left( 
\vec{p}-\FFr{e}{c}\vec{A}(\vec{R})
\right)^{2}-\mu_{0}.
$
Thus, we may
expand the quantities according to the degree of
inhomogenity somewhat like it we have done in equations (12.12.3) and 
eq.(12.12.4) of gap function $\Delta^{*}(\vec{r})$, which in mixed 
representation transforms to the following:
\begin{equation}
\label{eq: R12.17.8}
\Delta^{*}(\vec{R})=
T\S_{\omega}\IIn\FFr{d^{3}p}{(2\pi)^{3}}
F^{+}_{\omega}(\vec{p},\vec{R})
=T\S_{\omega}\IIn\FFr{d^{3}p}{(2\pi)^{3}}
\widehat{\gamma}_{A}(\vec{p},\vec{R})
F^{+}_{\Omega}(\vec{p},\vec{R}).
\end{equation}
The approximation was used to obtain the function $F^{+}_{\Omega}$
must be of one order higher 
$F^{+}_{\Omega}\simeq F^{(0)+}_{\Omega}+F^{(1)+}_{\Omega}+
F^{(2)+}_{\Omega}$ than that for function 
$\widetilde{G}_{\Omega}\simeq \widetilde{G}_{\Omega}^{(0)}+
\widetilde{G}_{\Omega}^{(1)}$. Applying an iteration method of solution
one replaces $K\rightarrow \widetilde{K},\quad 
G\rightarrow \widetilde{G}$ in eq.(12.17.5) and puts $\Theta^{(0)}=1,\quad
\widetilde{K}^{(1)}=0, \quad \widetilde{G}^{(1)}=0$. Hence 
$\widetilde{G}=\widetilde{G}^{(0)}$. 
The resulting equation
for gap function is similar to those occurring in [112], although 
not identical. The sole difference is that in the resulting equation we use
the expressions of $\Omega$ and $\xi$.

\subsection{Transmission of the electroweak symmetry breaking from 
the $W$-world to spacetime continuum; the two solid phenomenological 
implications of the MSM }
\label{simul}
A common feature of gauge theories is that to break the gauge symmetry 
down and leading to masses of the fields, one needs in general, several 
kinds of spinless Higgs mesons, with conventional Yukawa couplings to 
fermion currents and transforming by an irreducible representation of 
gauge group. The conventional Higgs theory like [121] involves these 
bosons as the ready made fundamental elementary fields defined on the 
Minkowski spacetime continuum $M_{4}$, which entails various difficulties. 
As it is seen in the previous subsections, the self-interacting isospinor 
scalar Higgs bosons arise in MSM as the collective modes of excitations of 
bound quasi-particle iso-pairs in the internal $W$-world. 
Whence, a first important phenomenological implication of the MSM ensues 
at once that 

$\bullet$ {\em such Higgs bosons never could emerge in spacetime continuum 
and, thus,  could not be discovered in experiments nor at any energy range}. 
\\
It just  remains to see how can these bosons break the gauge symmetry 
down in $M_{4}$ and lead to masses of the spacetime-components of the 
MW-fields? It is remarkable to see that the MSM, in contrast to the 
SM, predicts the transmission of electroweak symmetry breaking from the 
$W-$world to the $M_{4}$ spacetime continuum. Actually,
in standard scenario for the simplest Higgs sector, a gauge invariance of 
the Lagrangian is broken in the $W$-world when
the $H$-meson fields eq.(12.13.7) acquire a VEV $\quad\eta_{H}\neq 0$.
Thereby the mass $m_{H}$ and coupling 
constant $\lambda_{H}$ are in the form eq.(12.15.15). 
The spontaneous breakdown of symmetry is vanished at 
$\eta_{H}^{2}(\lambda, T > T_{c\mu}) < 0$.
When this doublet obtains a VEV, three of the
gauge fields $\Z1_{W}{}^{0}_{\mu},\,\W1_{W}{}^{\pm}_{\mu}$ acquire masses.
These fields are the $W$-components of the mesons mediating the weak 
interactions. Certainly, the derivative 
$$\D1_{W}{}_{\mu}H \equiv \left( \pr_{W}{}_{\mu}-
\FFr{i}{2} \,g\,
{\bf \tau \cdot \W1_{W}}{}_{\mu}-
\FFr{i}{2} \,g'\,\X1_{W}{}_{\mu}\right)H
$$
arisen in the eq.(12.16.9), in standard scenario leads to the masses  
$$
M_{W}=\FFr{g\,\eta_{H}}{2}, \quad 
M^{2}_{Z}=\FFr{\left(g^{2}+g'{}^{2}\right)^{1/2}\,\eta_{H}}{2}, \quad
\cos\,\theta_{W}=\FFr{M_{W}}{M_{Z}},
$$ 
respectively of the gauge field components
$$
\W1_{W}{}^{\pm}_{\mu}=\FFr{1}{\sqrt{2}}\left(\W1_{W}{}^{1}_{\mu} 
\pm \W1_{W}{}^{2}_{\mu}\right),
\quad
\Z1_{W}{}_{\mu}=\FFr{g\,\W1_{W}{}^{3}_{\mu}-
g'\,\X1_{W}{}_{\mu}}{\left(g^{2}+g'{}^{2}\right)^{1/2}}\equiv
\cos\,\theta_{W}\,\W1_{W}{}^{3}_{\mu}-\sin\,\theta_{W}\,\X1_{W}{}_{\mu}.
$$
Consequently, a remaining massless gauge field 
$$
\A1_{W}{}_{\mu}=\FFr{g'\,\W1_{W}{}^{3}_{\mu}+
g\,\X1_{W}{}_{\mu}}{\left(g^{2}+g'{}^{2}\right)^{1/2}}\equiv
\sin\,\theta_{W}\,\W1_{W}{}^{3}_{\mu}+\cos\,\theta_{W}\,\X1_{W}{}_{\mu}.
$$
will be identified as the $W$-component of the photon field coupled to the 
electric current.
Therewith the $x$-components of the fields above simultaneously 
acquire corresponding masses too, since, according to the specific MW scheme 
(see subsec.12.7), all the components of corresponding frame fields are 
defined on the MW mass shells, i.e.,
$$
\Bx_{x}\,\W1_{x}{}_{\mu}=M^{2}_{W}\,\W1_{x}{}_{\mu},\quad
\Bx_{x}\,\Z1_{x}{}_{\mu}=M^{2}_{Z}\,\Z1_{x}{}_{\mu},\quad
\Bx_{x}\,\A1_{x}{}_{\mu}=M^{2}_{A}\,\A1_{x}{}_{\mu},
$$
provided by
$$
M^{2}_{W}\,\W1_{W}{}_{\mu}\equiv \Bx_{W}\,\W1_{W}{}_{\mu},\quad
M^{2}_{Z}\,\Z1_{W}{}_{\mu}\equiv \Bx_{W}\,\Z1_{W}{}_{\mu},\quad 
M^{2}_{A}\,\A1_{W}{}_{\mu}\equiv \Bx_{W}\,\A1_{W}{}_{\mu}=0. 
$$
The microscopic structure
of these fields reads
$$
\begin{array}{l}
W^{+}\equiv {\phi}{}_{W}(\eta)\, 
(q_{1}q_{2}q_{3})^{Q}(q\bar{q})^{W}, \quad
W^{-}\equiv {\phi}{}_{W}(\eta)\, 
(\overline{q_{1}q_{2}q_{3}})^{Q}(\bar{q}q)^{W}, \\
Z^{0}\equiv {\phi}{}_{Z}(\eta)\, 
(q\bar{q})^{Q}(q\bar{q})^{W},\quad
A \equiv {\phi}{}_{A}(\eta)\, 
(q\bar{q})^{Q}\A1_{W}(0).
\end{array}
$$
The values of the masses $M_{W}$ and $M_{Z}$ are changed if the Higgs
sector is built up more compoundly. 
Due to Yukawa couplings eq.(12.9.2) the fermions acquire the masses after 
symmetry-breaking. The mass of electron reads
$
m_{e}=\FFr{\eta_{H}}{\sqrt{2}}f_{e}
$
etc. One gets for the leptons
$
f_{e}:f_{\mu}:f_{\tau}=m_{e}:m_{\mu}:m_{\tau}.
$
This mechanism does not disturb the
renormalizability of the theory  [122,123].
In approximation to lowest order
$
f=\Sigma_{Q}\simeq m_{Q}\ll \lambda^{-1/2}\quad
\left( \lambda^{-1}=\FFr{mp_{0}}{2\pi^{2}}
\ln\FFr{2\widetilde{\omega}}{\Delta_{0}}\right),
$
the Lagrangians
eq.(12.7.3) and eq.(12.7.4) produce the Lagrangian of phenomenological
SM, where at $f\sim 10^{-6}$ one gets $ \lambda\ll 10^{12}$.
In standard scenario the lowest pole $m_{Q}$ of the self-energy operator 
$\Sigma_{Q}$ has fixed the whole mass spectrum of the SM particles. But, in 
general, one must also take into 
account the mass spectrum of expected various collective excitations of 
bound quasi-particle pairs produced by higher-order interactions as a 
``superconductive'' solution obtained from a nonlinear spinor field 
Lagrangian of the $Q$-component possessed $\gamma_{5}$ invariance. 
These states must be considered as a direct effect of the same primary 
nonlinear fermion interaction which provides the mass of the $Q$-component of 
Fermi field, which itself is a collective effect. 
They would manifest themselves 
as stable or unstable states. The general features of mass spectrum of 
the collective excitations and their coupling with the fermions are 
discussed in [95] through the use of the Bethe-Salpeter 
equation handled in the simplest ladder approximation incorporated with the 
self-consistency conditions, when one is 
still left with unresolved divergence problem. 
One can reasonably expect that these results for the bosons of small masses at 
low energy compared to the unbound fermion states are essentially correct in 
spite of the very simple approximations.
Therein, some bound states are predicted too the obtained mass values 
of which are rather high, and these states should decay very quickly. 
The high-energy poles may in turn determine the low-energy resonances.
All this prompt us to expect that the other poles different from those of 
lowest one in turn will produce the new heavy SM family partners.
Hence one would expect a second important phenomenological implication of the 
MSM that: 

$\bullet$ {\em for each of the three SM families of quarks and 
leptons there are corresponding heavy family partners with the same 
$SU(3)_{c}\otimes SU(2)_{L}\otimes U(1)_{Y}$ quantum numbers at the energy 
scales related to next poles with respect to lowest one}.\\
To see its nature, now we may estimate the energy threshold of 
creation of such heavy family partners using the results far obtained in 
[95, 101]. It is therefore necessary under the simplifying assumption to 
consider in the $Q$-world a composite system of dressed fermion $(N_{*})$ 
made of the unbound fermion $(N)$ coupled with the different kind 
two-fermion bound states $(N\,\bar{N})$ at low energy, which all together 
represent the primary manifestation of the fundamental interaction. Such a 
dressed fermion would have a total mass $m_{*}\simeq m_{Q} + \mu$, where 
$m_{Q}$ and $\mu$ are the masses, respectively, of the unbound fermion and the 
bound state.
According to the general discussion of the mass spectrum of the collective 
excitations given in [95], here we are interested only in the following 
low-energy bound states written explicitly in spectroscopic notation 
$({}^{1}S_{0}){}_{N=0},\quad ({}^{1}S_{0}){}_{N=\pm 2},\quad 
({}^{3}P_{1}){}_{N=0}$ and $({}^{3}P_{0}){}_{N=0}$ with the expected masses 
$\mu=0,\,\,\,>\sqrt{2}m_{Q},\,\,
\sqrt{\FFr{8}{3}}m_{Q}$ and $2\,m_{Q}$, respectively, where the subscript $N$ 
indicates the nucleon number. One notes the peculiar symmetry existing 
between the pseudoscalar and the scalar states that the first has zero mass 
and binding energy $2m_{Q}$, while the opposite holds for the scalar state.
When the next pole $m_{*}$ to the lowest one $m_{Q}$ will be 
switched on, then due to the Yukawa couplings the all 
fermions will acquire the new masses with their common shift 
$\FFr{m_{*}}{m_{Q}}\equiv 1+k$ held upwards along the energy scale. To fix 
the energy threshold value all we have to do then is choose the heaviest 
member among the SM fermions, which is the top quark,  and to set up the 
quite obvious formula 
$$
E\geq E_{0}\equiv m_{t'}\,c^{2}=(1+k)\,m_{t}\,c^{2},
$$ 
where $m_{t}$ is the mass of the top quark. The top quark observed firstly in 
the two FNAL $p\,\bar{p}$ collider experiments in 1995, has the mass 
turned out to be startlingly large $m_{t}=(173.8 \pm 5.0)GeV/c^{2}$ 
compared to all the other SM fermion masses [125]. Thus, we get the most
important energy threshold scale estimate where the
heavy partners of the SM extra families of quarks and 
leptons likely would reside at:
$
E_{1}> (419.6 \pm 12.0)GeV,\quad E_{2}= (457.6 \pm 13.2)GeV$ and 
$
E_{3}=(521.4 \pm 15.0)GeV,
$
corresponded to the next nontrivial poles are written: 
$k_{1}>\sqrt{2}, \quad k_{2}=\sqrt{{8/3}}$ and $k_{3}= 2$, respectively.
We recognize well that the general results obtained in [95], however, 
model-dependent and may be considerably altered, especially in the high 
energy range by using better approximation. In present state of the theory it 
seemed to be a bit premature to get exact high energy results, which will be 
important subject for the future investigations. But, in the same time we 
believe that the approximation used in [95] has enough accuracy for the 
low-energy estimate made above . Anyhow, it is for one thing, the new scale 
where the family partners reside will be much higher than the 
electroweak scale and thus these heavy partners lie far above the electroweak 
scale.

\renewcommand{\theequation}{\thesection.\arabic{equation}}
\section{Emergence of two composite isospinor Higgs chiral supermultiplets}
\label {smult}
It is remarkable that just the two doublets of isospinor-scalar 
Higgs bosons arise (subsec.12.9), which are necessary in any 
self-consistent SUSY theory. One cannot write a SUSY version of 
Yukawa interactions of SM without at least a second Higgs doublet since 
there are three well known reasons for it [24-48].

$\bullet$ A model with a single Higgs doublet superfield suffers from quadratic 
divergences because the trace of the hypercharge generator does not vanish.

$\bullet$ Such a model has nonvanishing gauge anomalies associated with 
fermion triangle diagrams. 
The condition for cancellation of gauge anomalies include 
$Tr\left[Y^{3}\right]=Tr\left[T^{2}_{3}\,Y\right]=0$, where $T_{3}$ and 
$Y$ are the third component of weak isospin and the weak hypercharge, 
respectively, in a normalization where the electric charge is $Q=T_{3}+Y$. 
The trace run over all of the left-handed Weyl fermionic degrees of freedom in 
the theory. In the SM, these conditions are already satisfied by the known 
quarks and leptons, thus SM is anomaly-free. In SUSY version the definite 
chirality of the supersymmetric partner of Higgs boson carries
$U(1)$ hypercharge $Y=\FFr{1}{2}$ or $Y=-\FFr{1}{2}$. In either case alone, 
such a fermion upsets the anomaly cancellation condition by making a nonzero 
contribution to the trace. This can be avoided in the case of two Higgs 
supermultiplets, one with each of  $Y=\pm\FFr{1}{2}$.

$\bullet$ The masses of chiral fermions must be supersymmetric in 
conventional SUSY theory, i.e., they must originate from terms in the 
superpotential. In the SM 
the Higgs doublet (the complex conjugate of the doublet) can couple to the 
$T_{3}=+\FFr{1}{2}\,\,(T_{3}=-\FFr{1}{2})$ fermions in a gauge invariant way. 
In SUSY version, however, Yukawa interactions come from a superpotential, 
which cannot depend on a field as well as its complex conjugate. hence, any 
doublet can give mass either to a $T_{3}=+\FFr{1}{2}$ or $T_{3}=-\FFr{1}{2}$
fermion, but not both. Thus, to give masses to all the fermions one must 
introduce a second doublet $H_{d}$.

According to subsec.12.9 the Higgs bosons are composites
\begin{equation}
\label{eq: R19.1}
H = \gamma^{0}\,\ps1_{W}{}^{+}_{D\,L}\,\ps1_{W}{}_{D\,R}=
\gamma^{0}\,\bar{\ch1_{W}}\,\bar{\ps1_{W}},
\end{equation}
where the left $\bar{\ch1_{W}}$ and right-handed $\bar{\ps1_{W}}$ Weyl 
spinors are the members of chiral and anti-chiral superfields, respectively. 
Since they undergone SUSY transformations then Higgs bosons also have their 
SUSY partners. The $H$ belongs to chiral superfield, where its SUSY partner is 
Higgsino. Using eq.(13.1) and SUSY transformations of the components 
$\bar{\ch1_{W}}$ and $\bar{\ps1_{W}}$ one can obtain the explicit 
expression of the Higgsino. But it is much easier to handle it up, if one 
uses the SUSY transformation for the total Higgs boson itself as a member of 
chiral multiplet
\begin{equation}
\label{eq: R13.2}
\delta_{\xi}\,H =\sqrt{2}\,\xi\,\widetilde{H},
\end{equation}
where $\widetilde{H}$ is the Higgsino. By means of eq.(13.1) and
eq.(11.33) one gets
\begin{equation}
\label{eq: R13.3}
H + H^{+}= \gamma^{0}\,
\left(\bar{\ch1_{W}}\,\bar{\ps1_{W}}+\ch1_{W}\,\ps1_{W}\right)=
\gamma^{0}\,i\,\left(\FFr{2}{\lambda_{W}}\right)^{1/2}\,\F1_{W},
\end{equation}
namely
\begin{equation}
\label{eq: R13.4}
\delta_{\xi}\,\left(H + H^{+}\right)=
\gamma^{0}\,i\,\left(\FFr{2}{\lambda_{W}}\right)^{1/2}\,\delta_{\xi}\,\F1_{W},
\end{equation}
or
\begin{equation}
\label{eq: R13.5}
\xi\,\widetilde{H}+\left(\xi\,\widetilde{H}\right)^{+}=
i\,\gamma^{0}\,\left(\FFr{2}{\lambda_{W}}\right)^{1/2}\,
\left(
-i\,\bar{\xi}\,\bar{\sigma}{}^{m}\,\pr_{W}{}_{m}\,\bar{\ch1_{W}}.
\right)
\end{equation}
The Fiertz identity 
$\bar{\xi}\,\bar{\sigma}{}^{m}\,\pr_{W}{}_{m}\,\bar{\ch1_{W}}=-
\pr_{W}{}_{m}\,\bar{\ch1_{W}}\,\sigma^{m}\,\bar{\xi}$ gives
$$
\widetilde{H}{}^{+}=
-\gamma_{0}\,\pr_{W}{}_{m}\,\bar{\ch1_{W}}\,\sigma^{m}=-
\left(
\bar{\sigma}{}^{m}\,\pr_{W}{}_{m}\,\ch1_{W}\,\gamma_{0}
\right)^{+}.
$$ 
Thus, explicitly the spinor-$\FFr{1}{2}$ Higgsino doublets read
\begin{equation}
\label{eq: R13.6}
\widetilde{H}_{u}=-\bar{\sigma}{}^{m}\,
\left(
\pr_{W}{}_{m}\,\ch1_{W}{}_{u}
\right)\gamma_{0}, \quad 
\widetilde{H}_{d}=-\bar{\sigma}{}^{m}\,
\left(
\pr_{W}{}_{m}\,\ch1_{W}{}_{d}
\right)\gamma_{0}.
\end{equation}

\renewcommand{\theequation}{\thesection.\arabic{equation}}
\section{The superfield content of MSMSM and the resulting SUSY Lagrangian}
\label{cont}
The results obtained in the previous sections enable us to trace 
unambiguously rather general scheme of MSMSM, which is essentially a 
straightforward and viable supersymmetrization of the MSM  where
we want to keep the number of superfields and interactions as small as 
possible. To build up the MSMSM the major point is to define its superfield 
content. Below we recall some important features allowing us to write the 
resulting Lagrangian of MSMSM based on eq.(11.35). 

$\bullet$  Within the MSM, during the realization of MW 
connections of weak interacting fermions the P-violation compulsory 
occurred in W-world (subsec.12.8) incorporated with the symmetry 
reduction eq.(12.8.3). It has characterized by the Weinberg mixing angle 
with the fixed value at $30^{0}$. This gives rise to 
the local symmetry $SU(2)\otimes U(1)$, under which the
left-handed fermions transformed as six independent doublets, 
while the right-handed fermions transformed as twelve 
independent singlets.

$\bullet$   Due to vacuum rearrangement in Q-world the Yukawa couplings 
arise between the fermion fields and corresponding isospinor-scalar 
$H$-mesons in conventional form.

$\bullet$   In the framework of suggested mechanism, providing the effective 
attraction between the relativistic fermions caused by the exchange of 
the mediating induced gauge quanta in W-world, the two 
complex self-interacting isospinor-scalar Higgs doublets 
$\left(H_{u},\,H_{d}\right)$ as well as their spin-$\FFr{1}{2}$ SUSY 
partners $\left(\widetilde{H}_{u},\,\widetilde{H}_{d}\right)$ Higgsinos  
arise as the Bose-condensate. Taking into account this slight difference from 
the MSM arisen in the field content of MSMSM in the Higgs sector, we must 
explicitly write in the supersymmetric Lagrangian eq.(11.35) also the piece 
containing these fields coupled to the gauge fields in a gauge invariant way, 
when the symmetry-breaking Higgs bosons are counted off from the gap 
symmetry-restoring value as the point of origin (subsec.12.13).

$\bullet$   The gauge group of MSMSM is the same 
$SU_{c}(3)\otimes SU(2)_{L}\otimes U(1)$ (sec.12.8) as in the MSM, which 
requires a colour octet of vector superfields $V^{a},$ a weak triplet 
$V^{(\delta)}$ and a hypercharge singlet $V$. Thus, the kinetic terms of all 
the fields in eq.(11.17) now fixed by gauge invariance
\begin{equation}
\label{eq: R14.1}
\begin{array}{l}
L =\IIn\,d^{4}\,\theta\,
\widetilde{\Phi}^{+}_{ch}\,\left(e^{g_{1}\, V\,T+
g_{2}\, V^{(\delta)}\,T^{(\delta)}+g_{3}\, V^{a}\,T^{a}}\right)
\,\widetilde{\Phi}_{ch}\\
+
\left[
\IIn\,d^{2}\,\theta\,
\FFr{1}{4}\,
\left(W\, W + W^{(\delta)}\, W^{(\delta)} + W^{a}\, W^{a}
\right)+\mbox{h.c.}
\right],
\end{array}
\end{equation}
where $\widetilde{\Phi}_{ch}$ is the matter superfields, 
$T,\,T^{(\delta)},\,T^{a}$ are the generators of appropriate representations 
of the gauge group. The superpotential determines the scalar potential
\begin{equation}
\label{eq: R14.2}
 V(A,\,A^{*})=\FFr{1}{2}\,g^{2}_{1}\,D^{2}+
\FFr{1}{2}\,g^{2}_{2}\,D^{\delta}{}^{2}+
\FFr{1}{2}\,g^{2}_{2}\,D^{a}{}^{2}+
\left|P\right|^{2},
\end{equation}
where the functions $D$ and $P$ are given in eq.(11.17).

$\bullet$ By the index $I=1,2,3$ in the MW-SUSY Lagrangian eq.(11.35) will be 
labeled the three families of chiral quarks 
$q^{I}_{L},\,u^{I}_{R},\,d^{I}_{R},\,$ and 
chiral leptons $l^{I}_{L},\,e^{I}_{R}$, where all of them are Weyl fermions. 
A SUSY requires the presence of supersymmetric partners which form 
supermultiplets with known particles, i.e., for every field of SM there is a 
superpartner with the exact same gauge quantum numbers. Then, the quarks 
and leptons are promoted to chiral multiplets by adding scalar (spin-$0$) 
squarks $\left(\widetilde{q}{}^{I}_{L},\,\widetilde{u}{}^{I}_{R},\,
\widetilde{d}{}^{I}_{R}\right)$ and sleptons  
$\left(\widetilde{l}{}^{I}_{L},\,\widetilde{e}{}^{I}_{R}\right)$ to the 
spectrum. The gauge bosons are promoted to vector supermultiplet by adding 
their SUSY partners gauginos (spin-$\FFr{1}{2}$) 
$\left( \widetilde{G},\, \widetilde{W}\, \widetilde{B}\right)$ to the 
spectrum. \\
A content of superfields of MSMSM presents in Table 1:
\vskip 0.5truecm
\begin{center}
\begin{tabular}{|c||c||c||c|c|} \hline
  &\mbox{supermultiplet}  & $F \quad\quad B$ 
& $SU_{c}(3)\quad SU(2)_{L}\quad U(1)_{Y}$ & $U(1)_{em}$\\ \hline
& & & &\\
\mbox{quarks}  
& $Q^{I}_{L}=\left(\begin{array}{l} U^{I}_{L}\\D^{I}_{L}\end{array}\right) $
& $q^{I}_{L}\quad\quad \widetilde{q}{}^{I}_{L}$ 
&$3\quad \quad \quad 2\quad\quad\quad \quad 1/6$
& $\left(\begin{array}{l}2/3\\-1/3\end{array}\right)$ \\ 
&&&&\\
& $U^{I}_{R}$ & $u^{I}_{R}\quad\quad \widetilde{u}{}^{I}_{R}$ 
&$\bar{3}\quad \quad \quad 1\quad\quad \quad -2/3$ 
&$-2/3$\\ 
& $D^{I}_{R}$ & $d^{I}_{R}\quad\quad\widetilde{d}{}^{I}_{R}$ 
&$\bar{3}\quad \quad \quad 1\quad\quad \quad \quad 1/3$ 
&$1/3$\\ 
&&&&\\ \hline
&&&&\\
\mbox{leptons}  
& $L^{I}_{L}=\left(\begin{array}{l} {\cal N}^{I}_{L}\\E^{I}_{L}\end{array}
\right) $
& $l^{I}_{L}\quad\quad \widetilde{l}{}^{I}_{L}$ 
&$\quad \,\, 1\quad \quad \quad 2\quad\quad\quad  -1/2$
& $\left(\begin{array}{l} 0\\-1 \end{array}\right)$ \\ 
&&&&\\
& $E^{I}_{R}$ & $e^{I}_{R}\quad\quad \widetilde{e}{}^{I}_{R}$ 
&$1 \quad \quad \quad 1\quad\quad \quad \quad 1$ 
&$1$\\ 
&&&&\\ \hline
&&&&\\
\mbox{Higgs}  
& $\H1_{d}=\left(\begin{array}{l} H^{0}_{d}\\H^{-}_{d}\end{array}
\right) $
& $\left(\begin{array}{l} \widetilde{h}^{0}\\\widetilde{h}^{-}\end{array}
\right)\quad\quad 
\left(\begin{array}{l} h^{0}_{d}\\ h^{-}_{d}\end{array}\right)$
&$\quad 1\quad \quad \quad 2\quad\quad\quad  -1/2$
& $\left(\begin{array}{l} 0\\-1 \end{array}\right)$ \\ 
&&&&\\
& $\H1_{u}=\left(\begin{array}{l} H^{+}_{u}\\H^{0}_{u}\end{array}\right)$
& $\left(\begin{array}{l} \widetilde{h}^{+}\\\widetilde{h}^{0}\end{array}
\right)\quad\quad 
\left(\begin{array}{l} h^{+}_{u}\\ h^{0}_{u}\end{array}\right)$
&$1\quad \quad \quad 2\quad\quad\quad \quad 1/2$
& $\left(\begin{array}{l} 1\\ 0 \end{array}\right)$ \\ 
&&&&\\
&&&&\\ \hline
&&&&\\
\mbox{gauge} 
&G &$\widetilde{G}\quad \quad G$ 
&$8\quad \quad \quad 1\quad\quad\quad \quad 0$
&$0 $\\
\mbox{bosons} 
&W &$\widetilde{W}\quad \quad W$ 
&$1\quad \quad \quad 3\quad\quad\quad \quad 0$
&$(0,\,\pm 1)$ \\
&B &$\widetilde{B}\quad \quad B$ 
&$1\quad \quad \quad 1\quad\quad\quad \quad 0$
&$0$ \\
&&&&\\
&&&&\\ \hline
\end{tabular} 
\end{center}
\vskip 0.5truecm
Table 1. Field content of MSMSM. The column below F(B) denotes its fermionic 
(bosonic) content.\\ \\
Once the field content is fixed, putting it all together the most generic 
renormalizable MW-SUSY Lagrangian of MSMSM defined on the SMM:\, $SG_{N}$ 
eq.(9.1) ensues from the eq.(11.35), which is now invariant under local 
gauge symmetry $SU_{c}(3)\otimes SU(2)_{L}\otimes U(1)$, where a set of 
gauge fields are coupled to various superfields among which is also Higgs 
supermultiplets. Furthermore, we especially separated from the rest the piece 
containing only the $\eta$-components of the particles defined on the 
supermanifold $\SG1_{\eta}$, which, according to sec.9, is important for the
further discussion of a realistic realization of the MSMSM (next sec.). To 
facilitate writing we shall forbear here to write out the piece of 
Lagrangian containing only the terms of sparticles, as it is a somewhat 
lengthy and so standard. But, in the mean time, we shall retain the explicit 
terms of Higgs bosons arisen in the internal $W-$world to emphasize the 
specific mechanism of the electroweak symmetry breaking discussed in the 
subsec.12.14. The resulting Lagrangian reads
\begin{equation}
\label{eq: R14.3}
\begin{array}{l}
L_{SG_{N}}=-\FFr{1}{4}\S^{3}_{(a)=1}
\left(
\left(\F1_{\eta}{}^{b}_{mn}\,\F1_{\eta}{}^{mn\,b} \right)_{(a)}
\right)-\D1_{W}{}_{m}\,H_{u}\,\D1_{W}{}^{m}\,H_{u}{}^{*}-
\D1_{W}{}_{m}\,H_{d}\,\D1_{W}{}^{m}\,H_{d}{}^{*}\\
+\S^{3}_{I=1}\left(
-i\,\bar{q}{}^{I}_{L}\,\sig1_{\eta}{}^{m}\D1_{\eta}{}_{m}\,q^{I}_{L}
-i\,\bar{u}{}^{I}_{R}\,\sig1_{\eta}{}^{m}\D1_{\eta}{}_{m}\,u^{I}_{R}
-i\,\bar{d}{}^{I}_{R}\,\sig1_{\eta}{}^{m}\D1_{\eta}{}_{m}\,d^{I}_{R}
-i\,\bar{l}{}^{I}_{L}\,\sig1_{\eta}{}^{m}\D1_{\eta}{}_{m}\,l^{I}_{L}
\right.\\
\left.
-i\,\bar{e}{}^{I}_{R}\,\sig1_{\eta}{}^{m}\D1_{\eta}{}_{m}\,e^{I}_{R}
\right)
-\S^{3}_{I,\,J=1}\left(
\left(Y_{u}\right)_{IJ}\left(H_{u}{}^{*}+H_{d}{}^{*}\right)
\,q^{I}_{L}\,u^{J}_{R}
+\left(Y_{d}\right)_{IJ}\left(H_{u}+H_{d}\right)
\,q^{I}_{L}\,d^{J}_{R}\right.\\\left.
+\left(Y_{l}\right)_{IJ}\left(H_{u}+H_{d}\right)
\,l^{I}_{L}\,e^{J}_{R}+\mbox{h.c.}
\right)+\S^{3}_{(a)=1}i\,g\,\sqrt{2}
\left(
H_{u}{}^{*}\,T^{b}\,\widetilde{H_{u}}\,\lambda^{b}-
\bar{\lambda}{}^{b}\,\widetilde{H_{u}{}^{*}}\,T^{b}\,H_{u}
\right. \\ 
+H_{d}{}^{*}\,T^{b}\,\widetilde{H_{d}}\,\lambda^{b}-
\bar{\lambda}{}^{b}\,\widetilde{H_{d}{}^{*}}\,T^{b}\,H_{d}+
H_{u}{}^{*}\,T^{b}\,\widetilde{H_{d}}\,\lambda^{b}+
H_{d}{}^{*}\,T^{b}\,\widetilde{H_{u}}\,\lambda^{b}
-\bar{\lambda}{}^{b}\,\widetilde{H_{u}{}^{*}}\,T^{b}\,H_{d}
\\ \left.
-\bar{\lambda}{}^{b}\,\widetilde{H_{d}{}^{*}}\,T^{b}\,H_{u}
\right)_{(a)}+\S^{3}_{(a)=1}i\,g\,\sqrt{2}
\left( 
A^{*}_{J}\,T^{b\,J}_{I}\,\chi^{I}\,\lambda^{b}-
\bar{\lambda}{}^{b}\,\bar{\chi}{}^{I}\,T^{b\,I}_{J}\,A^{J}
\right)- V_{u}\left(H_{u},\,H_{u}{}^{*}\right)\\
-  V_{d}\left(H_{d},\,H_{d}{}^{*}\right)-
\FFr{1}{8}\,\left( g^{2}+g'{}^{2}\right)\,  
\left(\left|H_{u}\right|{}^{2}-\left|H_{d}\right|{}^{2}\right)^{2}+
\FFr{1}{2}\,g^{2}\,\left|H_{u}\,H_{d}{}^{*}\right|^{2}\\
+
\mbox{all the terms containing only the sparticles}.
\end{array}
\end{equation}
Here $\chi^{I}$ runs over all the particles, while $A^{J}$ runs over all 
the sparticles, the index $(a)$ labels the $3$ different features in the 
gauge group, $ V_{d}\left(H,\,H^{*}\right)$ is the scalar potential for 
each Higgs doublet
\begin{equation}
\label{eq: R14.4}
V_{u}\left(H_{u},\,H_{u}{}^{*}\right)=
-\FFr{1}{2}\,m^{2}_{u}\,\left|H_{u}\right|^{2}+
\FFr{1}{4}\,\lambda^{2}_{u}\,\left|H_{u}\right|^{4}
\quad
V_{d}\left(H_{d},\,H_{d}{}^{*}\right)=
-\FFr{1}{2}\,m^{2}_{d}\,\left|H_{d}\right|^{2}+
\FFr{1}{4}\,\lambda_{d}^{2}\,\left| H_{d}\right|^{4}.
\end{equation}
A contribution of the ``$D$'' term to the Higgs potential has also taken 
into account in last term in eq.(14.3)
\begin{equation}
\label{eq: R14.5}
 V_{D}=\FFr{1}{2}\,D^{(a)}\,D^{(a)},\quad 
D^{(a)}=-g\,A^{I}{}^{*}\,T^{a}_{IJ}\,A^{J},
\end{equation}
or
\begin{equation}
\label{eq: R14.6}
V_{D}=\FFr{g^{2}+g'{}^{2}}{8}\,
\left(\left|H_{u}\right|{}^{2}-\left|H_{d}\right|{}^{2}\right)^{2}+
\FFr{1}{2}\,g^{2}\,\left|H_{u}\,H_{d}{}^{*}\right|^{2}.
\end{equation}
The number of major free parameters in the Lagrangian eq.(14.3) are the 
primary coupling constants $\lambda_{Q}$ and $\lambda_{W}$ of nonlinear 
fermion interaction of the internal MW-components $i=Q,W$ and  gauge 
couplings $g_{1}\,g_{2}\,g_{3}$. The SM relation 
$Q_{e}=g_{1}\,\cos \,\theta_{W}$ holds, where 
$\theta_{W}$ is the weak mixing angle 
$\cos^{2} \,\theta_{W}=g^{2}\left/(g^{2}+g'{}^{2})\right.$ (sec.12.8). The 
Yukawa couplings $(Y_{l}\, Y'_{l})$ are given in eq.(12.9.2)$:\,\, 
Y=f_{Q}=Z_{Q}$.

\section{Realistic realization of the MW-SUSY: M\rlap/SMSM}
\label{real}
The MW-SUSY cannot be an exact symmetry of nature and has to be realized 
in its broken phase (sec.9). The major point of our strategy is a 
realistic realization of the supersymmetric extension of the MSM. Thus, the 
test of the theory will depend on its ability to account for the breaking of 
the MW-SUSY as well. Here, suggested approach creates a particular incentive 
for its study. In previous sections we have made a headway of 
reasonable framework of exact MW-SUSY defined on the exact MW-supermanifold 
$SG_{N}$. Therefore, one will be able to verify its virtues manifested, 
first of all, in the power of boson-fermion cancellations. One of the two 
principal offshoots of the supersymmetrization of the MSM is the solution of 
the zero point energy problem. Also, in its unbroken form it solves the 
technical aspects of the naturalness and hierarchy problem (sec.1), 
when in non-SUSY theories scalar fields receive large mass corrections even 
if the bare mass is set to zero, and small masses are ``unnatural'' 
[124-126]. This applied to the Higgs bosons of the SM (as well as MSM) yields 
a difficulty in understanding of the smallness of $M_{Z}$ and how it can be 
kept stable against quantum corrections in some extensions of the SM 
containing apart from the weak scale $M_{Z}$ also a second larger scale 
$M_{GUT}>>M_{Z}$ [126,127], which holds in Grand Unified theories. 
The cancellation of quadratic divergences in SUSY theories is a consequence 
of general non-renormalization theorem [128,129] or the ``taming'' of the 
quantum corrections, which stabilizes the Higgs mass 
and thus weak scale $M_{Z}$ without fine-tuning. It is remarkable that these 
attractive features of the unbroken MSMSM can be maintained as well in the 
broken M\rlap/SMSM. Achieving it one should perform an inverse passage 
($\SG1_{\eta}\rightarrow \G1_{\eta}$) to the \rlap/SMM:\, $\rlap/SG_{N}$ 
eq.(8.5). It is due to the fact that the most powerful boson-fermion 
cancellation can be regarded as a direct consequence of a constraint 
stemming from holomorphy, therefore, it should be held even in  the
$M\rlap/SMSM$. 
Then the Lagrangian $L_{\rlap/SG_{N}}$ of the  M\rlap/SMSM ensues from the 
Lagrangian  $L_{SG_{N}}$ eq.(14.3) of the MSMSM (eq.(9.11)-eq.(9.13))
\begin{equation}
\label{eq: R15.1}
 L_{\rlap/SG_{N}}=L_{SG_{N}}+L_{soft},
\end{equation}
where, according to the sec.9, one has
\begin{equation}
\label{eq: R15.2}
\begin{array}{l}
L_{soft}=\left(
-m^{2}_{IJ}\,A^{I}\,A^{J}-\FFr{1}{2}\,\widetilde{m}_{ab}\,\lambda^{a}
\,\lambda^{a}-\FFr{1}{2}\,m_{u}
\,\widetilde{\H1_{u}}\,\widetilde{\H1_{u}}-
\FFr{1}{2}\, m_{d}\,\widetilde{\H1_{d}}\,\widetilde{\H1_{d}}
+ \mbox{h.c.}\right)\\
+ b\,\epsilon_{ij}\,\left(\H1_{u}{}^{i}\H1_{d}{}^{j}+
\mbox{h.c.}\right).
\end{array}
\end{equation}
Here $m^{2}_{IJ}$ is the mass matrix for all the scalars of the chiral 
multiplets, $m\equiv (\widetilde{m}_{ab},\,m_{u},\,m_{d})$ is the mass 
matrix respectively for the gauginos of each factor of the gauge group, and 
Higgsinos. The  last term of the interaction is 
induced for the following reason: according to eq.(14.4) and subsec.12.16, 
these doublets above in free states imply 
\begin{equation}
\label{eq: R15.3}
\Delta\,\hat{m}{}^{2}_{u}=-m^{2}_{u}+\lambda^{2}_{u}\,v^{2}_{u},=0,\quad
\Delta\,\hat{m}{}^{2}_{d}=-m^{2}_{d}+\lambda^{2}_{d}\,v^{2}_{d}=0,
\end{equation}
where $m^{2},\,\lambda^{2},\,v^{2}$ are respectively the mass, the coupling 
constant and VEV of given doublet. In the case at hand, certainly, there is 
an interaction between the bosons $H_{u}$ and $H_{d}$ described by the 
last term in eq.(15.2), when the strength of interaction $b$ will be fixed 
through the minimization conditions of the total Higgs potential. This 
can be used to derive a more physical relationship among the physical 
parameters. As it will be seen in sec.16 the case 
\begin{equation}
\label{eq: R15.4}
\Delta\,\hat{m}{}^{2}_{u}=-\Delta\,\hat{m}{}^{2}_{d}\neq 0
\end{equation}
corresponds to the situation when the axion $A^{0}\,(m^{2}_{A^{0}}=0)$ has 
arisen after the breaking of electroweak gauge symmetry. But the other case
of
\begin{equation}
\label{eq: R15.5}
\left(\hat{m}{}^{2}_{u}>0,\quad \hat{m}{}^{2}_{d}>0\right)\quad
\mbox{or}\quad
\left(\hat{m}{}^{2}_{u}<0,\quad \hat{m}{}^{2}_{d}<0\right),
\end{equation}
implies an existence  of the neutral physical particle of the mass 
\begin{equation}
\label{eq: R15.6}
m^{2}_{A^{0}}=\hat{m}{}^{2}_{u}+\hat{m}{}^{2}_{d}\neq 0.
\end{equation}
Note that such Higgs doublets arisen on equal footing have counted off 
from the same point of origin for the same vacuum eq.(12.13.5), then we will 
be interested physically in the most important simplest case when the 
electroweak symmetry breaking is parametrized just only by the single Higgs 
VEV
\begin{equation}
\label{eq: R15.7}
v_{u}=v_{d}\,, \quad \hat{m}{}^{2}_{u}=\hat{m}{}^{2}_{d}>0.
\end{equation}
Of course, we shall carry out a computation in the generic case of 
eq.(15.6), but in the aftermath we shall turn to the case of eq.(15.7). The 
non-supersymmetric breaking terms do not spoil a condition of cancellation 
of quadratic divergences, i.e., a mass-squared sum rule [94]  
\begin{equation}
\label{eq: R15.8}
Str\,M^{2}\equiv \S^{1}_{J=0}(-1)^{2J}(2J+1)Tr\,M^{2}_{J}= const.
\end{equation}
where $\vec{J}$ is the spin of the particles.
It holds independently of the values of the scalar fields. Eventually 
the mass terms for the scalars contribute a constant, field independent 
piece in eq.(15.8), while a generic mass matrix for the fermions reads 
\begin{equation}
\label{eq: R15.9}
M_{1/2}= M_{1/2}^{S}+\delta\,M_{1/2},
\end{equation}
where $ M_{1/2}^{S}$ is the supersymmetric part of $M_{1/2}$, when 
$\delta\,M_{1/2}$ is given 
\begin{equation}
\label{eq: R15.10}
\delta\,M_{1/2}=\left(\begin{array}{l}\delta\,P_{IJ}\quad 
\delta \, D^{b}_{I}\\\delta \, D^{a}_{J}\quad \delta \, \widetilde{m}
\end{array}\right).
\end{equation}
A computation for the considered fields gives $\delta \,P=0=\delta \,D,$ while 
$\delta \, \widetilde{m}$ can be arbitrary.

\section{The generating functional of the viable microscopic theory of the 
standard model (VMSM)}
\label{elb}
In previous sections we have systematically build up the  M\rlap/SMSM. In 
what follows we shall be motivated by the purpose to derive a final 
generating functional of the VMSM defined on the four-dimensional 
Minkowski space $M_{4}$, which is free of all the problems of MSM. 
The relevant steps are as follows: we start with 
generic functional $Z_{real}\left[{\cal J}\right]$ eq.(9.11) exploring the 
Lagrangian eq.(15.1), then we extract the pertinent piece containing 
only the $\eta$-field components defined on $\G1_{\eta}:\quad 
\left(Ext_{\G1_{\eta}}\right)$. In the aftermath, by passing to 
$M_{4},\quad \left(\G1_{\eta}\rightarrow M_{4} \right)$ (sec.2.1) we get 
\begin{equation}
\label{eq: R16.1}
\begin{array}{l}
Z_{VMSM}\left[{\cal J}\right]\equiv Z_{M_{4}}\left[{\cal J}\right]=
Z_{\G1_{\eta}}\left[{\cal J}\right]_{\left(\G1_{\eta}\rightarrow 
M_{4} \right)} 
\equiv Ext_{\G1_{\eta}}\,
\left(Z_{real}\left[{\cal J}\right]\right)_{\left(\G1_{\eta}
\rightarrow M_{4} \right)}\\
=\left( 
\IIn\, {\cal D}\,\left[ \varphi\right]\,\exp\,
\left\{
i\,\IIn\,d^{4}\,x\, \left( L_{VMSM}+{\cal J} \,\varphi \right) 
\right\}
\right)_{ \left( R\,;\,\A1_{x}{}_{sp}\rightarrow 1 \right) },
\end{array}
\end{equation}
where, as before, with retained terms of Higgs bosons arisen in the 
$W$-world (subsec.12.14), put forth only in illustration of a point at issue, 
the Lagrangian of the VMSM can be written down 
\begin{equation}
\label{eq: R16.2}
\begin{array}{l}
L_{VMSM}=-\FFr{1}{4}\S^{3}_{(a)=1}
\left(\left(\F1_{x}{}^{b}_{mn}\,\F1_{x}{}^{mn\,b} \right)_{(a)}\right)-
\D1_{W}{}_{m}\,H_{u}\,\D1_{W}{}^{m}\,H_{u}{}^{*}-
\D1_{W}{}_{m}\,H_{d}\,\D1_{W}{}^{m}\,H_{d}{}^{*}\\
+\S^{3}_{I=1}\left(-i\,\bar{q}{}^{I}_{L}\,\rlap/D \,q^{I}_{L}
-i\,\bar{u}{}^{I}_{R}\,\rlap/D\,u^{I}_{R}
-i\,\bar{d}{}^{I}_{R}\,\rlap/D\,d^{I}_{R}
-i\,\bar{l}{}^{I}_{L}\,\rlap/D\,l^{I}_{L}
\right.\\\left.
-i\,\bar{e}{}^{I}_{R}\,\rlap/D\,r^{I}_{R}
\right)
-\S^{3}_{I,\,J=1}\left(
\left(Y_{u}\right)_{IJ}\left(H_{u}{}^{*}+H_{d}{}^{*}\right)
\,q^{I}_{L}\,u^{J}_{R}
+\left(Y_{d}\right)_{IJ}\left(H_{u}+H_{d}\right)
\,q^{I}_{L}\,d^{J}_{R}\right.\\
\left.
+\left(Y_{l}\right)_{IJ}\left(H_{u}+H_{d}\right)
\,l^{I}_{L}\,e^{J}_{R}+\mbox{h.c.}\right)+
\left[\S^{3}_{(a)=1}i\,g\,\sqrt{2}
\left(
H_{u}{}^{*}\,T^{b}\,\widetilde{H_{u}}\,\lambda^{b}-
\bar{\lambda}{}^{b}\,\widetilde{H_{u}{}^{*}}\,T^{b}\,H_{u}
\right. \right.\\ 
+H_{d}{}^{*}\,T^{b}\,\widetilde{H_{d}}\,\lambda^{b}-
\bar{\lambda}{}^{b}\,\widetilde{H_{d}{}^{*}}\,T^{b}\,H_{d}+
H_{u}{}^{*}\,T^{b}\,\widetilde{H_{d}}\,\lambda^{b}+
H_{d}{}^{*}\,T^{b}\,\widetilde{H_{u}}\,\lambda^{b}
-\bar{\lambda}{}^{b}\,\widetilde{H_{u}{}^{*}}\,T^{b}\,H_{d}
\\ \left.\left.
-\bar{\lambda}{}^{b}\,\widetilde{H_{d}{}^{*}}\,T^{b}\,H_{u}+
A^{*}_{J}\,T^{b\,J}_{I}\,\chi^{I}\,\lambda^{b}-
\bar{\lambda}{}^{b}\,\bar{\chi}{}^{I}\,T^{b\,I}_{J}\,A^{J}\right)_{(a)}
\right]_{loop}- V_{u}\left(H,\,H^{*}\right),
\end{array}
\end{equation}
where $\rlap/D=\sigma^{m}\,\D1_{x}{}_{m}$, and
the scalar potential reads
\begin{equation}
\label{eq: R16.3}
\begin{array}{l}
V(H,\,H^{*})=V_{u}\left(H_{u},\,H_{u}{}^{*}\right)+
V_{d}\left(H_{d},\,H_{d}{}^{*}\right)+
b\,\left(\epsilon_{ij}\,H_{u}{}^{i}\,H_{d}{}^{j}+\mbox{h.c.}\right)\\
+
\FFr{1}{8}(g^{2}+g'{}^{2})\left(\left|H_{u}\right|^{2}-
\left|H_{d}\right|^{2}\right)^{2}+ \FFr{g^{2}}{2}
\left|H_{u}\,H_{d}{}^{*}\right|^{2}.
\end{array}
\end{equation}
The magnitudes of the quartic potential terms are not arbitrary, i.e., 
they are constrained by the supersymmetry to be of magnitude $g^{2}$ and 
$g'{}^{2}$. Since at the limit $\A1_{x}{}_{sp}\rightarrow 1$ in eq.(16.1)
all the sparticles become independent of the $x$-field component as well of 
$x$-coordinate $(x\in M_{4})$, therefore the sparticles $A^{\pm}$ 
involved in the brackets may finally contribute only in the radiative loop 
corrections to the fermion and boson mass of all the particles maintaining 
the boson-fermion cancellations. Here, as before, $\chi^{\pm}$ runs over all 
the particles, while $A^{\pm}$ runs over all the sparticles. In the VMSM 
electroweak symmetry breaking slightly complicated due to the two complex 
Higgs doublets instead of just one in the MSM. 
This completes the definition of the VMSM. We are now ready to investigate 
some of its properties in more details.
It is reasonable to 
start discussion of the phenomenology of VMSM with a treatment of its Higgs 
sector. As before, the $SU(2)_{L}\otimes U(1)_{Y}$ should be broken 
spontaneously down to electromagnetism $U(1)_{em}$ simultaneously in both the 
$W$- world and $M_{4}$ (subsec.12.14), for which the scalar potential 
eq.(16.3) should have its absolute minimum away from the origin. 
Using the $SU(2)$ gauge transformations we can ignore the charged components 
without loss of generality when minimizing the potential and establish the 
conditions necessary for $h^{0}_{u}$ and $h^{0}_{d}$ to get nonzero VEVs. 
Furthermore, they can be chosen to be real: $<h^{0}_{u}>= v_{u}$, and  
$<h^{0}_{d}>= v_{d}\,\,$ while $<h^{+}_{u}>=<h^{+}_{d}>=0$.
In standard manners these VEVs can be connected to the known mass of the 
$Z^{0}$ boson and the electroweak gauge couplings:
\begin{equation}
\label{eq: R16.4}
M^{2}_{Z}=\FFr{1}{2}(g^{2}+g'{}^{2})v^{2}\equiv
\FFr{1}{2}\,\widetilde{g}{}^{2}\,v^{2},
\end{equation}
where $v_{u}=v\,\sin \,\beta,\quad v_{}=v\,\cos \,\beta $.
The VEVs are directly related to the parameters of the Higgs potential 
\begin{equation}
\label{eq: R16.5}
\begin{array}{l}
V=-\FFr{1}{2}\,m^{2}_{u}
\left(\left|h_{u}^{0}\right|^{2}+\left|h_{u}^{+}\right|^{2}\right)
+
\FFr{1}{4}\, \lambda^{2}_{u}
\left(\left|h_{u}^{0}\right|^{2}+\left|h_{u}^{+}\right|^{2}\right)
^{2} -\FFr{1}{2}\,m^{2}_{d}
\left(\left|h_{d}^{0}\right|^{2}+\left|h_{d}^{+}\right|^{2}\right)\\
+
\FFr{1}{4}\, \lambda^{2}_{d}
\left(\left|h_{d}^{0}\right|^{2}+\left|h_{d}^{+}\right|^{2}\right)
^{2} 
+ \left( b \, \left( h_{u}^{+} \, h^{-}_{d}- 
h^{0}_{u}\,h^{0}_{d}\right)+\mbox{h.c.}\right)+
\FFr{1}{4}\, \FFr{M^{2}_{Z}}{v^{2}}
\left(\left|h_{u}^{0}\right|^{2}+\left|h_{u}^{+}\right|^{2}\right.\\\left.
-
\left|h_{d}^{0}\right|^{2}-\left|h_{d}^{-}\right|^{2}\right)
^{2} +
\FFr{M^{2}_{W}}{v^{2}}\, 
\left|h_{u}^{+}\,h_{d}^{0}{}^{*}+h_{u}^{0}\,h_{d}^{-}{}^{*}\right|^{2}
\end{array}
\end{equation}
via the minimization condition 
\begin{equation}
\label{eq: R16.6}
\begin{array}{l}
\FFr{\partial\,V}{\partial\,h_{u}^{0}}=
\left(-m^{2}_{u} + \lambda^{2}_{u}\, v^{2}_{u} - 
M^{2}_{Z} \,\cos\,2\beta\right)\, v_{u}-2\,b\, v_{d}=0,\\\\
\FFr{\partial\,V}{\partial\,h_{d}^{0}}=
\left(-m^{2}_{d} + \lambda^{2}_{d}\, v^{2}_{d} + 
M^{2}_{Z} \,\cos\,2\beta\right)\, v_{d}-2\,b\, v_{u}=0,
\end{array}
\end{equation}
or
\begin{equation}
\label{eq: R16.7}
\hat{m}{}^{2}_{u}=b\,\cot\,\beta +\FFr{1}{2}\,M^{2}_{Z} \,\cos\,2\beta,\quad
\hat{m}{}^{2}_{d}=b\,\tan\,\beta -\FFr{1}{2}\,M^{2}_{Z} \,\cos\,2\beta,
\end{equation}
provided by (eq.(12.12.7)) 
\begin{equation}
\label{eq: R16.8}
\begin{array}{l}
m^{2}_{u}=\FFr{24}{7\,\zeta (3)}\left(\FFr{\hbar}{\xi_{0}}\right)^{2}\,
\left(\FFr{\mu_{u}}{m}\right)^{2}\,
\left[ 1-\FFr{T}{T_{c\,\mu}}-\left(\FFr{\mu_{u}}{m}-1\right)\, 
\ln \,\FFr{2\,\widetilde{\omega}}{\Delta_{0}}\right],\\
m^{2}_{d}=\FFr{24}{7\,\zeta (3)}\left(\FFr{\hbar}{\xi_{0}}\right)^{2}\,
\left(\FFr{\mu_{d}}{m}\right)^{2}\,
\left[ 1-\FFr{T}{T_{c\,\mu}}-\left(\FFr{\mu_{d}}{m}-1\right)\, 
\ln \,\FFr{2\,\widetilde{\omega}}{\Delta_{0}}\right],\\
\lambda^{2}_{u}=\FFr{96}{7\,\zeta (3)}\left(\FFr{\hbar}{\xi_{0}}\right)^{2}\,
\FFr{1}{N_{u}},\quad 
\lambda^{2}_{d}=\FFr{96}{7\,\zeta (3)}\left(\FFr{\hbar}{\xi_{0}}\right)^{2}\,
\FFr{1}{N_{d}}.
\end{array}
\end{equation}
Here the Higgs mechanism works in the following way:
Before the symmetry was broken in the $W$-world, the $2$ complex 
$SU(2)_{L}$ Higgs doublets had $8$ degrees of freedom. 
Three of them were the would-be Nambu-Goldstone 
bosons $G^{0},\,G^{\pm},$ which were absorbed to give rise the longitudinal 
modes of the massive $W$-components of the $Z^{0}$ and $W^{\pm}$ vector 
bosons, which simultaneously give rise the corresponding $x$- components too, 
leaving $5$ physical degrees of freedom. The latter consists of a charged 
Higgs boson pairs $H^{\pm},$ a CP-odd neutral Higgs boson $A^{0}$, and 
CP-even neutral Higgs bosons $h^{0}$ and $H^{0}$. 
The mass eigenstates and would-be 
Nambu-Goldstone bosons are made of the original gauge-eigenstate fields, 
where the physical pseudoscalar Higgs boson $A^{0}$ is made of from the 
imaginary parts of $h^{0}_{u}$ and $h^{0}_{d}$, and is orthogonal to 
$G^{0}$; while the neutral scalar Higgs bosons are mixtures of the real 
parts of $h^{0}_{u}$ and $h^{0}_{d}$. The CP conservation in the Higgs 
sector is automatic [19,23]. The mass of any physical Higgs boson that is 
SM-like is strictly limited, as are the radiative corrections to the 
quartic potential terms. The tree-level masses for these Higgs states 
are calculated from the mass matrices of second derivatives of the Higgs 
potential:
\begin{equation}
\label{eq: R16.9}
\begin{array}{l}
M^{2}_{\cal J}=\FFr{1}{2}\FFr{\partial^{2}\,V}
{\partial\left({\cal J}m \left[h^{0}_{i}\right]\right)\,
\partial\left({\cal J}m\left[h^{0}_{j}\right]\right)}=
\FFr{1}{2}\,m^{2}_{A}\,\sin\,2\beta \left(\begin{array}{ll}
\tan\, \beta \quad 1 \\ 1 \quad \cot\, \beta\end{array}\right),\\\\
M^{2}_{\cal R}=\FFr{1}{2}\FFr{\partial^{2}\,V}
{\partial\left({\cal R}e\left[h^{0}_{i}\right]\right)\,
\partial\left({\cal R}e\left[h^{0}_{j}\right]\right)}=
\FFr{1}{2}\,m^{2}_{A}\,\sin\,2\beta \left(\begin{array}{ll}
\tan\, \beta +\tau_{d}\,\cot\,\beta\quad -1 \\ 
-1 \quad \cot\, \beta + \tau_{u}\,\tan\, \beta\end{array}\right),\\\\
+ \FFr{1}{2}\,M^{2}_{Z}\,\sin\,2\beta \left(\begin{array}{ll}
\cot\, \beta \quad -1 \\ -1 \quad \tan\, \beta\end{array}\right),\\\\
M^{2}_{\pm}=\FFr{1}{2}\FFr{\partial^{2}\,V}
{\partial\,h^{-}_{i}\,\partial\,h^{+}_{j}}=
\FFr{1}{2}\,M^{2}_{H^{\pm}}\,\sin\,2\beta \left(\begin{array}{ll}
\tan\, \beta \quad 1 \\ 1 \quad \cot\, \beta\end{array}\right),
\end{array}
\end{equation}
where
\begin{equation}
\label{eq: R16.10}
\begin{array}{l}
m^{2}_{A}=\hat{m}{}^{2}_{u}+\hat{m}{}^{2}_{d}=\FFr{2\,b}{\sin\,2\beta },
\quad \tau_{d}=\FFr{\lambda^{2}_{d}\, v^{2}}{m^{2}_{A}},
\quad \tau_{u}=\FFr{\lambda^{2}_{u}\, v^{2}}{m^{2}_{A}},\\
M^{2}_{H^{\pm}}=M^{2}_{W}+m^{2}_{A},\quad i,j=u,d.
\end{array}
\end{equation}
Once one has adopted the standard parameters then all the physical 
observables can be expressed in terms of them.  
The eigenvalues of $M^{2}_{\cal J}$ are $m^{2}_{G^{0}}=0$ and $m^{2}_{A}$. 
The $m^{2}_{A}$ and $\tan\, \beta$ can be chosen as the theory parameters. 
The eigenvalues of the $M^{2}_{\cal R}$ are written
\begin{equation}
\label{eq: R16.11}
\begin{array}{l}
m^{2}_{H^{0}}=\FFr{1}{2}\,\left[m^{2}_{A_{1}}+M^{2}_{Z}+
\sqrt{\left(m^{2}_{A_{1}}+M^{2}_{Z}\right)^{2}-4\,m^{2}_{A}\,M^{2}_{Z}
\left(\cos^{2}\,2\,\beta+
\theta_{\tau}\,\FFr{m^{2}_{A}}{M^{2}_{Z}}\right)}\right],\\\\
m^{2}_{h^{0}}=\FFr{1}{2}\,\left[m^{2}_{A_{1}}+M^{2}_{Z}-
\sqrt{\left(m^{2}_{A_{1}}+M^{2}_{Z}\right)^{2}-4\,m^{2}_{A}\,M^{2}_{Z}
\left(\cos^{2}\,2\,\beta+
\theta_{\tau}\,\FFr{m^{2}_{A}}{M^{2}_{Z}}\right)}\right],
\end{array}
\end{equation}
where
\begin{equation}
\label{eq: R16.12}
\begin{array}{l}
m^{2}_{A_{1}}=m^{2}_{A}\left(1+\tau_{d}\,\cos^{2}\,\beta+
\tau_{u}\,\sin^{2}\,\beta\right),\\
\theta_{\tau}=\tau_{d}\,\cos^{4}\,\beta+
\tau_{u}\,\sin^{4}\,\beta+\FFr{1}{4}\,\sin^{2}\,2\,\beta\,
\left[\tau_{d}\, \tau_{u}+\left(\tau_{d}+\tau_{u}\right)\,
\FFr{M^{2}_{Z}}{m^{2}_{A}}\right].
\end{array}
\end{equation}
It is then easy by ordinary manipulations to investigate the region of 
definition of the $m^{2}_{h^{0}}$ in several cases. But, as it is mentioned 
in previous section we are interested in the physically most important case 
when the electroweak symmetry breaking is parametrized just only by the 
single Higgs VEV $v_{u}=v_{d}=\FFr{v}{\sqrt{2}}$ and $\hat{m}^{2}_{u}=
\hat{m}^{2}_{d}>0$ (eq.(15.7)), namely
\begin{equation}
\label{eq: R16.13}
\begin{array}{l}
b=\hat{m}{}^{2}_{u}=\hat{m}{}^{2}_{d}>0,\quad m^{2}_{A}=2\,b>0,\\
\lambda^{2}_{u}\,v^{2}_{u}=\FFr{m^{2}_{A}}{2}+m^{2}_{u},\quad
\lambda^{2}_{d}\,v^{2}_{d}=\FFr{m^{2}_{A}}{2}+m^{2}_{d}.
\end{array}
\end{equation} 
Hence, the $M^{2}_{\cal R}$ is written down
\begin{equation}
\label{eq: R16.14}
M^{2}_{\cal R}=
\left(\begin{array}{l}
M^{2}+m^{2}_{d}\quad -M^{2}\\
-M^{2}\quad M^{2}+m^{2}_{u}
\end{array}\right),
\end{equation}
where $M^{2}=\FFr{1}{2}\,\left(M^{2}_{Z}+m^{2}_{A}\right)$. The eigenvalues 
of the $M^{2}_{\cal R}$ then will be 
\begin{equation}
\label{eq: R16.15}
\begin{array}{l}
m^{2}_{H^{0}}=\FFr{1}{2}\,\left[m^{2}_{u}+m^{2}_{d}+2\,M^{2}+
\sqrt{\left(m^{2}_{u}-m^{2}_{d}\right)^{2}+4\,M^{4}}\right],\\\\
m^{2}_{h^{0}}=\FFr{1}{2}\,\left[m^{2}_{u}+m^{2}_{d}+2\,M^{2}-
\sqrt{\left(m^{2}_{u}-m^{2}_{d}\right)^{2}+4\,M^{4}}\right].
\end{array}
\end{equation}
In the limit $\left| m^{2}_{u}-m^{2}_{d}\right| \ll 2\,M^{2}$ it gives
\begin{equation}
\label{eq: R16.16}
m^{2}_{h^{0}}\simeq \FFr{1}{2}\,\left(m^{2}_{u}+m^{2}_{d}\right)^{2}, \quad
m^{2}_{H^{0}}\simeq m^{2}_{h^{0}}+M^{2}_{Z}, 
\end{equation}
while at $\left|m^{2}_{u}-m^{2}_{d}\right|\gg 2\,M^{2}$ one has
\begin{equation}
\label{eq: R16.17}
m^{2}_{H^{0}}\simeq m^{2}_{u}+m^{2}_{A}+M^{2}_{Z},\quad 
m^{2}_{h^{0}}\simeq m^{2}_{d}+m^{2}_{A}+M^{2}_{Z}.
\end{equation}  
In generic, if $\left|m^{2}_{u}-m^{2}_{d}\right| \leq 2\,M^{2}$ then the 
lower bound of $m^{2}_{h^{0}}$ can be written
\begin{equation}
\label{eq: R16.18}
m^{2}_{h^{0}}\geq \FFr{1}{2}\,\left[m^{2}_{u}+m^{2}_{d}-
(\sqrt{2}-1) \left(m^{2}_{A}+M^{2}_{Z}\right)\right],
\end{equation}   
and if $\left|m^{2}_{u}-m^{2}_{d}\right| \geq 2\,M^{2}$, then 
\begin{equation}
\label{eq: R16.19}
m^{2}_{h^{0}}\geq \FFr{1}{2}\,\left[m^{2}_{A}+M^{2}_{Z}+
m^{2}_{d}\,(1+\sqrt{2})- 
m^{2}_{u}\,(\sqrt{2}-1)\right].
\end{equation}   
In the same manner one gets the eigenvalues of the $M^{2}_{\pm}$
\begin{equation}
\label{eq: R16.20}
m^{2}_{G^{\pm}}=0,\quad m^{2}_{H^{\pm}}=M^{2}_{W}+m^{2}_{A}.
\end{equation}   
It is important to mention  that unlike the conventional MSSM models 
a suggested VMSM does not predict the existence of any light neutral 
Higgs boson $h^{0}$ in the supersymmetric two-doublets Higgs sector. 
In contrary, both of the $H^{0}$ and $h^{0}$ bosons have enough 
large masses, which is straightforward to see from the estimates 
eq.(16.15)-eq.(16.19). Especially in the case of particular interest 
$m^{2}_{u}\sim m^{2}_{d}\gg M^{2}_{Z}$, the  $H^{0}$ and $h^{0}$ are 
much heavier than the $Z^{0}$ boson. Furthermore, it is well known fact 
for the Higgs bosons that the one-loop radiative corrections for some 
of their masses can push up the upper bound even significantly. 

\section{Three solid testable predictions of VMSM}
\label {impl}
Discussing now the relevance of our present approach to the physical 
realities we should attempt to provide some ground for checking the 
predictions of the VMSM against experimental evidence. It is remarkable that
the resulting theory makes plausible following three 
testable implications for the current experiments at LEP2, at the Tevatron and
LHC discussed below, which are drastically different from the predictions of 
conventional models:\\
1. At first recall that, in conventional scenario there is a great 
belief for the $h^{0}$ boson having the mass 
$m_{h^{0}}\leq M_{Z}$, which would be 
the only Higgs boson that can be discovered at the next round of colliders. 
Therefore, searches for this boson (if the mass is below 
$150$ GeV or so [39]) would be one of the crucial points in 
testing of MSSM models in particular, as well of conventional SUSY in general. 
This prediction remains one of characteristic features of such theories 
and it even holds in the limit that all masses of the supersymmetric 
particles are sent to infinity when one recovers the non-supersymmetric 
Standard Model. In the same time searches for this boson will 
invalidate the whole framework of VMSM  or will serve as the direct 
indication of 
its validity, where, in contrast to the formers, we are led to reject 
such an expectation due to the specific mechanism 
of the electroweak symmetry breaking. While the two important 
phenomenological implication of the MSM given in the subsec.12.14 just are 
the first two testable predictions of the VMSM for the current experiments. 
\\
2. It is well known that once $SU(2)_{L}\otimes U(1)_{Y}$ is broken, the 
fields with different $SU(2)_{L}\otimes U(1)_{Y}$ quantum numbers can mix 
if they have the same $SU(3)_{c}\otimes U(1)_{em}$ quantum numbers. Such a 
phenomenon occurs in the sfermion sector of the M\rlap/SMSM. If one ignore 
mixing between sfermions of different generations but will include the 
mixing between $SU(2)$ doublet and singlet sfermions then the sfermion mass 
matrix decomposes into a series of $2\times 2$ matrices of the sfermions of a 
given flavour. The charginos are mixtures of the charged Higgsinos and the 
charged gauginos, and neutralinos are the mixture of neutral Higgsinos and 
the neutral gauginos, etc.. We can readily obtain the resulting explicit 
forms of corresponding mass matrices within standard technique. But shall 
forbear to write them out here as the sfermions are no longer of 
consequence for discussion of the final fields defined on $M_{4}$. The 
sparticles could never emerge in $M_{4}$ (eq.16.1) and will be of no 
interest for the future experiments. By this we arrived to the second 
principle point of drastic deviation of M\rlap/SMSM from the conventional 
MSSM models. In MSSM models as well in any conventional SUSY theory the 
supersymmetry was implemented in the Minkowski space $M_{4}$ by adding a 
new four odd dimensions, and there are two major motivation for SUSY to be 
realized in the TeV range, i.e., the masses of sparticles are of the 
order of a few TeV or less. First one is a solution of the hierarchy 
problem, when in order to introduce no new fine-tunning all soft terms 
should be of the same order of magnitude at most in the TeV range-weak 
scale (e.g. [65]). The second motivation for low energy SUSY comes from the 
view point of gauge unification (a supersymmetric GUT). Since the
current experiments at LEP2, at the Tevatron and at LHC will explore this 
energy range, then, the second great expectation of such theories arise that 
at least some of the sparticles can be found and their parameters like 
masses and coupling constants will also be measured (the precise 
measurements). Reflecting upon the 
results far obtained here, in a strong contrast to such theories the 
unbroken MW-SUSY is implemented on the MW-SMM:\, $SG_{N}$ by, at first, 
lifting up $\G1_{\eta}\rightarrow \SG1_{\eta}$ 
and consequently making an inverse 
passage to the $\rlap/SG_{N}$ \,\,($\SG1_{\eta}\rightarrow\G1_{\eta}$) 
on which the resulting theory M\rlap/SMSM is defined (sec.16). 
Applying the final passage $(\G1_{\eta}\rightarrow M_{4})$ 
(eq.(16.1)) we arrive to the final VMSM, where only the particles will survive 
on the $M_{4}$ at the real physical limit under the R-parity conservation 
(eq.(16.1)). Then, 

$\bullet$ {\em all the sparticles 
never could emerge in the $M_{4}$ neither at TeV range nor at any energy 
range at all}. \\
From the view point of achieving the final potentially 
realistic supersymmetric field theory this will be third crucial test in 
experiments above for verifying the efforts made either in MSSM model 
building (the conventional SUSY theories) or in suggested VMSM 
(the MW-SUSY), which are based on two quite different approaches. 
To sum up the discussion thus far, we have argued that, in strong contrast 
to conventional SUSY theories, if the VMSM given here proves 
viable it becomes an crucial issue to hold in experiments the 
above-mentioned three tests.

\renewcommand{\theequation}{\thesubsection.\arabic{equation}}
\subsection{Quark flavour mixing and the Cabibbo angles}
\label{Cabib}
An implication of quark generations into general scheme will be carried out
in the same way of the leptons (subsec.12).
But before proceeding further that it is profitable to enlarge it by the 
additional assumption without asking the reason behind it:

$\bullet$ The MW components imply
\begin{equation}
\label{eq: R16.1.3}
{}^{i}{\bar{\ps1_{u}}}^{A}(\cdots,\theta_{i_{1}},\cdots\theta_{i_{n}},\cdots)\,\,
{}^{j}{\ps1_{u}}^{B}(\cdots,\theta_{i_{1}},\cdots\theta_{i_{n}},\cdots)=
\delta_{ij}\S_{l=i_{1},\ldots,i_{n}}f^{AB}_{il}{}\,\,^{i}\left(
\bar{q}_{l}q_{l}\right),
\end{equation}
namely, the contribution of each individual subquark ${}^{i}q_{l}$, into the 
component of given world ($i$) is determined by the 
{\em partial formfactor} $f^{AB}_{il}$.
Under the group $SU(2)\otimes U(1)$ the 
left-handed quarks transform as three doublets, while the right-handed
quarks transform as independent singlets except of following 
differences:\\
1. The values of weak-hypercharge of quarks are changed due to their
fractional electric charges
$
q_{L}:Y^{w}=\FFr{1}{3},\quad u_{R}:Y^{w}=\FFr{4}{3},\quad
d_{R}:Y^{w}=-\FFr{2}{3}$
etc.\\
2. All Yukawa coupling constants have nonzero values.\\
3. An appearance of quark mixing and Cabibbo angle, which
is unknown in the scope of standard model.\\
4. An existence of CP-violating phase in unitary matrix of quark
mixing. We shall discuss it in the next 
section.\\
In [2] we attempt to give an explanation to quark mixing and Cabibbo angle.
We consider this problem, for simplicity, on the example of four 
quarks $u,d,s,c$. The further implication of all quarks would complicate 
the problem only in algebraic sense.
Instead of mixing of the $d'$ and $s'$ it is convenient to consider a quite 
equivalent mixing of $u'$ and $c'$.
Similar formulas can be worked out for the other mixings.
Hence, the nonzero value of Cabibbo angle arises due to nonzero coupling
constant $f_{u'c'}$. The problem is to calculate all coupling
constants $f_{u'c'}$,$f_{c't'}$, and $f_{t'u'}$ generating three Cabibbo 
angles
$$
\tan 2\theta_{3}=\FFr{2f_{u'c'}}{f_{c'}-f_{u'}},\quad
\tan 2\theta_{1}=\FFr{2f_{c't'}}{f_{t'}-f_{c'}},\quad
\tan 2\theta_{2}=\FFr{2f_{t'u'}}{f_{u'}-f_{t'}}.
$$
Since the Q-components of the quark fields $u',c'$ 
and $t'$ contain at least one identical subquark, 
the partial formfactors $\bar{f}_{i}$, as well then all coupling
constants, acquire nonzero values causing a
quark mixing with the Cabibbo angles ([2]).
The resulting expressions are as follows:
\begin{equation}
\label{eq: R21.16}
\begin{array}{l}
\tan 2\theta_{3}=
\FFr{\bar{f}_{3}
\left( \bar{\Sigma}{}_{Q\,u}^{2}+\bar{\Sigma}{}_{Q\,u}^{3}+
\bar{\Sigma}{}_{Q\,c}^{3}+\bar{\Sigma}{}_{Q\,c}^{1}\right)}
{\left(\bar{\Sigma}{}_{Q\,c}^{3}+\bar{\Sigma}{}_{Q\,c}^{1}-
\bar{\Sigma}{}_{Q\,u}^{2}-\bar{\Sigma}{}_{Q\,u}^{3}\right)},\quad
\tan 2\theta_{1}=
\FFr{\bar{f}_{1}
\left( \bar{\Sigma}{}_{Q\,c}^{3}+\bar{\Sigma}{}_{Q\,c}^{1}+
\bar{\Sigma}{}_{Q\,t}^{1}+\bar{\Sigma}{}_{Q\,t}^{2}\right)}
{\left(\bar{\Sigma}{}_{Q\,t}^{1}+\bar{\Sigma}{}_{Q\,t}^{2}-
\bar{\Sigma}{}_{Q\,c}^{3}-\bar{\Sigma}{}_{Q\,c}^{1}\right)},\\
\tan 2\theta_{2}=
\FFr{\bar{f}_{2}
\left( \bar{\Sigma}{}_{Q\,t}^{1}+\bar{\Sigma}{}_{Q\,t}^{2}+
\bar{\Sigma}{}_{Q\,u}^{2}+\bar{\Sigma}{}_{Q\,u}^{3}\right)}
{\left(\bar{\Sigma}{}_{Q\,u}^{2}+\bar{\Sigma}{}_{Q\,u}^{3}-
\bar{\Sigma}{}_{Q\,t}^{1}-\bar{\Sigma}{}_{Q\,t}^{2}\right)}.
\end{array}
\end{equation}
Therefore, the unimodular orthogonal group of global rotations arises, and the quarks 
$u',c'$ and $t'$ come up in doublets 
$(u',c')$,$(c',t')$, and $(t',u')$. For the leptons
these formfactors equal zero 
$\bar{f}_{i}^{lept}\equiv 0$, because of
eq.(12.4.1), namely the lepton mixing is absent.
In conventional notation 
$
\left(\matrix{
u'\cr
d\cr}\right)_{L}, 
\left(\matrix{
c'\cr
s\cr}\right)_{L},
\left(\matrix{
t'\cr
b\cr}\right)_{L}\rightarrow
\left(\matrix{
u\cr
d'\cr}\right)_{L}, 
\left(\matrix{
c\cr
s'\cr}\right)_{L},
\left(\matrix{
t\cr
b'\cr}\right)_{L},
$
which gives rise to
$
f_{u'c'}\rightarrow f_{d's'},\quad
f_{c't'}\rightarrow f_{s'b'},\quad
f_{t'u'}\rightarrow f_{b'd'},\quad
f_{u'}\rightarrow f_{d'},\quad 
f_{c'}\rightarrow f_{s'},\quad 
f_{t'}\rightarrow f_{b'},\quad 
f_{d}\rightarrow f_{u},\quad
f_{s}\rightarrow f_{c},\quad
f_{b}\rightarrow f_{t}.
$
\subsection{The CP-violating phase}
\label{Phase}

The required magnitude of the CP-violating complex parameter $\varepsilon$
depends upon the specific choice of theoretical model for
explaining the $ K^{0}_{2}\rightarrow 2\pi $ decay [130, 131].
From the experimental data it is somewhere
$
| \varepsilon |\simeq 2.3\times 10^{-3}.
$
In the framework of Kobayashi-Maskawa (KM) parametrization of unitary matrix
of quark mixing [132], this parameter may be expressed in terms of 
three Eulerian angles of global rotations in the three dimensional quark 
space and one phase parameter. We attempt to derive the 
KM-matrix with an explanation given to an appearance of the CP-violating 
phase ([2]).
Recall that during the realization of MW- structure 
the P-violation compulsory occurred in the W-world provided by the spanning
eq.(12.8.1). The three dimensional effective space 
$W^{loc}_{v}(3)$ arises as follows:
\begin{equation}
\label{eq: R22.3}
\begin{array}{l}
W^{loc}_{v}(3)\ni q^{(3)}_{v}=
\left( \matrix{
q^{w}_{R}(\vec{T}=0)\cr
\cr
q^{w}_{L}(\vec{T}=\FFr{1}{2})\cr
}\right)\equiv\\ \\
\equiv
\left( \matrix{
u_{R},d_{R}\cr
\cr
\left( \matrix{
u'\cr
d\cr
}\right)_{L}
\cr
}\right),
\left( \matrix{
c_{R},s_{R}\cr
\cr
\left( \matrix{
c'\cr
s\cr
}\right)_{L}
\cr
}\right),
\left( \matrix{
t_{R},b_{R}\cr
\cr
\left( \matrix{
t'\cr
b\cr
}\right)_{L}
\cr
}\right)
\equiv
\left( \matrix{
q^{w}_{3}\cr
\cr
\left( \matrix{
q^{w}_{1}\cr
q^{w}_{2}\cr
}\right)
\cr
}\right),
\left( \matrix{
q^{w}_{1}\cr
\cr
\left( \matrix{
q^{w}_{2}\cr
q^{w}_{3}\cr
}\right)
\cr
}\right),
\left( \matrix{
q^{w}_{2}\cr
\cr
\left( \matrix{
q^{w}_{3}\cr
q^{w}_{1}\cr
}\right)
\cr
}\right),
\end{array}
\end{equation}
where the subscript $(v)$ formally specifies a vertical direction of
multiplet, the subquarks $q^{w}_{\alpha} (\alpha=1,2,3)$ associate with 
the local rotations around corresponding axes of three dimensional
effective space $W^{loc}_{v}(3)$. The local gauge transformations
$f^{v}_{exp}$ are implemented upon the multiplet 
${q'}^{(3)}_{v}=f^{v}_{exp} q^{(3)}_{v}$, where 
$f^{v}_{exp}\in SU^{loc}(2)\otimes U^{loc}(1)$. 
If for the moment we leave it intact and make a closer examination of the 
content of the middle row in eq.(17.2.1), then we distinguish the other 
symmetry arisen along the horizontal line $(h)$. 
Hence, we may expect a situation similar to those of subsec.12.8 will be held
in present case. The procedure just explained therein can be followed again.
We have to realize that due to the specific structure of 
W-world implying the condition of realization of the MW connections 
eq.(12.2.5) with $\vec{T}\neq 0, \quad Y^{w}\neq 0$, the subquarks 
$q^{w}_{\alpha}$ tend to be compulsory involved into triplet. They form
one ``doublet''  $\vec{T}\neq 0$ and one singlet $Y^{w}\neq 0$. Then the
quarks $u'_{L},c'_{L}$ and $t'_{L}$ form
a $SO^{gl}(2)$ ``doublet'' and a $U^{gl}(1)$ singlet 
\begin{equation}
\label{eq: R22.4}
\begin{array}{l}
\left( \left( u'_{L},c'_{L}\right)t'_{L}\right)\equiv
\left( \left( q^{w}_{1},q^{w}_{2}\right)q^{w}_{3}\right)\equiv
q^{(3)}_{h}\in W^{gl}_{h}(3),\\ 
\left( u'_{L},\left( c'_{L},t'_{L}\right)\right)\equiv
\left( q^{w}_{1},\left( q^{w}_{2},q^{w}_{3}\right)\right),\quad
\left( \left( t'_{L},u'_{L}\right)c'_{L}\right)\equiv
\left( \left( q^{w}_{3},q^{w}_{1}\right),q^{w}_{2}\right).
\end{array}
\end{equation}
Here $W^{gl}_{h}(3)$ is the three dimensional effective space in
which the global rotations occur. They are implemented upon the
triplets through the transformation matrix $f^{h}_{exp}$:\\ 
$
{q'}^{(3)}_{h}=f^{h}_{exp} q^{(3)}_{h},
$
which reads (eq.(17.2.2))
$$
f^{h}_{exp}=\left( \matrix{
f_{33} &0 &0\cr
0 & c & s\cr
0 & -s & c\cr
}\right)
$$
in the notation $c=\cos \theta, \quad s=\sin \theta$. This 
implies the incompatibility relation eq.(2.5.2), namely
\begin{equation}
\label{eq: R22.7}
\|f^{h}_{exp}\|=f_{33}(f_{11}f_{22}-f_{12}f_{21})=
f_{33}\varepsilon_{123}\varepsilon_{123}\|f^{h}_{exp}\|f^{*}_{33}.
\end{equation}
That is
$
f_{33}f^{*}_{33}=1,
$
or
$
f_{33}=e^{i\delta}
$
and $\|f^{h}_{exp}\|=1$.
The general rotation in $W^{gl}_{h}(3)$ is described by Eulerian three angles
$\theta_{1},\theta_{2},\theta_{3}$. If we put the arisen phase only in the 
physical sector then a final KM-matrix of quark flavour mixing would result.
The CP-violating parameter $\varepsilon$ 
approximately is written [130]
$
\varepsilon\sim s_{1}s_{2}s_{3}\sin \delta\neq 0.
$
Thus, while the spanning
$W^{loc}_{v}(2)\rightarrow W^{loc}_{v}(3)$ eq.(17.2.1) underlies 
the P-violation and the expanded symmetry
$G^{loc}_{v}(3)=SU^{loc}(2)\otimes U^{loc}(1)$, the 
CP-violation stems from the similar
spanning $W^{gl}_{h}(2)\rightarrow W^{gl}_{h}(3)$ eq.(17.2.2) 
with the expanded global symmetry group.

\subsection{The mass-spectrum of leptons and quarks}
\label{Mass}
The mass-spectrum of leptons and quarks
stems from their internal MW-structure eq.(12.4.1) and eq.(12.5.1)
incorporated with the quark mixing eq.(17.1.2). We start a discussion 
with the leptons. It is worthwhile 
to adopt a simple viewpoint on Higgs sector.
Following the subsec.12.14, the explicit expressions of the lepton masses 
read
$m_{i}=\FFr{\eta}{\sqrt{2}}f_{i}$ and 
$m_{i}^{\nu}=\FFr{\eta}{\sqrt{2}}f_{i}^{\nu}$, that 
$m_{e}:m_{\mu}:m_{\tau}=f_{e}:f_{\mu}:f_{\tau}=
L_{1}^{2}:L_{2}^{2}:L_{3}^{2}$ provided by 
$L_{1}^{2}=\FFr{m_{i}}{M}$ and $\sqrt{M}=\S_{i}\sqrt{m_{i}}$. Thus, 
$L_{1}=(8.9;7.8)\times 10^{-3},\quad L_{2}= (0.13; 0.11),\quad L_{3}=
(0.9;0.88)$. \\
Taking into account the 
eq.(12.5.1) and eq.(17.1.2) the coupling constants of the quarks $d, s$ 
and $b$ can be written
\begin{equation}
\label{eq: R23.8}
\begin{array}{l}
f_{d}=L_{1}\,tr(\rho_{d}\,\Sigma_{Q})\equiv L_{1}\,\widetilde{f}_{d}, \quad
f_{s}=L_{2}\,tr(\rho_{s}\,\Sigma_{Q})\equiv L_{2}\,\widetilde{f}_{s}, \quad
f_{b}=L_{3}\,tr(\rho_{b}\,\Sigma_{Q})\equiv L_{3}\,\widetilde{f}_{b}, \\
\rho_{d}=\rho^{Q}\rho_{d}^{B},\quad
\rho_{s}=\rho^{Q}\rho_{s}^{B}\rho^{s},\quad
\rho_{b}=\rho^{Q}\rho_{b}^{B}\rho^{b}.
\end{array}
\end{equation}
Hence
$m_{d}=\FFr{\eta}{\sqrt{2}}f_{d}, \quad
m_{s}=\FFr{\eta}{\sqrt{2}}f_{s}, \quad m_{b}=\FFr{\eta}{\sqrt{2}}f_{b}, \quad$
and 
$m_{d}:m_{s}:m_{b}=(L_{1}\,\widetilde{f}_{d}):(L_{2}\,\widetilde{f}_{s}):
(L_{3}\,\widetilde{f}_{b}).$
According to the sec.17, we derive the masses of the $u,c$ and $t$ quarks
\begin{equation}
\label{eq: R23.8}
\begin{array}{l}
m_{u}=\FFr{\eta}{\sqrt{2}}\left\{ 
\left( \bar{\Sigma}{}^{2}_{Q\,u}+ \bar{\Sigma}{}^{3}_{Q\,u} \right)\cos^{2} 
\theta_{3}
+\left( \bar{\Sigma}{}^{3}_{Q\,c}+ \bar{\Sigma}{}^{1}_{Q\,c}\right)\sin^{2} 
\theta_{3} -\FFr{\bar{f}_{3}}{2}
\left(  \bar{\Sigma}{}^{2}_{Q\,u}+ \bar{\Sigma}{}^{3}_{Q\,u}+ 
\right.\right.\\ \left.\left.
\bar{\Sigma}{}^{3}_{Q\,c}+ \bar{\Sigma}{}^{1}_{Q\,c}\right)
\sin 2\theta_{3}
\right\}=\FFr{\eta}{\sqrt{2}}\left\{ 
\left( \bar{\Sigma}{}^{2}_{Q\,u}+ \bar{\Sigma}{}^{3}_{Q\,u} \right)\cos^{2} 
\theta_{2}
+\left( \bar{\Sigma}{}^{1}_{Q\,t}+ \bar{\Sigma}{}^{2}_{Q\,t}\right)\sin^{2} 
\theta_{2} +
\right.\\ \left.
\FFr{\bar{f}_{2}}{2}
\left( \bar{\Sigma}{}^{1}_{Q\,t}+ \bar{\Sigma}{}^{2}_{Q\,t} +
\bar{\Sigma}{}^{2}_{Q\,u}+ \bar{\Sigma}{}^{3}_{Q\,u}+ \right)
\sin 2\theta_{2}
\right\},
\end{array}
\end{equation}
\begin{equation}
\label{eq: R23.8}
\begin{array}{l}
m_{c}=\FFr{\eta}{\sqrt{2}}\left\{ 
\left( \bar{\Sigma}{}^{2}_{Q\,u}+ \bar{\Sigma}{}^{3}_{Q\,u} \right)\sin^{2} 
\theta_{3}
+\left( \bar{\Sigma}{}^{3}_{Q\,c}+ \bar{\Sigma}{}^{1}_{Q\,c}\right)\cos^{2} 
\theta_{3} +\FFr{\bar{f}_{3}}{2}
\left(  \bar{\Sigma}{}^{2}_{Q\,u}+ \bar{\Sigma}{}^{3}_{Q\,u}+ 
\right.\right.\\ \left.\left.
\bar{\Sigma}{}^{3}_{Q\,c}+ \bar{\Sigma}{}^{1}_{Q\,c}\right)
\sin 2\theta_{3}
\right\}=\FFr{\eta}{\sqrt{2}}\left\{ 
\left( \bar{\Sigma}{}^{3}_{Q\,c}+ \bar{\Sigma}{}^{1}_{Q\,c} \right)\cos^{2} 
\theta_{1}
+\left( \bar{\Sigma}{}^{1}_{Q\,t}+ \bar{\Sigma}{}^{2}_{Q\,t}\right)\sin^{2} 
\theta_{1} -
\right.\\ \left.
\FFr{\bar{f}_{1}}{2}
\left( \bar{\Sigma}{}^{3}_{Q\,c}+ \bar{\Sigma}{}^{1}_{Q\,c} +
\bar{\Sigma}{}^{1}_{Q\,t}+ \bar{\Sigma}{}^{2}_{Q\,t}+ \right)
\sin 2\theta_{1}
\right\},
\end{array}
\end{equation}
\begin{equation}
\label{eq: R23.8}
\begin{array}{l}
m_{t}=\FFr{\eta}{\sqrt{2}}\left\{ 
\left( \bar{\Sigma}{}^{1}_{Q\,t}+ \bar{\Sigma}{}^{2}_{Q\,t} \right)\cos^{2} 
\theta_{1}
+\left( \bar{\Sigma}{}^{3}_{Q\,c}+ \bar{\Sigma}{}^{1}_{Q\,c}\right)\sin^{2} 
\theta_{1} +\FFr{\bar{f}_{1}}{2}
\left(  \bar{\Sigma}{}^{1}_{Q\,t}+ \bar{\Sigma}{}^{2}_{Q\,t}+ 
\right.\right.\\ \left.\left.
\bar{\Sigma}{}^{3}_{Q\,c}+ \bar{\Sigma}{}^{1}_{Q\,c}\right)
\sin 2\theta_{1}
\right\}=\FFr{\eta}{\sqrt{2}}\left\{ 
\left( \bar{\Sigma}{}^{1}_{Q\,t}+ \bar{\Sigma}{}^{2}_{Q\,t} \right)\cos^{2} 
\theta_{2}
+\left( \bar{\Sigma}{}^{2}_{Q\,u}+ \bar{\Sigma}{}^{3}_{Q\,u}\right)\sin^{2} 
\theta_{2} -
\right.\\ \left.
\FFr{\bar{f}_{2}}{2}
\left( \bar{\Sigma}{}^{1}_{Q\,t}+ \bar{\Sigma}{}^{2}_{Q\,t} +
\bar{\Sigma}{}^{2}_{Q\,u}+ \bar{\Sigma}{}^{3}_{Q\,u}+ \right)
\sin 2\theta_{2}
\right\}.
\end{array}
\end{equation}

\section{The physical outlook and concluding remarks}
\label{Conc}
The physical outlook on suggested approach and concluding remarks are given 
in this section in order to resume once again a whole 
physical picture and to provide a sufficient background for its understanding 
without undue hardship. 
To complete the MSM [1,2], here we attempted to develop its realistic, 
viable, minimal SUSY extension in order to solve the zero point energy and 
hierarchy problems standing before it. 

$\bullet$
Our scheme based on the OM formalism(sec.2), which is
the mathematical framework for our physical outlook embodied in the idea 
that the geometry and fields, with the internal symmetries and all 
interactions, as well the four major principles of relativity (special and 
general), quantum, gauge and colour confinement, are derivative. They come 
into being simultaneously in the stable system of the underlying 
``primordial structures'' involved in the ``linkage'' establishing 
processes. 
The OM formalism is the generalization of secondary 
quantization of the field theory with appropriate expansion over the 
geometric objects leading to the quantization of geometry different from all 
existing schemes.

$\bullet$
We have chosen a simple setting and considered the primordial  
structures, which are designed to posses certain physical properties 
satisfying the general rules stated briefly in subsec.2.3, and have involved 
in the linkage establishing processes. The processes of their creation and 
annihilation in the lowest state (the regular structures) just are described 
by the OM formalism. In all the higher states the primordial structures are 
distorted ones, namely they have undergone the distortion transformations 
(subsec.2.4). 
These transformations yield the ``quark'' and ``antiquark'' 
fields defined on the simplified geometry (one $u$-channel) given in the 
subsec.2.4, and skeletonized for illustrative purposes. Due to geometry 
realization conditions held in the stable systems of primordial structures 
they emerge in confined phase. This scheme still should be 
considered as the preliminary one, which is further elaborated in the 
subsec.3.2 to get the physically more realistic picture.

$\bullet$
The distortion transformation 
functions are the operators acting in the space of the internal degrees of 
freedom (colours) and imply the incompatibility relations eq.(2.5.2), which 
hold for both the local and the global distortion rotations. They underly the 
most important symmetries such as the internal symmetries $U(1), SU(2), 
SU(3)$, the $SU(2)\otimes U(1)$ symmetry of electroweak interactions, etc. 
(see sec.12, 17).
We generalize the OM formalism via the concept of the OMM yielding the
MW geometry involving the spacetime continuum and the internal worlds of the 
given number. In an enlarged framework of the OMM we define and 
clarify the conceptual basis of subquarks and their
characteristics stemming from the various symmetries of the internal worlds. 
They imply subcolour confinement and gauge principle (subsec.3.2). 
By this we have arrived at an entirely satisfactory answer to the question of 
the physical origin of 
the geometry and fields, the internal symmetries and interactions, as well 
the principles of relativity, quantum, gauge and subcolour confinement.
The value of the present version of hypothesis of existence of the
MW-structures defined on the MW-geometry resides in solving of some key 
problems of the SM, wherein we attempt to suggest a microscopic approach to 
the properties of particles and interactions.

$\bullet$ We derive the MW-SUSY (sec.8,9), which has an algebraic origin in 
the sense that it has arisen from the subquark algebra defined on the internal 
worlds, while the nilpotent supercharge operators are derived (sec.4). 
Therefore, the MW-SUSY realized only on the internal worlds but not on the 
spacetime continuum. 
Thus, it cannot be an exact symmetry of nature and has to be realized in its 
broken phase (sec.9). Our purpose above is much easier to handle, by 
restoring in the first the ``exact'' MW-SUSY. It can be achieved by lifting 
up each sparticle to corresponding particle state (sec.9). This enables the 
sparticle to be included in the same supermultiplet with corresponding 
particle. Due to different features of particles and sparticles when passing 
back to physically realistic limit eq.(9.11) one must have always to 
distinguish them by introducing an additional discrete internal symmetry, 
i.e., the multiplicative $Z_{2}\quad R$-parity (sec.9). 

$\bullet$ We write then the most 
generic renormalizable MW-SUSY action eq.(11.17) involving gauge and 
supersymmetric matter frame fields, and, thus, the corresponding generating 
functional. Therein, we are led to the principal point of drastic change of 
the standard SUSY scheme to specialize the superpotential to be in such a 
form eq.(11.28)-eq.(11.34), which enables the microscopic approach to
the key problems of particle phenomenology (sec.12).

$\bullet$
Within this approach, due to the symmetry of Q-world of electric charge
the condition of realization of the MW connections has arisen embodied in 
the Gell-Mann-Nishijima relation (subsec.12.2). 
For the MW-structures the symmetries of corresponding internal worlds are 
unified into higher symmetry including also the operators of isospin and 
hypercharge. We conclude that the
possible three lepton generations consist of six lepton fields
with integer electric and leptonic charges being free of confinement
condition (subsec.12.4). As well, the three quark 
generations exist composed of six possible quark fields (subsec.12.5), which
carry fractional electric and baryonic charges realized in the 
confined phase. The global group unifying all global symmetries of the 
internal worlds of quarks is the flavour group $SU_{f}(6)$ (subsec.12.3).
The whole complexity of leptons, quarks and other composite particles, 
and their interactions arises from the MW-frame
field, which has nontrivial MW internal structure
and involves nonlinear fermion self-interaction of the components. 
This Lagrangian contains only two major free parameters, which are the 
coupling constants of nonlinear fermion and gauge interactions (subsec 12.6). 
To realize the MW-connections of the weak interacting fermions the 
P-violation compulsory occurred in W-world (subsec.12.8) incorporated with 
the gauge symmetry reduction. It has characterized by the Weinberg 
mixing angle with the fixed value at $30^{0}$. This gives rise to 
the local symmetry $SU(2)\otimes U(1)$, under which the
left-handed fermions transformed as six independent doublets, 
while the right-handed fermions transformed as twelve independent singlets. 

$\bullet$
Due to vacuum rearrangement in Q-world the Yukawa couplings 
arise between the fermion fields and corresponding isospinor-scalar 
$H$-mesons in conventional form. In the framework of suggested specific 
mechanism providing the effective attraction between the relativistic 
fermions caused by the exchange of the mediating induced gauge quanta in the 
W-world, the two complex self-interacting isospinor-scalar Higgs doublets 
$\left(H_{u},\,H_{d}\right)$ as well as their spin-$\FFr{1}{2}$ SUSY 
partners $\left(\widetilde{H}_{u},\,\widetilde{H}_{d}\right)$ Higgsinos  
arise as the Bose-condensate. Taking into account this slight difference 
from the MSM arisen in the field content of MSMSM in the Higgs sector the 
supersymmetric Lagrangian eq.(11.35) now also contains  these fields coupled 
to the gauge fields in a gauge invariant way, when the 
symmetry-breaking Higgs bosons are counted off from the gap 
symmetry-restoring value as the point of origin (subsec.12.13).

$\bullet$
The Higgs mechanism does work in the following way:
Before the symmetry was broken in the $W$-world, the $2$ complex 
$SU(2)_{L}$ Higgs doublets had $8$ degrees of freedom. 
Three of them were the would-be Nambu-Goldstone 
bosons $G^{0},\,G^{\pm},$ which were absorbed to give rise the longitudinal 
modes of the massive $W$-components of the $Z^{0}$ and $W^{\pm}$ vector 
bosons, which simultaneously give rise the corresponding $x$- components 
too, leaving $5$ physical degrees of freedom. 
The latter consists of a charged 
Higgs boson pairs $H^{\pm},$ a CP-odd neutral Higgs boson $A^{0}$, and 
CP-even neutral Higgs bosons $h^{0}$ and $H^{0}$. 
The mass eigenstates and would-be Nambu-Goldstone bosons are made of the 
original gauge-eigenstate fields, where the physical pseudoscalar Higgs 
boson $A^{0}$ is made of from the imaginary parts of $h^{0}_{u}$ and 
$h^{0}_{d}$, and is orthogonal to $G^{0}$; while the neutral scalar Higgs 
bosons are mixtures of the real parts of $h^{0}_{u}$ and $h^{0}_{d}$. 
The mass of any physical Higgs boson that is SM-like is strictly limited, as 
are the radiative corrections to the quartic potential terms. 
We calculated the tree-level masses for these Higgs states (sec.16) and shown
that the $h^{0}$ Higgs boson arisen in the internal $W$-world is much 
heavier of that $Z^{0}$ boson.

$\bullet$
In contrast to the SM, the suggested microscopic approach predicts the 
electroweak symmetry breakdown in the $W$-world by the VEV of spin zero 
Higgs bosons and the transmission of electroweak symmetry breaking from the 
$W-$world to the $M_{4}$ spacetime continuum (subsec.12.14). 
The resulting Lagrangian of unified electroweak 
interactions of leptons and quarks ensues, which 
in lowest order approximation
leads to the Lagrangian of phenomenological SM. In general,
the self-energy operator underlies the Yukawa coupling constant, which takes 
into account a mass-spectrum of all expected collective excitations of bound 
quasi-particle pairs. If the MSM proves viable it becomes an crucial 
issue to hold in experiments the two testable predictions given in 
subsec.12.14.

$\bullet$
The realistic generating functional should be derived by passing back to the 
physical limit eq.(9.11). Such a breaking of the MW-SUSY can be implemented 
by subtracting back all the explicit soft mass terms formerly introduced for 
the sparticles eq.(15.2). These terms do not reintroduced the quadratic 
diagrams which motivated the introduction of SUSY framework. Therewith, the 
boson-fermion cancellation in the above-mentioned problems can be regarded 
as a consequence of a constraint stemming from holomorphy of the observables, 
therefore it will be held at the limit eq.(9.11) too. Thus, we extract the 
pertinent piece containing only the $\eta$-field components and then in 
afterwards pass to $M_{4}$ (subsec.2.1) to get the final VMSM yielding the 
realistic particle spectrum. 

$\bullet$
Thus, if the VMSM proves viable it becomes an crucial issue to hold 
in the experiments at LEP2 and at the Tevatron three testable solid 
implications given in sec.17, which are drastically different from those of 
conventional MSSM models.

$\bullet$
The implication of quarks into the VMSM is carried out in the same way
of leptons except that of appearance of quark mixing with Cabibbo angle 
(subsec.17.1) and the existence of CP-violating complex phase in unitary 
matrix of quark mixing (subsec.17.2). 
The Q-components of the quarks contain at least 
one identical subquark, due to which the partial formfactors 
gain nonzero values. This underlies the quark mixing with 
Cabibbo angles. In lepton's case these formfactors
are vanished  and lepton mixing is absent.
The CP-violation stems from the spanning eq.(17.2.2).
Adopting a simple viewpoint on Higgs sector the masses of leptons and quarks 
are given in subsec.17.3.
\\\\
We hope that the outlined VMSM, if it proves viable in the  experiments at 
LEP2 and at the Tevatron, will be an attractive basis for the future 
theories. As yet no direct signal has been found in them, the absent of which 
has been cleared up the lower limits on Higgs bosons and sparticles masses.

\vskip 1truecm
\centerline {\bf\large Acknowledgements}
\vskip 0.1\baselineskip
\noindent
This work was supported in part by the exchange program of 
'Jumelage France-Armenie'. It 
is pleasure to thank for their hospitality the 'Centre de 
Physique Theorique' (CNRS Division 7061, Marseille) and 
'D.A.R.C.' (Observatoir de Paris-CNRS). I am grateful to 
R.Triay for useful comments and suggestions, and acknowledge discussions 
with G.Sigle and M.Lemoine.
I'm indebted to A.M.Vardanian and K.L.Yerknapetian
for support.
\section * {Appendix A}
\label {form}
\setcounter{section}{0}
\renewcommand {\theequation}{A.\thesection.\arabic {equation}}
\section{The field equations}
The  state vectors are in the form
$$
\begin{array}{lll}
\chi^{0}(\nu_{1},\nu_{2},\nu_{3},\nu_{4})=
\mid 1,1>^{\nu_{1}}\cdot\mid 1,2>^{\nu_{2}}\cdot
\mid 2,1>^{\nu_{3}}\cdot\mid 2,2>^{\nu_{4}},\\
\nu_{i}= \left\{ \begin{array}{ll}
                   1   & \mbox{if $\nu=\nu_{i}$}\quad  \mbox{for some $i$,} \\
                   0   & \mbox{otherwise},
                   \end{array}
\right. 
\\
\mid\chi_{-}(1)>= \chi^{0}(1,0,0,0),\quad   
\mid\chi_{+}(1)>=\chi^{0}(0,0,0,1),\quad
<\chi_{\pm}(\lambda)\mid\chi_{\pm}(\mu)>=\delta_{\lambda\mu}, \\
\mid\chi_{-}(2)>  = \chi^{0}(0,0,1,0),\quad   
\quad
\mid\chi_{+}(2)>= \chi^{0}(0,1,0,0),\quad
<\chi_{\pm}(\lambda)\mid\chi_{\mp}(\mu)>=0,
\end{array}
$$
provided
$
<\chi_{\pm}\mid A\mid \chi_{\pm}>\equiv
\S_{\lambda}<\chi_{\pm}(\lambda)\mid A\mid \chi_{\pm}(\lambda)>. 
$
The free field defined on
the multimanifold 
$G_{N}=\G1_{\eta}\oplus
\G1_{u_{1}}\oplus \cdots\oplus\G1_{u_{N}}$ is written
$$
\Psi =\ps1_{\eta}(\eta)\ps1_{u}(u),\quad 
\ps1_{u}(u)=\ps1_{u_{1}}(u_{1})\cdots\ps1_{u_{N}}(u_{N}), 
$$
where $\ps1_{u_{i}}$ is  the bispinor
defined on the internal manifold $\G1_{u_{i}}$.
A Lagrangian of free field reads
\begin{equation}
\label{eq: RC.22.1}
\widetilde{L}_{0}(D)=
\FFr{i}{2} \{ \bar{\Psi}_{e}(\zeta)\,
{}^{i}\gamma^{(\lambda,\mu,\alpha)}
{\pr_{i}}{}_{(\lambda,\mu,\alpha)}\Psi_{e}(\zeta)-
{\pr_{i}}{}_{(\lambda,\mu,\alpha)}\bar{\Psi}_{e}(\zeta)\,
{}^{i}\gamma^{(\lambda,\mu,\alpha)}
\Psi_{e}(\zeta) \}.
\end{equation}
We adopt the following conventions:
\begin{equation}
\label{eq: RC.22.2}
\begin{array}{l}
\Psi_{e}(\zeta)=e\otimes\Psi(\zeta)=
\left( \matrix{
1 &1\cr
1 &1\cr
} 
\right)  
\otimes\Psi(\zeta), 
\quad
\bar{\Psi}_{e}(\zeta)=e\otimes \bar{\Psi}(\zeta),\quad
\bar{\Psi}(\zeta)=\Psi^{+}(\zeta)\gamma^{0}, \\ 
{}^{i}\gamma^{(\lambda,\mu,\alpha)}=
{}^{i}{\widetilde{O}}^{\lambda,\mu}
\otimes{\widetilde{\sigma}}^{\alpha}, \quad
{}^{i}{\widetilde{O}}^{\lambda,\mu}=\FFr{1}{\sqrt{2}}
\left( \nu_{i}\xi_{0}\otimes
{\widetilde{O}}^{\mu}+\varepsilon_{\lambda}\xi \otimes{}^{i}
{\widetilde{O}}^{\mu}\right),\\ 
\varepsilon_{\lambda}=\left\{ \matrix{
1 &\lambda=1\cr
-1 &\lambda=2\cr
}\right., \quad
<\nu_{i},\nu_{j}>=\delta_{ij},\quad
\left\{ {}^{i} {\widetilde{O}}^{\lambda},{}^{j}{\widetilde{O}}^{\mu}
\right\}=\delta_{ij}{}^{*}\delta^{\lambda\mu},\\ 
{\widetilde{O}}^{\mu}=\FFr{1}{\sqrt{2}}
\left( \xi_{0}+
\varepsilon_{\mu}\xi \right),\quad
{\widetilde{O}}^{\lambda}=
{}^{*}\delta^{\lambda\mu}{\widetilde{O}}_{\mu}=
{({\widetilde{O}}_{\lambda})}^{+},\quad 
{}^{i}{\widetilde{O}}^{\mu}=\FFr{1}{\sqrt{2}}
\left( \xi_{0i}+
\varepsilon_{\mu}\xi_{i} \right),
\\
{\pr_{i}}{}_{(\lambda,\mu,\alpha)}=\partial/\partial\,{}^{i}\zeta^
{(\lambda,\mu,\alpha)}, \quad 
\xi_{0}=\left( \matrix{
1 &0 \cr
0 &-1\cr
}\right)
\quad
\xi=\left( \matrix{
0 &1 \cr
-1 &0\cr
}\right),\\ 
{\xi_{0}}^{2}=-\xi^{2}=-{\xi_{0i}}^{2}=\xi^{2}_{i}=1,\quad
\{\xi_{0},\xi\}=\{\xi_{0},\xi_{0i}\}=\{\xi_{0},\xi_{i}\}=\\ 
=\{\xi,\xi_{0i}\}=\{\xi,\xi_{i}\}=\{\xi_{0i},\xi_{j}\}_{i\neq j}=
\{\xi_{0i},\xi_{0j}\}_{i\neq j}=\{\xi_{i},\xi_{j}\}_{i\neq j}=0.
\end{array}
\end{equation}
Field equations are written
\begin{equation}
\label{eq: RC.22.3}
\begin{array}{l}
(\hp1_{\eta}-m)\ps1_{\eta}(\eta)=0,\quad       
\bar{\ps1_{\eta}}(\eta)(\hp1_{\eta}-m)=0,\\ 
(\hp1_{u}-m)\ps1_{u}(u)=0, \quad \bar{\ps1_{u}}(u)(\hp1_{u}-m)=0,
\end{array}
\end{equation}
where
\begin{equation}
\label{eq: RC.22.4}
\begin{array}{ll}
\hp1_{\eta}=i\hpr_{\eta}, \quad \hp1_{u}=i\hpr_{u},\quad 
\hpr_{u}={}^{i}\gamma^{(\lambda\alpha)}{\pr_{u_{i}}}{}_{(\lambda\alpha)},
\quad
{\pr_{\eta}}{}_{(\lambda\alpha)}=\partial/\partial\eta^{(\lambda\alpha)},
\quad{\pr_{u_{i}}}{}_{(\lambda\alpha)}=\partial /
\partial u^{(\lambda\alpha)}_{i},\\ 
{}^{i}{\gam_{\eta}}^{(\lambda\alpha)}=
{}^{i}{\widetilde{\O1_{\eta}}}^{\lambda}\otimes
{\widetilde{\sigma}}^{\alpha}=
\nu_{i}\xi_{0}\otimes\gamma^{(\lambda\alpha)}=
\nu_{i}\xi_{0}\otimes{\widetilde{O}}^{\lambda}\otimes
{\widetilde{\sigma}}^{\alpha},\\ 
{}^{i}{\gam_{u}}^{(\lambda\alpha)}=
{}^{i}{\widetilde{\O1_{u}}}^{\lambda}\otimes
{\widetilde{\sigma}}^{\alpha}=
\xi\otimes{}^{i}\gamma^{(\lambda\alpha)}=
\xi\otimes{}^{i}{\widetilde{O}}^{\lambda}\otimes
{\widetilde{\sigma}}^{\alpha},\\ 
\left(\gamma^{(\lambda\alpha)}\right)^{+}=
{}^{*}\delta^{\lambda\tau}\delta^{\alpha\beta}\gamma^{(\tau\beta)}=
\gamma_{(\lambda\alpha)},\quad 
\left({}^{i}{\gam_{u}}{}^{(\lambda\alpha)}\right)^{+}=
-{}^{i}{\gam_{u}}{}_{(\lambda\alpha)}.
\end{array}
\end{equation}

\end{document}